\documentclass[12 pt]{article}
\pdfoutput=1 
\usepackage{mathrsfs}
\usepackage{jheppub}
\usepackage{mathtools,leftindex,tensor}
\usepackage[version=4]{mhchem}
\usepackage{amssymb}
\usepackage[labelfont=bf]{caption}
\usepackage{dsfont}
\usepackage{tikz}
\usetikzlibrary{calc}
\newcommand{\Cont}{\operatorname{Cont}}
\newcommand{\Aut}{\operatorname{Aut}}

\newcommand{\Tr}{\operatorname{Tr}}
\newcommand{\id}{\mathbf{1}}
\DeclareMathOperator{\sn}{sn}
\DeclareMathOperator{\cn}{cn}
\DeclareMathOperator{\dn}{dn}
\DeclareMathOperator{\am}{am}
\newcommand{\eA}{e_1^{\wp}}
\newcommand{\eB}{e_2^{\wp}}
\newcommand{\eAB}{e_3^{\wp}}
\usepackage{amsmath}
\usepackage{bbm}
\usepackage{amsfonts}
\usepackage{orcidlink}
\usepackage{multicol} \usepackage{multirow}
\usepackage{physics}
\usepackage[most]{tcolorbox}
\usepackage{amsmath}
\usepackage{caption}
\usepackage{subcaption,longtable,stmaryrd}
\usepackage{slashed}
\usepackage{bigints}
\usepackage{cancel}
\usepackage{multicol}
\usepackage{blindtext}
\usepackage{tikz}
\usepackage{enumitem}
\usepackage[customcolors,shade]{hf-tikz}
\usepackage{graphicx}
\usepackage{cancel}

\DeclareMathAlphabet\mathbfcal{OMS}{cmsy}{b}{n}
\DeclareSymbolFont{usualmathcal}{OMS}{cmsy}{m}{n}
\DeclareSymbolFontAlphabet{\mathcal}{usualmathcal}
\DeclareSymbolFont{rmlargesymbols}{OMX}{mdbch}{m}{n}
\DeclareMathSymbol{\rmintop}{\mathop}{rmlargesymbols}{82}
\DeclareMathSymbol{\rmointop}{\mathop}{rmlargesymbols}{72}

\definecolor{mygray}{gray}{0.5}

\title {Worldsheets reconstructed: Duals to two-cut matrix models}
\author[a]{Sounak Pal\,\orcidlink{0000-0002-2250-0466}}
\affiliation[a]{\it Indian Institute of Technology, Gandhinagar, Gujarat-382355, India}
\emailAdd{palsounak@iitgn.ac.in}

\abstract{We compute the higher-rank translation ($\mathbfcal{T}$) and rotation ($\mathbfcal{R}$) matrices associated with a genus-one spectral curve. Using these data, we reconstruct the worldsheet correlators of vertex operators in the corresponding supersymmetric Landau--Ginzburg (LG) theory. This provides a generalization of the genus-zero, one-cut spectral-curve construction to higher-genus geometries. The matrix-model correlators are mapped to integrals over the moduli space of Riemann surfaces for the symmetric sextic double-well potential, which has a double-cut solution. We further identify the rational and elliptic contributions to the worldsheet integrated vertex operators and show that each of the sector precisely matches the corresponding correlators of single-trace operators in the matrix model. Altogether, our construction showcases an explicit dictionary between the higher-genus spectral-curve data, the associated Cohomological field theory (CohFT), and the worldsheet description of the corresponding matrix-model observables.}

\makeatletter
\begin{document}

\maketitle
\section{Introduction}
Recent studies of worldsheet string theory and its connection to matrix models have achieved renewed interest. There is a vast literature on studying matrix models for several string theories \cite{Collier:2026pxi,McGreevy:2003kb}. The top-down approach is to figure out worldsheet string theory from the matrix model  \cite{Giacchetto:2026gpq}. Recent computations have shown explicit matching of the matrix models with worldsheet string theory \cite{Giacchetto:2026gpq}. These are also very important in the field of studying gravitational theories. Gravitational chiral partition functions on $AdS_3$ have been studied and shown to have a direct relation with the recently developed Virasoro Minimal string theories \cite{Collier:2023cyw}. Their supersymmetric cousins have also opened a direction for a variety of studies \cite{Rangamani:2025wfa,Eberhardt:2026hfh,Johnson:2026ugu}. Starting from the determinantion of the matrix model, the full UV worldsheet theory contains many non-perturbative components such as FZZT, ZZ type branes \cite{Gaiotto:2005gd,Ishibashi:2005zf,Eniceicu:2022nay,Okuyama:2021eju,Marino:2008vx,Kutasov:2004fg,Chakrabhavi:2024szk,Eniceicu:2022xvk}. These are quite important for studying beyond the realm of semiclassical correspondence.\\ \par  

In this paper, we will focus, however, on a bottom-up approach to determine the worldsheet models for a given matrix model. Depending on the cut structure of the matrix model, the worldsheet theories change from the one-cut Landau Ginzburg (LG) theory \cite{Vafa:1990mu,Dijkgraaf:1991qh,Lazaroiu:2003zi,Gholson:2018zrl,Morozov:1994hh,Gu:2020nub,Giacchetto:2026gpq,Kapustin:2003ga}. LG theories with boundaries involving branes are also considered in detail \cite{Lazaroiu:2003zi,Kapustin:2003ga,Knapp:2020oba}. It extends the idea of gauge-string duality. Though the computations from the worldsheet side is subtle and non-trivial already at genus-1, for one-cut matrix models, the calculations for two-cut models are much more involved. Primarily, the two cut models with four distinct edges has elliptic uniformization map to the complex plane. While for genus-0 (one-cut) spectral curves it is the standard Joukowsky map, which renders much simpler uniformization \cite{Eynard:2007kz,Eynard:2015aea}. Our calculations, hence, involve elliptic and rational sectors to match precisely on both sides. We show a perfect agreement of the genus-1 spectral data for a sextic matrix model with the integrated vertex operators on the moduli space of Riemann surfaces. The integrals are performed order by order for our case, as finding an analytic closed form for all order data, including elliptic functions, is non-trivial in the first place. Apart from the computed B-twisted sigma model with Calabi-Yau target space, there are cousins like A-twisted sigma models, and there is a recent surge of interest in computing the matrix model version of it \cite{Guffin:2008kt,Closset:2015rna,Garavuso:2022nac,Guffin:2008pi}.\\ \par
Using the techniques of the cohomological field theory (CohFT) \cite{Giacchetto:2026gpq,Witten:1990bs,2017arXiv171202528P,Vonk:2005yv,Moore:2025tmt,Jiang:2022awu,Jiang:2024nnk,Szabo:1996md,Cordes:1994fc,Borot:2024zwn} we compute the rotation matrix and the translation matrices\footnote{Though equivariant versions are also important for localization aspects. See \cite{Atiyah:1984px}.} \cite{Giacchetto:2026gpq}. Using these data(s) we also figure the edge matrices, important for non-seperating degenerations of the worldsheet Riemann surface. Recent developments of cohomological field theory demonstrate a new way out to compute the twisted intersection numbers from the input of the spectral data. The matrix models have long occupied a special place in the development of large-$N$ gauge theory and string theory \cite{Gopakumar:2011ev,Gopakumar:2012ny,Gopakumar:2022djw}. Already in the early work on the ’t Hooft expansion, matrix integrals provided a particularly transparent way in which the organization of Feynman diagrams by topology could be seen explicitly. In the large-\(N\) limit, the perturbative expansion is naturally reorganized according to the genus of the corresponding ribbon graphs, with planar diagrams dominating at leading order. Though computing the moment integral for a given spectral density only gives the contribution of the planar diagrams. Non-planar diagrams are extensively computed from the free energy, by taking derivatives with respect to the ’t Hooft parameters.\\ \par

For simple matrix integrals, all the planar diagrams can often be resummed non-perturbatively by solving an associated saddle-point problem for the eigenvalue density. In this sense, matrix models constitute one of the cleanest arenas for the first part of the large-$N$ gauge--string correspondence: the emergence of a topological expansion suggests a perturbative closed-string genus expansion, as advocated in \cite{Dubrovin:1994hc}.\\ \par

What has remained considerably more subtle is the second, and conceptually sound, step of this program. The topological classification of the $1/N$ expansion strongly suggests that the matrix integral should admit a description in terms of a two-dimensional worldsheet theory whose perturbative expansion reproduces the matrix-model genus expansion \cite{Gopakumar:2024jfq}. However, constructing such a worldsheet theory directly in the ordinary ’t Hooft limit has proved difficult. In particular, while the combinatorial problem of constructing the discretized surface is kind of immediate, proving a duality of the matrix model genus expansion with the continuum worldsheet theory, together with their worldsheet action and observables, is a much more delicate problem. The reason can be somewhat cast in the way, i.e how one constructs the worldsheet vertex operators in the supersymmetric Landau-Ginzburg theory, corresponding to a matrix model of specific cut structure. We would like to precisely extend the calculation of the genus-0 spectral curve here, to spectral curves having higher genus structure. \\ \par

Historically, the most successful route toward a continuum string interpretation of matrix models has involved taking a double-scaling limit \cite{Bertoldi:2006eh}. In this limit, the matrix-model couplings are adjusted to special critical values at the same time as $N$ is taken to infinity. Near such critical points, the discrete random surfaces generated by the matrix integral develop a continuum limit and can be identified with non-critical string theories, most famously minimal conformal field theories coupled to Liouville gravity.\\ \par

At the same time, the double-scaling limit fundamentally changes the nature of the problem. The microscopic couplings of the original matrix potential are driven toward critical values, and the independent expansion parameter of the resulting continuum theory is not simply the original \(1/N\). Much of the detailed dependence of matrix-model observables on the couplings of the potential is therefore reorganized, and in many cases effectively lost, in favor of a universal description associated with the chosen critical point.\\ \par

This is qualitatively different from the structure familiar from standard examples of gauge--string duality \cite{tHooft:1973alw,Gopakumar:1998ki,Gopakumar:2012ny,Gopakumar:2024jfq}. In AdS/CFT \cite{Maldacena:1997re,Witten:1998qj,Aharony:1999ti,Maldacena:1998im,Beisert:2010jr}, for example, the gauge-theory parameters are not discarded in taking the large--$N$ limit, instead they are translated into geometric or stringy data on the dual side. The ’t Hooft coupling determines quantities such as the curvature scale in string units, while $1/N$ controls the string-loop expansion. One therefore obtains a family of string backgrounds continuously parametrized by the couplings of the boundary theory. From this viewpoint, a genuinely developed worldsheet description of matrix integrals should ideally retain the dependence on the matrix-model couplings already in the ordinary large-$N$ regime, rather than only after tuning the theory to a critical point.\\ \par

The challenge is thus to go beyond the usual double-scaled perspective and to understand whether the full large-$N$ matrix model possesses an intrinsic worldsheet description at finite, generic values of its couplings. Such a construction should reproduce not only the genus expansion, but also the detailed dependence of matrix-model observables on the underlying spectral curve and on the parameters entering the matrix potential. Establishing this correspondence would provide a more direct realization of the original large-\(N\) intuition: the matrix model would not merely generate a formal sum over discretized surfaces, but would be equivalent, order by order in the topological expansion, to a continuum worldsheet theory defined on those surfaces. Recently, several matrix model trialities has been proposed in \cite{Gopakumar:2022djw,Mazenc:2026yrm}.

\subsection*{Main claim.} The main focus of this paper is to extend the genus-0, one-cut spectral curve computation  to genus-1, two-cut spectral curves. While, the genus-$g$ contribution of single trace operator correlators is claimed to be equal to integrated correlator of vertex operators of LG theory at genus-$g$, the computation has been performed for genus-0 genera in \cite{Giacchetto:2026gpq}. Mathematically,
\begin{align}
    \bigg\langle\prod_{i=1}^n \text{Tr}M^{k_i}\bigg\rangle^{\text{M.M}}_{g}=\int_{\overline{\mathcal{M}}_{g,n}}\bigg\langle\!\!\!\bigg\langle\prod_{i=1}^n \mathcal{V}_{k_i}\bigg\rangle\!\!\!\bigg\rangle^{\text{LG+grav}}_{g}
\end{align}
We extend this computation to higher genus (genus-1) spectral curves and match both the contributions of the elliptic and the rational sector.\\

Our paper is organised as follows,
In Sec.~\eqref{sec22} we motivate the two-cut extension and discuss the uniformization map for sphere.
In Sec.~\eqref{sec33} we discuss the spectral curve data and solve the relevant correlators for single trace operators from the matrix model side. Then, in Sec.~\eqref{sec44} we give a brief review of the Landau-Ginzburg theory. In Sec.~\eqref{sec55} we use Givental's description of the cohomological field theory (CohFT) and computed the rotation and translation matrices.
In Sec.~\eqref{sec66} we define the matter primary dictionary.
Sec.~\eqref{sec77} is devoted to discussion of the main construction relevant to our case and we also perfrom cross checks between matrix model and worldsheet data. In Sec.~\eqref{sec88} we computed the integral over the characteristic classes to show the matching with the spectral data. Sec.~\eqref{sec99} talks about different limiting regimes and checks. In Appendix~\eqref{appA}--\eqref{appe} we derive $\mathbfcal{R}$ (rotation) and $\mathbfcal{T}$ (translation) matrices along with the edge matrix ($E$) relevant to our discussion using the language of the cohomological field theory.

\section{Matrix model Construction}
\label{sec22}
\subsection{Perspective on the two-cut extension}

The two-cut problem \cite{Akemann:1995im} is a natural testing ground for the proposed
reconstruction of the matrix model and the dual worldsheet construction because it is the first case in which the local branch-point analysis must be supplemented by global information. In a one-cut solution, rational uniformization makes many steps appear elementary \cite{Giacchetto:2026gpq}, which is highly intricate for matrix models admitting spectral curves of higher genus.\footnote{
There are a few holographic models which realise these two-cut structures \cite{Johnson:2021owr}.}\\

The simplicity is useful for establishing the basic correspondence \cite{Giacchetto:2026gpq}, but it can
also conceal many distinct subtleties involved in higher genus calculations. The presence of higher genus curves removes the simplification, which happens for the one-cut model. Its period lattice forces the calculation to retain the information needed to specify a globally normalized object, and therefore, tests the algebraic geometry intuitions on a stronger footing. The symmetric sextic potential \eqref{3.4e} is particularly well suited to this purpose. It has both single-cut and double-cut solutions in different parameter regimes. Symmetry also supplies selection rules for moments and provides immediate checks on the parity of intermediate expressions. From a broader point of view, the result separates two questions that are occasionally conflated. One question asks whether a matrix integral has a large-$N$ expansion organized by Riemann surfaces. This follows directly from ribbon graphs. The second asks whether the coefficients of that expansion admit a continuum worldsheet description with local closed string vertex operator insertions. The CohFT construction (described in Sec.~\eqref{sec55}) addresses the second question. It converts the spectral curve data into classes on moduli space, and after performing the integral over the relevant moduli space of Riemann surfaces, one can prove the duality between the matrix model single trace correlators and the integral of the closed string vertex operators. We will demonstrate how this works in the later sections. The explicit agreement with the matrix-model correlator demonstrates how this conversion works when the spectral curve itself has genus-one, similar to our case \eqref{3.58i}.\\ \par

\begin{figure}
    \centering
    \includegraphics[width=0.6\linewidth]{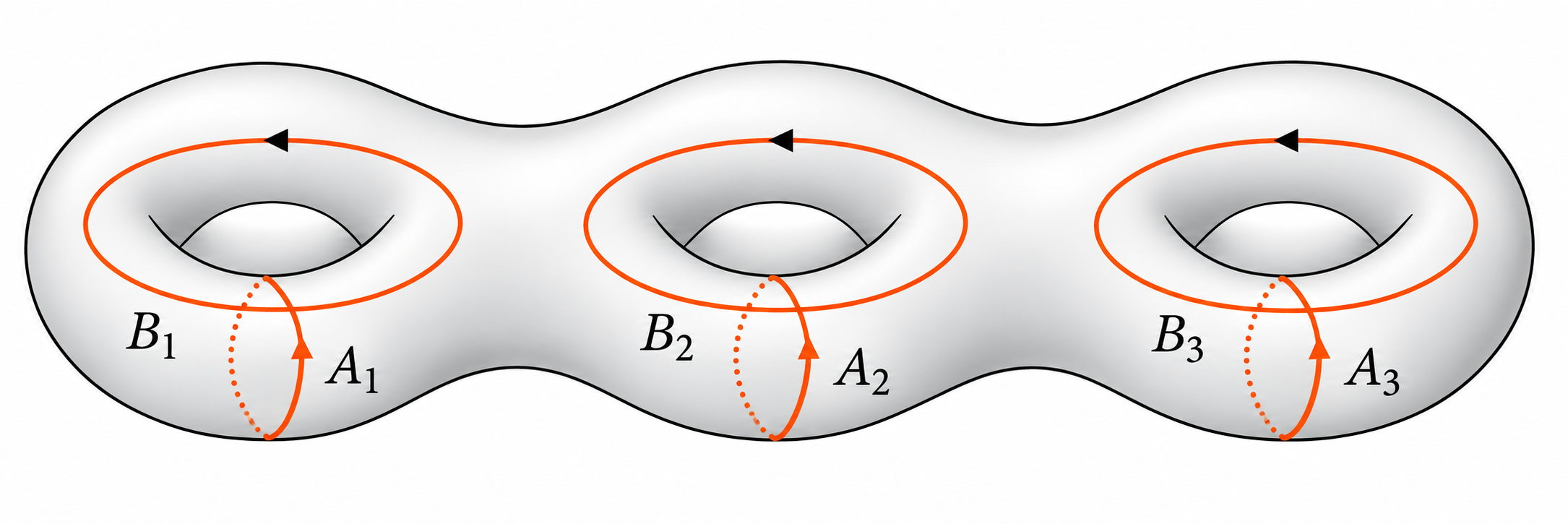}
    \caption{For a multi-cut ($4$-cut) matrix model this is the corresponding Riemann surface it maps to. }
    \label{fig:1}
\end{figure}

The method also makes clear what data a future generalization must preserve. The branch points and their local coordinates are required for local recursion; filling fractions and normalized cycles are required for global periods, and the metric, together with the rotation data, is required for
gluing. Changing the potential changes these model-dependent inputs, but does not alter the universal structure of the stable-graph sum. This division of labor is what makes it realistic to extend the analysis to less symmetric potentials, additional cuts, and correlators with more marked points.

\begin{figure}[htb!]
    \centering
\includegraphics[width=.99\linewidth]{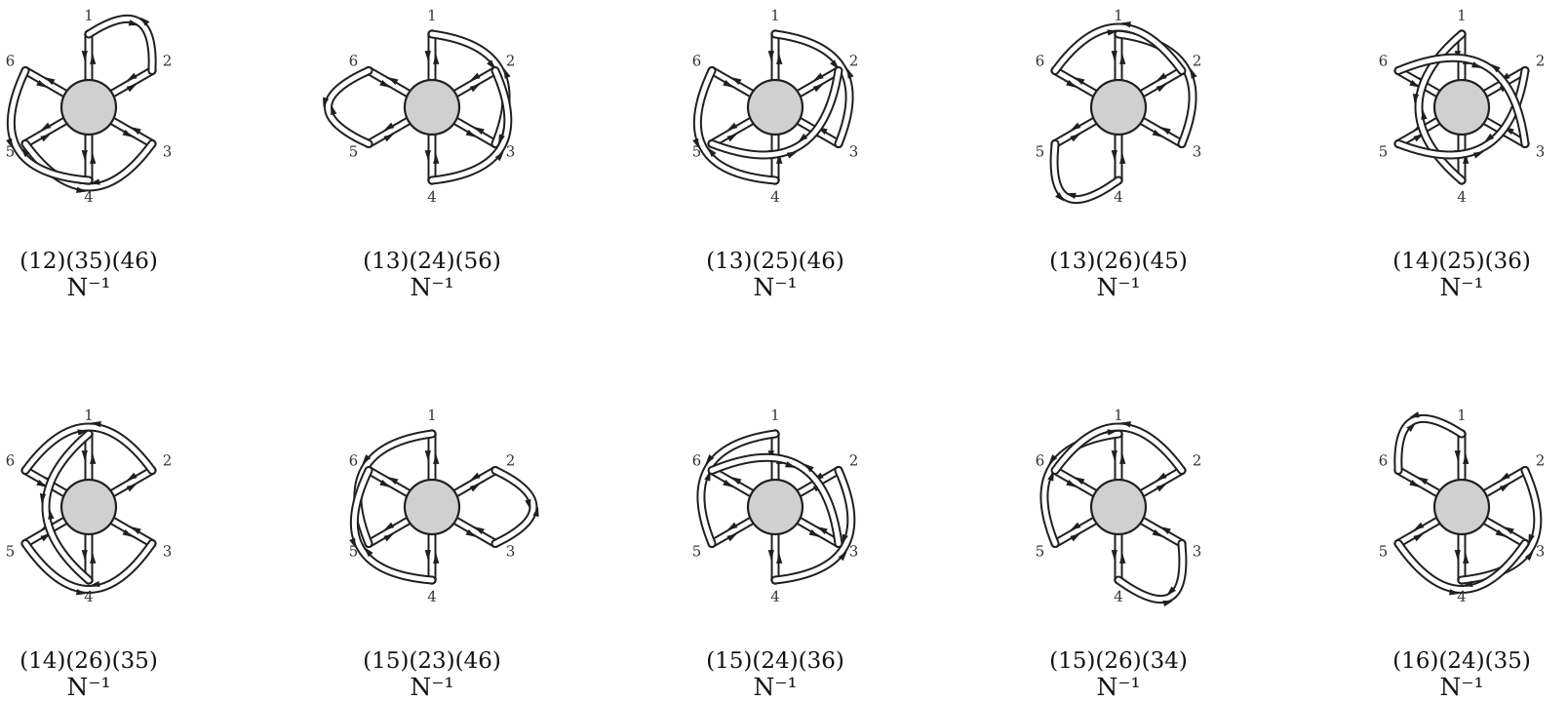}
    \caption{Diagrams of contractions relevant for the $\text{Tr}(M^6)$ correlator at genus-1. Note that in each pannel the scaling is $N^{-1 }$ because we did not conider vertex factor of N in $N\, \text{Tr} V(M)$. Each diagram is non planar type ribbon diagram. }
\label{fig:55}
\end{figure}

Now, in this section, we demonstrate the construction of the matrix model which embeds a double cut structure (In the Fig.~\eqref{fig:1}, it entails the information of a 4-cut spectral curve.). For matrix models of Gaussian type and along with that having all sorts of ’t Hooft couplings of the form $t_{2k}M^{2k}$(for $k\geq 3$), the dual string theory construction has been done in the one-cut phase. In this paper, we initiate the study of generic two-cut models (Spectral curve with genus-1). We start with a matrix potential of type,
\begin{align}
    V(\mathbf{M}):=-\frac{g}{2}\mathbf{M}^2+\sum_{k\geq 2}t_{2k}\mathbf{M}^{2k}.
\end{align}
where $g$ is the coupling parameter for the Gaussian term. We are specifically interested in computing correlators of single trace operators ($\text{Tr}(M^{2k})$) and then we will show the example of matching with the relevant string theory by turning on two of the ’t Hooft couplings i.e $t_4$ and $t_6$. These are important for computing resolvents of planar and non-planar type (as shown in Fig.~\eqref{fig:55}). The planar diagrams are dominating, and the non-planar ones are sub-dominating in the large-N limit. 
The partition function for these types of models is given by,
\begin{align}
\begin{split}
    Z:&=\int \underbrace{[d\mathbf{M}]}_{\text{harr measure}}\,e^{-N \,\text{Tr}V(\mathbf{M})},\\&
    =\int \prod_id\lambda_i \,\Delta^2(\lambda)e^{-N \,\sum_{i}V(\lambda_i)}.
    \end{split}
\end{align}
In the second line we changed the measure over the product of eigenvalues of the matrices. Alongwith, $\Delta(\lambda)$ is the Vandermonde determinant factor and it is given by,
\begin{align}
    \Delta(\lambda)=\prod_{i<j}|\lambda_i-\lambda_j|.
\end{align}
Now, we proceed to compute the correlators of single trace operators similar to the gaussian model. \textit{In case of the gaussian model, the moment integral formula (demonstrated in \eqref{2.42e}) correctly reproduce all the matrix correlators specifically in the planar limit.} For large-$N$ the ribbon diagrams having genus are supressed in the powers of $N^{2-2g}$. Below, we proceed for computing the matrix model single trace operator correlators for sextic coupling turned on.\\ \par
\subsection{Example correlators for the symmetric double well potential alongwith sextic coupling }
The matrix correlators for the genus-0 spectral data is found already by switching on the ’t Hooft couplings \cite{Giacchetto:2026gpq}. We now proceed to generalise it for the genus-1 case. In this section, we will solve the correlators of matrices for the following type of potential,
\begin{align}
    V(\mathbf M)=\frac{t_6}{6}\mathbf M^6+\frac{t_4}{4}\mathbf M^4-\frac{\mu}{2}\mathbf M^2\label{3.4e}
\end{align}
As we will show, this potential has double-cut structure rendering the uniformization map on the complex plane involving elliptic functions. But before proceeding to compute the spectral curve data related to this potential, let us comment on the unimormization maps for different cut structures. \textit{For the genus-0 spectral data (one-cut) the uniformization map from the Riemann sphere to the plane is given by the Joukwosky map.} The Joukowsky map is given by \cite{Eynard:2015aea},\\
\begin{align}
    x(z)=\frac{a+b}{2}+\gamma(z+\frac{1}{z}), \,\,\,\,\,\text{with} \,\,\,\,\,\,\gamma=\frac{a-b}{4}\label{3.5t}
\end{align}
where, $[a,b]$ are the spectral edges of the one-cut model. The inverse of the Joukowsky map is given by,
\begin{figure}[t!]
    \centering
\includegraphics[width=0.85\linewidth]{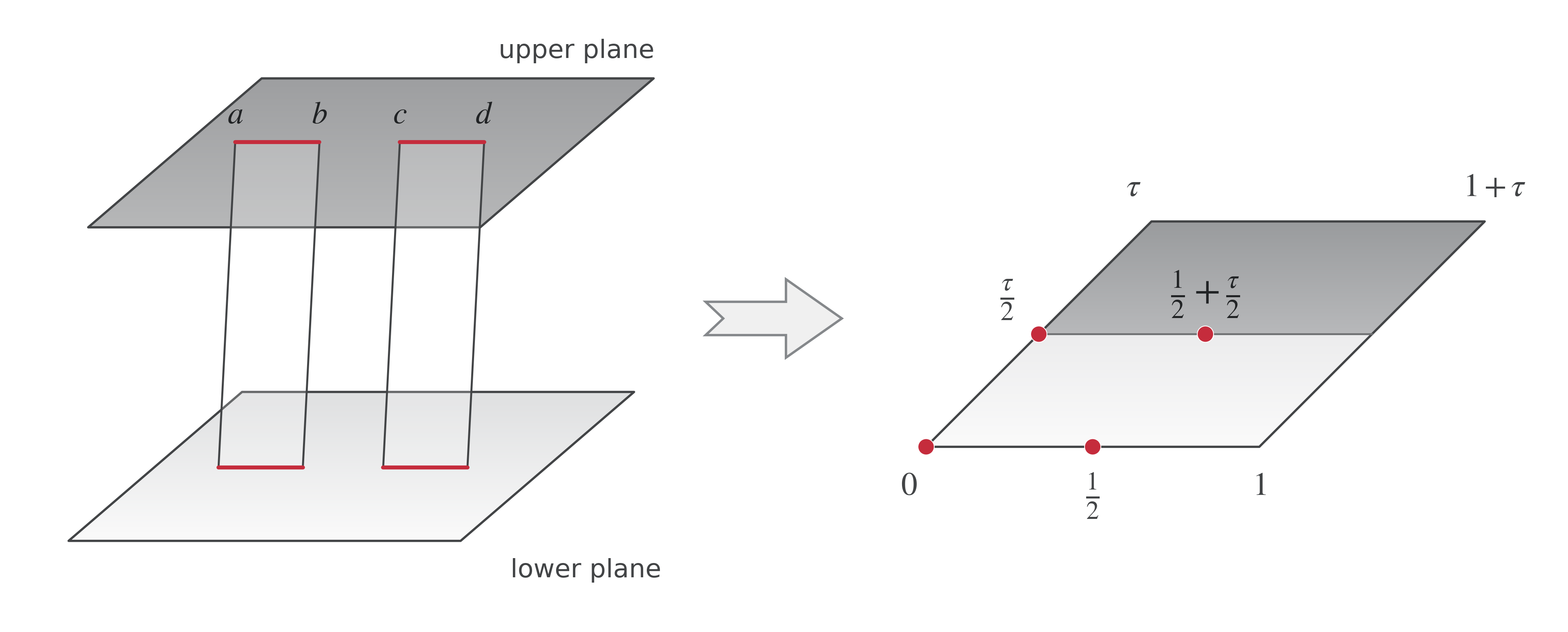}
    \caption{Figure depicting the uniformization for the genus-1 spectral data. }
    \label{fig:22}
\end{figure}

\begin{align}
    z=\frac{1}{2\gamma}\bigg(x-\frac{a+b}{2}\pm \sqrt{\bigg(x-\frac{a+b}{2}\bigg)^2-4\gamma^2\bigg)}\bigg)
\end{align}
where the $+/-$ sign corresponds to the upper branch/lower branch respectively. Now, we proceed to highlight the uniformization for the higher genus case. \\ \par
\noindent
\textbf{{Uniformization for genus-1 curve.}}\\
The uniformization map for the genus-1 (two-cut) case involves elliptic functions and the torus uniformizing coordinate $h$ in terms of the complex plane coordinates is given by,
\begin{align}
    h(x)=\frac{1}{2\int_{a}^{b} \frac{dx'}{\sqrt{\sigma(x')}}}\int_a^x\frac{dx'}{\sqrt{\sigma(x')}},\,\,\,\,\text{with}\,\,\,  \sigma(x):=(x-a)(x-b)(x-c)(x-d)
\end{align}
where, $h((a,b,c,d))=(0,\frac{1}{2},\frac{1}{2}+\frac{\tau}{2},\frac{\tau}{2})$
Now, the modulus of the torus is given by,\\
\begin{align}
    \tau=2h(d)=i \frac{_2{F}_1(1/2,1/2,1;1-r)}{_2{F}_1(1/2,1/2,1;r)}, \,\,\text{with} \,\,\,\,\,r:=\frac{(a-b)(c-d)}{(a-d)(c-b)}
\end{align}
Here, $r$ is the cross-ratio of the four points of the spectral edges. The spectral edges of two-interval $[a,b]$ and $[c,d]$ are shown in the Fig.~\eqref{fig:22}.
For the two-cut model, the uniformization can also be written in terms of Jacobi amplitudes in cases where the endpoints of the matrix model can be mapped to ${(-\frac{1}{\ell},-1,1,\frac{1}{\ell}})$ by using M\"obius transformation, which changes the matrix potential from the polynomial to the rational function. The uniformization map in this case is non-trivial and complicated than the genus-0 analogs. It is given by the following map \cite{Eynard:2015aea},
\begin{align}
    x=\text{sn}(u,\ell), \,\,\,\,\text{with}\,\,\,\,\, \sqrt{\sigma(x)}= \text{cn}(u,\ell) \,\text{dn}(u,\ell)
\end{align}
In the next section, we discuss the spectral curve and describe the general procedure to compute the data for the spectral curve.
\section{Spectral curve and elliptic data}
\label{sec33}

The purpose of this section is to make explicit a number of intermediate ideas which have been used throughout the calculation. It is useful to separate these remarks from the main derivation because the latter is naturally organized by the data that have to be computed, whereas the conceptual route from the
matrix integral to the worldsheet answer proceeds through several equivalent descriptions. \textcolor{black}{In particular, the spectral curve is simultaneously the planar solution of the loop equations, the input to topological recursion, and the geometric data from which the semisimple CohFT is reconstructed.  Keeping these three roles in view helps explain why the same branch-point quantities enter the resolvent, the rotation matrix, and the integrated correlators}.\\

Consider a hermitian one-matrix integral with potential $V(\mathbf M)$. At large $N$, the eigenvalues condense on a collection of intervals and the planar resolvent is the Stieltjes transform of the corresponding density of state. With the
conventional normalization one writes the planar resolvents as with the following asymptote \cite{Eynard:2015aea},
\begin{equation}
 W_{0,1}(x)=\lim_{N\to\infty}\frac{1}{N}\left\langle
 \operatorname{Tr}\frac{1}{x-\mathbf{M}}\right\rangle,
 \qquad W_{0,1}(x)\sim\frac{1}{x}\quad (x\to\infty).
\end{equation}
The discontinuity of this function across a cut determines the density. The asymptotic condition is more than a normalization convention: it selects the physical saddle and provides one of the constraints that specify the endpoints. For an even double-well potential, symmetry reduces the number of independent
endpoint parameters, but the two-cut geometry still retains a non-trivial period. That period is the origin of the elliptic quantities which distinguish the present construction from the one-cut case, which in  turn contains only the rational counterparts.\\ \par
\begin{figure}
    \centering
    \includegraphics[width=0.65\linewidth]{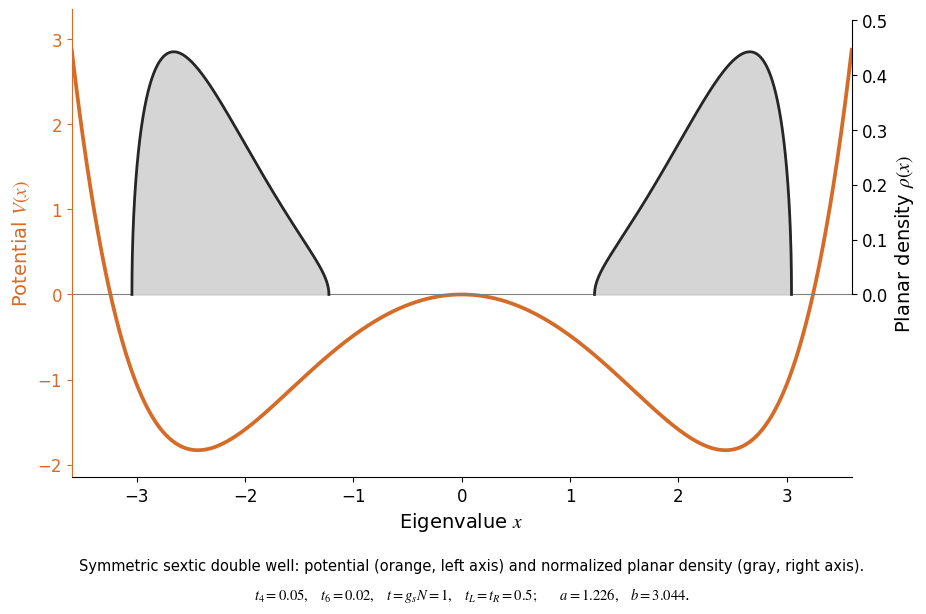}
    \caption{We plot the sextic symmetric double well potential along with the density of state. For small 't-Hooft parameters the eigenvalues are supported on two cuts.}
    \label{fig:33}
\end{figure}
The algebraic form of the planar curve may be summarized as,
\begin{equation}
 y(x)=V'(x)-2W_{0,1}(x),\qquad y(x)^2=\hat\Pi(x),
\end{equation}
\textit{where $\hat \Pi$ is fixed by the potential together with filling-fraction data. In a two-cut phase it has four simple zeros (as shown in Fig.~\eqref{fig:33}) relevant to the support.}
The local square-root coordinate at a zero is important. It turns the apparently singular behavior in the $x$-coordinate into a regular expansion on the curve, and it is this regular expansion that is used in the local topological-recursion kernels. Apart from having universality in the matrix model spectral curve, the planar resolvents also show universality near those points.

The filling fractions provides a global constraint. If $A$ is a cycle encircling one cut, it is customary to describe it by an integral of the spectral differential. Schematically,
\begin{equation}
 \tilde\delta=\frac{1}{2\pi i}\oint_A \underbrace{y(x)\,dx.}_{\text{spectral differential}}
\end{equation}
Fixing $\tilde\delta$ fixes the relative distribution of eigenvalues between the two cuts. It is interesting to study transitions of eigenvalues bewteen two different cuts of the matrix model and henceforth study the exponentially small but finite instanton contributions. In the symmetric situation this condition is particularly natural, but it remains essential even when the potential itself is odd. The local behavior near the four edges does not by itself determine the normalized elliptic curve, \textit{the period condition supplies the missing information, which turns out to be global}.\\

\newpage
\noindent
\textit{Why rational and elliptic contributions appear together?}\\

The separation into rational and elliptic pieces is best understood from the uniformization problem. Local residues depend only on expansions in the neighborhood of individual branch points. Such terms contribute rational functions. On top of this we have complex period lattice of the torus, which is necessarily responsible for elliptic contributions. The genus one curve has one remaining even moduli $\tau$, and the integral over the moduli space is always over the fundamental domain of the torus.
The rational and elliptic, two kind of terms are consequently not competing answers, they are the local and global components of a single geometric calculation\footnote{One useful way to state this is to introduce a normalized holomorphic
differential $\omega$ satisfying
\begin{equation}
 \oint_A\omega=1,\qquad \tau=\oint_B\omega.
\end{equation}
The modulus $\tau$ characterizes the complex torus. Changes of local
coordinates modify intermediate expansions, while the normalized periods keep
track of the global geometry.  Expressions involving $\mathcal K$,$\mathcal E$, or the Weierstrass functions encode precisely this dependence. They cannot generally be removed by an algebraic reparametrization of the endpoints. This also clarifies why an equality between matrix-model and worldsheet quantities must be checked sector by sector: an agreement only in the rational part would not test the normalisation of the period.}.\\ \par
In practice, it is efficient first to keep the modulus implicit and express the answer in terms of periods. Only after all residue calculations
have been performed, one should reduce the result to a chosen set of elliptic functions. This order avoids spurious simplifications. For example, a relation between the quasi-period and the ratio $\frac{\mathcal E}{\mathcal K}$ is a global identity, so applying it before the local residues are organized can obscure cancellations that are transparent in the period language, and we face it in our computations later.

\subsection{Loop insertion and operator insertions}

The connection between resolvents and single-trace observables follows from
expanding the resolvent at infinity.  Formally, one has
\begin{equation}
 \frac{1}{x-\mathbf M}=\sum_{k\geq 0}\frac{\mathbf M^k}{x^{k+1}},
 \qquad
W_{0,1}(x)=\sum_{k\geq0}\frac{\langle\operatorname{Tr}\mathbf M^k\rangle_0/N}{x^{k+1}}.
\end{equation}
This simple expansion explains why a calculation of the resolvent contains an infinite tower of moments. Connected multi-resolvents similarly generate connected correlators. The loop-insertion operation differentiates with respect to the potential couplings and thereby turns a one-point quantity into higher-point data. The worldsheet vertex operator chiral rings are formed to reproduce the same tower after integration over the relevant moduli space\footnote{This point also gives a useful interpretation of the labels on vertex operators. The vertex operators should also be seen as closed string insertions on the worldsheet on which the topological LG theory is defined. The equality of correlators is therefore an equality of generating functions after the appropriate coefficient extraction. The worldsheet computation has the advantage that the genus dependence is encoded in the geometry of moduli space, whereas the matrix-model computation contains it through the topological expansion of the free energy. One can also show that the corrections in $N^{2-2g}$, with higher number of `genus' of the ribbon graph, determines the strebel differentials \cite{Maity:2021wsi,Gaberdiel:2020ycd,Gopakumar:2024jfq} on the Riemann surface of genus-$g$.}.

\subsubsection {Derivation of the loop equation}
For the sake of completeness, we review the derivation of the solution to loop equations for multicut and hermitian matrix models \footnote{The double scaled limits are also discussed in detail in literature. See \cite{Bertoldi:2006eh} for example.} \cite{Chekhov:2006cb,Bertoldi:2006eh,Eynard:2007kz,Chekhov:2005rr}.
Consider the Hermitian one-matrix model, with the following form of the partition function, having a generic potential as,
\begin{equation}
Z=\int d\Phi\,e^{-N\Tr V(\Phi)},
\qquad
V(x)=\sum_{j\geq 1}\frac{t_j}{j}x^j,
\end{equation}
and define the matrix resolvent and its expectation value:
\begin{equation}
R_p=(p\id-\Phi)^{-1},
\qquad
W(p)=\frac{1}{N}\left\langle\Tr R_p\right\rangle.
\end{equation}
\noindent
Now, if one performs the infinitesimal change of variables,
\begin{equation}
\Phi\longrightarrow\Phi+\epsilon R_p.
\end{equation}
and choose an orthonormal basis $T^a$, $a=\{1,\ldots,N^2\}$,
of all Hermitian matrices, satisfying the following,
\begin{equation}
\Phi=\sum_a\phi_aT^a,
\qquad
\Tr(T^aT^b)=\delta^{ab},
\qquad
\sum_a T^aAT^a=(\Tr A)\id.
\end{equation}
Then using the variation of an inverse matrix,
\begin{equation}
\delta R_p=R_p\,\delta\Phi\,R_p,
\end{equation}
and $\delta\phi_a=\epsilon\Tr(T^aR_p)$, the divergence
of the vector field is
\begin{align}
\begin{split}
\sum_a\frac{\partial}{\partial\phi_a}\Tr(T^aR_p)
&=\sum_a\Tr(T^aR_pT^aR_p) \\
&=\Tr\left[
\left(\sum_aT^aR_pT^a\right)R_p
\right] \\
&=(\Tr R_p)^2.
\end{split}
\end{align}
Consequently, the integration measure transforms as,
\begin{equation}
d\Phi\longrightarrow
d\Phi\left[
1+\epsilon(\Tr R_p)^2+O(\epsilon^2)
\right].
\end{equation}
Meanwhile, the action varies as,
\begin{equation}
N\Tr V(\Phi+\epsilon R_p)
=
N\Tr V(\Phi)
+\epsilon N\Tr\!\left[V'(\Phi)R_p\right]
+O(\epsilon^2).
\end{equation}
Invariance of the partition function, assuming vanishing boundary terms, therefore implies at a specific linear order in $\epsilon$,
\begin{equation}
0=
\left\langle
(\Tr R_p)^2
-N\Tr\!\left[V'(\Phi)R_p\right]
\right\rangle.
\label{eq:ward}
\end{equation}

\noindent
Dividing by $N^2$ and separating the connected part gives the following expression,\footnote{where, the connected component of the correlator is given by,
\begin{equation}
\langle AB\rangle_c
=
\langle AB\rangle-\langle A\rangle\langle B\rangle.
\end{equation}}
\begin{equation}
\frac{1}{N}
\left\langle
\Tr\frac{V'(\Phi)}{p\id-\Phi}
\right\rangle
=
W(p)^2+\frac{1}{N^2}
\left\langle\Tr R_p\,\Tr R_p\right\rangle_c,
\label{eq:connected}
\end{equation}

The loop-insertion operator is defined by
\begin{equation}
\frac{d}{d V(q)}
\equiv
-\sum_{j\geq 1}
\frac{j}{q^{j+1}}
\frac{\partial}{\partial t_j}.
\end{equation}
This operator differentiates the potential's couplings. Now, differentiating the normalized expectation value yields the following,
\begin{align}
\begin{split}
\frac{\partial W_{0,1}(p)}{\partial t_j}
&=
-\frac{1}{j}\left(
\left\langle\Tr R_p\,\Tr\Phi^j\right\rangle
-\left\langle\Tr R_p\right\rangle
 \left\langle\Tr\Phi^j\right\rangle
\right) \\
&=
-\frac{1}{j}
\left\langle\Tr R_p\,\Tr\Phi^j\right\rangle_c.
\end{split}
\end{align}
Using the large-$q$ expansion,
\begin{equation}
R_q=\frac{\id}{q}
+\sum_{j\geq 1}\frac{\Phi^j}{q^{j+1}},
\end{equation}
\begin{figure}[htb!]
    \centering
    \includegraphics[width=0.7\linewidth]{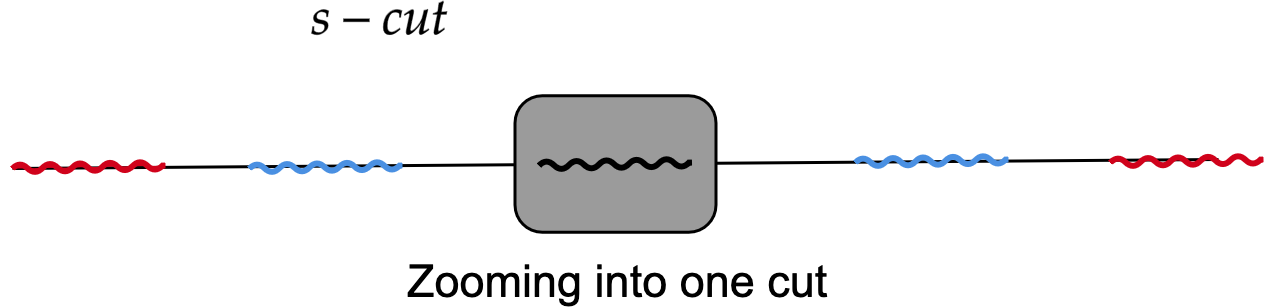}
    \caption{For a multi-cut($s$-cut) matrix model zooming into a specific cut of many possible choices. }
    \label{fig:4}
\end{figure}
we obtain the following expression after inserting the loop operator acting on the planar resolvent as,
\begin{align}
\begin{split}
\frac{d}{d V(q)}W_{0,1}(p)
&=
\sum_{j\geq 1}\frac{1}{q^{j+1}}
\left\langle\Tr R_p\,\Tr\Phi^j\right\rangle_c \\
&=
\left\langle
\Tr R_p\,
\Tr\left(R_q-\frac{\id}{q}\right)
\right\rangle_c \\
&=
\left\langle\Tr R_p\,\Tr R_q\right\rangle_c.
\end{split}
\end{align}
The constant $N/q$ drops out of the connected correlator. The equality extends by analytic continuation. Setting $q=p$, we find that,
\begin{equation}
\frac{d}{d V(p)}W_{0,1}(p)
=\left\langle\Tr R_p\,\Tr R_p\right\rangle_c.
\label{eq:insertion}
\end{equation}

\noindent
To obtain the contour representation, let
$\mathcal C_D$ encircle the eigenvalue support
counterclockwise (as shown in Fig.~\eqref{fig:4}), with $p$ outside the contour.
For each enclosed eigenvalue $\lambda_i$,
Cauchy's formula gives,
\begin{equation}
\oint_{\mathcal C_D}
\frac{d\omega}{2\pi i}\,
\frac{V'(\omega)}
     {(p-\omega)(\omega-\lambda_i)}
=
\frac{V'(\lambda_i)}{p-\lambda_i}.
\end{equation}
Summing over eigenvalues and taking the expectation value,
\begin{align}
\begin{split}
\oint_{\mathcal C_D}
\frac{d\omega}{2\pi i}\,
\frac{V'(\omega)}{p-\omega}W_{0,1}(\omega)
&=
\frac{1}{N}
\left\langle
\sum_{i=1}^N
\frac{V'(\lambda_i)}{p-\lambda_i}
\right\rangle=
\frac{1}{N}
\left\langle
\Tr\frac{V'(\Phi)}{p\id-\Phi}
\right\rangle.
\label{eq:contour}
\end{split}
\end{align}

\noindent
Combining Eqs.~\eqref{eq:connected},
\eqref{eq:insertion}, and \eqref{eq:contour},
we arrive at the loop equation:
\begin{equation}
\oint_{\mathcal C_D}
\frac{d\omega}{2\pi i}\,
\frac{V'(\omega)}{p-\omega}W_{0,1}(\omega)
=
W_{0,1}(p)^2+
\frac{1}{N^2}
\frac{d}{d V(p)}W_{0,1}(p),
\qquad p\notin D.
\end{equation}

\noindent
For a multi-cut solution,
\begin{equation}
D=
\bigcup_{\alpha=1}^{n}
[\mu_{2\alpha-1},\mu_{2\alpha}],
\qquad
\mathcal C_D=
\sum_{\alpha=1}^{n}
\mathcal C_{[\mu_{2\alpha-1},\mu_{2\alpha}]}.
\end{equation}
The number of cuts enters through the contour surrounding the spectral support. The change-of-variables argument applies for any number of cuts. The Schwinger--Dyson identity \eqref{eq:ward} is exact at finite $N$. For the usual unconstrained Hermitian
integral, the exact finite-$N$ eigenvalue density has
tails outside the compact limiting cuts. Accordingly, the contour around only those limiting cuts is understood in the formal genus expansion. Now, after solving the loop equation for genus-0 at large-N, we get,
\begin{align}
    W_{0,1}(p)^2=V'(p)W_{0,1}(p)+\oint_{\mathcal C_{\infty}}\frac{d\omega}{2\pi i}\frac{V'(\omega)}{p-\omega}W_{0,1}(\omega).
\end{align}
The solution is given by the following expression \cite{Eynard:2015aea},
\begin{align}
\begin{split}
    &W_{0,1}(p)=\frac{1}{2}V'(p)-\frac{1}{2}\sqrt{V'(p)^2+4P(p)}\\&
    P(p)=\oint_{\mathcal C_{\infty}}\frac{d\omega}{2\pi i}\frac{V'(\omega)}{p-\omega}W_{0,1}(\omega)\label{3.30p}
    \end{split}
\end{align}
Throughout our computation we will choose the following,
\begin{align}
    y_{\text{LG}}=-\frac{1}{2}y_{\text{M.M}}, \,\,\,\,\,\,\,\,\,\,\underbrace{\Omega}_{\text{Calabi-Yau top form}}=dy_{\text{LG}}.
\end{align}
Now, after figuring out the genus-0 large-$N$ solution, we shift to genus-one free energy and then we will proceed to discuss the general procedure to compute the data relevant for the sextic double well potential. 
\subsection{Genus-one free energy as a consistency}
The genus-one free energy occupies a special place in this comparison. Its derivatives with respect to 't-Hooft couplings give genus-one correlators of the matrix model, while its dependence on the discriminant reflects the behavior when branch points approach one another. Near such a degeneration, the local coordinate ceases
to be uniformly good and the logarithmic structure of the free energy records the resulting singular behavior. From the worldsheet perspective, the corresponding phenomenon is a boundary stratum of the compactified moduli space. A non-separating degeneration pinches a handle, and the edge contribution in the CohFT graph sum accounts
for this boundary. This is why the edge matrix is indispensable for a genus-one check. The vertex and leg contributions alone describe local insertions, but
they do not reconstruct the contribution of a pinched cycle (non-separating degenerations). The comparison with the genus-one matrix-model answer consequently tests not only the translation and rotation matrices, but also their compatibility with gluing property as a definition of the CohFT.\\

\noindent
\textbf{A guide to the subsequent formulas.} The formulas in the main text can be read according to the following logic. First, the matrix potential fixes the planar curve and its branch points.
Second, local expansions at these points determine the coefficients entering the $\mathbfcal R$-matrix. Third, the regular part of the same expansions,
together with normalization data, determines the translation.  Finally, the bilinear combination required by gluing determines the edge matrix. The
resulting graph expansion is integrated against tautological classes on $\overline{\mathcal M}_{g,n}$. At every stage the initial spectral curve is
the common input, so no additional dynamical assumption is being made when passing from matrix-model language to worldsheet language\footnote{It is helpful to keep track of what is local and what is global in this chain. Branch-points are local data. Periods, filling fractions, and quasi-periods are a part of global data. Stable-graphs are universal, its numerical weights are model dependent through the spectral data. The claimed duality states that
after these ingredients are assembled, the resulting moduli-space integral reproduces the coefficient of the matrix-model topological expansion. The
explicit genus-one calculation is therefore both a non-trivial example and a test of the proposed general mechanism.}.

\subsection{General formalism for computing data}
Now, for a generic potential $V(M)$, the planar one-loop resolvents are given by,
\begin{align}
    W_{0,1}(x):=\frac{1}{2}\bigg(V'(x)-y(x)\bigg), \text{where} \,\,x\in \text{supp}\bar{\rho}
\end{align}
where, $y(z)$ is the spectral differential for defining a matrix model spectral curve $(\Sigma, x(z),y(z),B(z_1,z_2))$ with $B(z_1,z_2)$ is the bi-symmetric differential also known as the \textit{Bergmann kernel}. $\Sigma$ denotes the Riemann surface. Now, for a dirac comb like eigenvalue distribution,
\begin{align}
    \rho(x)=\frac{1}{N}\sum_{i}\delta(x-\lambda_i),
\end{align}
the one-loop resolvent is given by its Stieltjes transform \cite{Eynard:2015aea},
\begin{align}
    W_{0,1}(x)=\frac{1}{N}\sum_{i}\frac{1}{x-\lambda_i}
\end{align}
Now, introducing the polynomial 
\begin{align}
    P(x)=\frac{1}{N}\sum_{i=1}^{N}\frac{V'(x)-V'(\lambda_i)}{x-\lambda_i}
\end{align}
one should solve for the Riccati equation,
\begin{align}
    W_{0,1}(x)^2+\underbrace{\frac{1}{N}W'(x)}_{\text{dropped at large $N$}}=V'(x)W_{0,1}(x)-P(x).
\end{align}
\noindent
Once, \(P(x)\) is specified for a generic matrix model potential, the equation provides a closed nonlinear differential equation for $W_{0,1}(x)$. By construction, $P(x)$ is a rational function of degree $d-1$, whose asymptotic behaviour as $x \to \infty$ matches that of
$\frac{V'(x)}{x}$.
In the special case $d=1$, these conditions determine $P(x)$ uniquely, giving
$$
P(x)=V''(x).$$
For a general potential of degree \(d\), however, \(d-1\) coefficients remain undetermined, which may be fixed using the symmetries of the potential. Now, after solving the loop equation in the planar limit, the first order derivative of the potential is given by,
\begin{align}
    V'(x)=\overline{W}(x+i0)+\overline{W}(x-i0); \,\,\text{for}\,\, x\in \text{supp}(\bar\rho)
\end{align}
where, 
\begin{align}
    \overline{W}(x)=\lim_{N\to \infty}W_{0,1}(x),\,\,\,\,\overline{P}(x)=\lim_{N\to \infty}P(x)
\end{align}
and the density of the eigenvalues is given by,
\begin{align}
    \bar{\rho}(x)=\textbf{disc}\,(\overline{W}(x))=-\frac{1}{2\pi i}\bigg(\overline{W}(x+i0)-\overline{W}(x-i0)\bigg)
\end{align}

\noindent
Now, after writing down the loop equations explicitly they look like,
\begin{align}
\begin{split}
    &V'(x)^2-4\overline{P}(x)=M^2(x)\sigma(x)\\&
    \overline{P}(x)=V'(x)\overline{W}(x)-\overline{W}(x)^2
    \end{split}
\end{align}
where, $M(x)$ contains no zeroes of the spectral edge.
We finally express the planar one-loop resolvent as,
\begin{align}
    \overline{W}(x)=\frac{1}{2}\bigg(V'(x)-M(x)\sqrt{\sigma(x)}\bigg)
\end{align}
 Also, using the \textit{Tricomi relation} the planar resolvent can be solved as (which we also derived in \eqref{3.30p}),\\
\begin{align}
\begin{split}
    \overline{W}(x)=\frac{1}{2\pi i}\int_{\text{supp}\overline{\rho}} dx' \frac{\sqrt{\sigma(x)}}{\sqrt{\sigma(x')}}\,\frac{V'(x)}{x-x'}
\,\,\,\,\,\,\text{with},
   \,\,\,\,\,\sigma(x):=\prod_{i=1}^{2s}(x-a_i) 
   \end{split}
\end{align}
for a $s$-cut matrix model.
Now, to find the spectrum or the eigenvalue density one can compute the discontinuity as advocated above. In the next part, we solve and apply the above formalism for our chosen potential turning on the ’t Hooft couplings ($t_4,t_6$). 
\noindent
\subsubsection{{Data for genus-0 curve}} In this section, we compute the spectrum of the genus-0 spectral data for the above matrix model with the specific ’t Hooft couplings turned on.
We have,
\begin{align}
    V'(x)=t_6x^5+t_4x^3-\mu x\label{2.20i}
\end{align}
Now, for symmetric endpoints and one-cut solutions we choose $[a,b]$ spectral edges as $(-2\gamma,2\gamma)$.\footnote{To find the spectral edges $[a,b]$ (one-cut) in the usual scenario one should solve the following set of equations,
\begin{align}
    \int_{a}^bdx' \frac{V(x')}{\sqrt{(x'-a)(x'-b)}}=0 \,\,\,\text{and }\,\,\,\,
   \frac{1}{2\pi i} \int_{a}^bdx' \frac{x'V(x')}{\sqrt{(x'-a)(x'-b)}}=1
\end{align}
}
In the next section, we will describe the two-cut (genus-1) spectral curve data, which is also a solution but for different parameter regime.
By solving the Riemann-Hilbert problem and plugging the uniformization using the Joukowsky map \eqref{3.5t}, we get,
\begin{align}
    V'\bigg(\frac{a+b}{2}+\gamma(z+\frac{1}{z})\bigg)=\sum_{k=0}^dv_k(z^{k}+z^{-k})
\end{align}
where, $d$ is the degree and determined by the following equation,
\begin{align}
    \sum_{m}k_m=2d.
\end{align}
where, $k_m$ is the number of zeroes of order $m$ of $M^2(x)\sigma(x)$\footnote{The prefactor polynomial can also be computed using,
\begin{align}
    M(x):=\oint_{\mathcal{C}_\infty} \frac{d\omega}{2\pi i}\frac{1}{\omega-x}\frac{V'(\omega)}{\sqrt{\sigma(x)}}
\end{align}
where, $\mathcal{C}_\infty$ denotes the contour encircling all the $s$-cuts.}. The zeroes of the spectral edges are differently counted and not included in the above equation. Now, the differential one forms are given by,
\begin{align}
    \overline{\omega}(z):=\sum_{k}{v}_kz^{-k}, \,\,\text{with} \,\,\,v_0=1,v_1=\frac{1}{\gamma}
\end{align}
For the chosen potential, 
the non zero entries of $(v_i)$ are given as,
\begin{align}
    v_1=3t_4\gamma^3-\mu,\,\,v_3=5t_6\gamma^5+\gamma^3t_4,\,\,v_5=t_6\gamma^5
\end{align}
Now, plugging in $v_1=\frac{1}{\gamma}$, one can solve $\gamma$ in terms of the ’t Hooft couplings.
Now the data for the differential and the spectral curve is given by,\\
\begin{align}
\begin{split}
    &\overline{\omega}(z):=\frac{1}{\gamma z}+(5t_6\gamma^5+\gamma^3t_4)\frac{1}{z^3}+\frac{1}{z^5}t_6\gamma^5\\&
    \overline{W}(x)=\frac{1}{2}\sqrt{x^2-4\gamma^2}\,\bigg[t_4\bigg(x^2+2\gamma^2-\frac{x^3}{\sqrt{x^2-4\gamma^2}}\bigg)\\&\hspace{1 cm}+t_6\bigg(x^4+2x^2\gamma^2+6\gamma^4-\frac{x^5}{\sqrt{x^2-4\gamma^2}}\bigg)-\mu+\frac{\mu x}{\sqrt{x^2-4\gamma^2}}\bigg],\\&
    \rho^{(0)}(x)=\frac{1}{2}\bigg[\bigg(t_4(x^2+2\gamma^2)+t_6(x^4+2x^2\gamma^2+6\gamma^4)-\mu\bigg)\sqrt{x^2-4\gamma^2}\bigg]
    \end{split}
\end{align}
For $\overline{W}(x)$ and spectral density $x$ is just a variable. To express it terms of the $z$ variable, one should use the Joukowsky map defined earlier. Now, we found all the data. To specify whether it is a genus-0 data we need to fix the regime and one can easily check that the condition is as follows. For, \begin{align}
&
\begin{cases}
&\mu>\mu_c=2t_4\gamma_c^2+6t_6\gamma_c^4, \,\,\,\,\text{two-cut}\\&
\mu<\mu_c=2t_4\gamma_c^2+6t_6\gamma_c^4, \,\,\,\,\text{One-cut}
\end{cases}
\end{align}
\noindent
Next, we proceed to show the construction for the genus-1 spectral curve and generalise the constraints for the moments.

\noindent
\subsubsection{{Data for genus-1 curve}}
Now, for the genus-1 case, using the relation,
\begin{align}
    y(x)=V'(x)-2W_{0,1}(x)
\end{align}
one should note that, the derivative of the potential in \eqref{2.20i} has the highest weight $5$ in the polynomial. Therefore, as $\sqrt{\sigma(x)}\sim x^2$ for large-$x$ for a two-cut model, $M(x)$ must be a polynomial of degree $3$. Now, as $V'(x)$ is odd in $x$ and $W(-x)=-W_{0,1}(x)$.
\subsection*{Finding the spectral edges.}
\noindent
The planar one-loop resolvent and the density of states for the two-cut model with the above potential, are given by,
\begin{align}
\begin{split}
    \overline{W}_{0,1}(x)=\frac{1}{2\pi i}\int_{\text{supp}\overline{\rho}} dx' \frac{\sqrt{\sigma_2(x)}}{\sqrt{\sigma_2(x')}}\,\frac{V'(x)}{x-x'}
\,\,\,\,\,\,\text{with},
   \,\,\,\,\,\sigma_2(x):=\prod_{i=1}^{4}(x-a_i). \label{2.29i}
   \end{split}
\end{align}\\
Now, to find the spectral edges which are dependent on the ’t Hooft parameter, we calculate the moments as follows,
\begin{align}
   \mathcal{M}_k:=\int_{[a,b]\cup [c,d]}dx\, \frac{x^k\,V'(x)}{\sqrt{\sigma_2(x)}}.
\end{align}
Now, these moments encodes information about the spectral edges. As we know,
\begin{align}
    \sqrt{\sigma_2(x)}=x^2\bigg(1-\frac{s_1}{2x}+\mathcal{O}(x^{-2})+\cdots\bigg), \,\,\,\,\, \text{and}\,\,\frac{1}{x-x'}=\frac{1}{x}\bigg(1+\frac{x'}{x}+\frac{x'^2}{x^2}+\cdots\bigg)
\end{align}
We can express and club terms plugging in these expressions into \eqref{2.29i}, and rewrite the resolvent in the following way,
\begin{align}
    \overline{W}_{0,1}(x)=\frac{1}{2\pi i}\bigg[x\mathcal{M}_0+(\mathcal{M}_1-\frac{s_1}{2}\mathcal{M}_0)+\frac{1}{x}(\mathcal{M}_2-\frac{s_1}{2}\mathcal{M}_1+\cdots+\mathcal{M}_0)+\mathcal{O}(x^{-2})\bigg]{\label{2.32l}}
\end{align}
As, we already have the asymptotic behaviour (analytic at infinity) as,
\begin{align}
    \overline {W}_{0,1}(x)\xrightarrow[x\to \infty]{}\frac{1}{x}+\mathcal{O}(x^{-2})
\end{align}
it renders the following moments as\footnote{The terms in the ellipsis in \eqref{2.32l} are combinations of first and second moments. As they individually vanish, the terms in the ellipsis are not important. },
\begin{align}
    \mathcal{M}_0=0,\,\,\,\mathcal{M}_1=0,\,\,\,\frac{\mathcal{M}_2}{2\pi i}=1,\cdots
\end{align}
\noindent
One can notice that, there is one non trivial and non vanishing moment by comparison. It leads to, the four symmetric spectral edges which are consequently,
\begin{equation}
a_1=-b,\qquad a_2=-a, \qquad a_3=a, \qquad
a_4=b,
\end{equation}
where,
\begin{equation}
a=\sqrt{\frac{q-t_4}{2t_6}-\frac{2}{\sqrt{q}}},
\qquad
b=\sqrt{\frac{q-t_4}{2t_6}+\frac{2}{\sqrt{q}}}.
\end{equation}
Hence, as we stated ealier, the endpoints are dependent on the 't-Hooft parameters. The physical roots must satisfy,
\begin{equation}
q>0,\qquad a^2>0.
\end{equation}
The derivation of the end points is given in appendix~\eqref{appA}. Now, we proceed to compute one important ingredient of the matrix model, i.e the density of state. The density of state for this two-cut matrix model is given by the discontinuity of the planar one-loop resolvent and can be written as,
\begin{align}
    \rho_0(x)=y_{\text{M.M}}=\frac{|x|}{2\pi}(t_6x^2+t_4+\frac{t_6}{2}\mathbfcal{S})\sqrt{(b^2-x^2)(x^2-a^2)}.\label{3.58i}
\end{align}
So, one can check that the polynomial multiplying the branch-cut structure is cubic as mentioned earlier. Now, one can also verify that,
\begin{align}
       \bigg( \int_{a}^b+\int_{-b}^{-a}\bigg)\,\rho^{(0)}(x)dx=\frac{1}{32} \left(a^2-b^2\right)^2 \left({t_6} \left(a^2+b^2+\mathbfcal{S}\right)+2 {t_4}\right)=1
\end{align}\\
using $\mathbfcal{D}^2{\mathcal{L}}=16$, with $\mathcal{L}=t_4+t_6\mathbfcal{S}$ and $\mathbfcal{D}$ (defined in \eqref{eq:D-def}).
We also derive in \eqref{A.10i} the genus-1 planar one-loop correlator as,
\begin{align}
\overline{W}_{0,1}(z)
=\frac{1}{2}\Bigg[&
\left(t_4-c+\frac{t_6\mathbfcal{S}}{2}\right)z^3
+\left(-\mu+\frac{c\mathbfcal{S}}{2}+\frac{t_6\mathbfcal{D}}{8}\right)z+\frac{\mathbfcal{D}}{16}\left(t_6\mathbfcal{S}+2c\right)\frac{1}{z}
+O(z^{-3})
\Bigg].
\end{align}Hence, after sorting out the spectral curve data for the genus-1 spectral curve, we proceed to compute correlators of single trace operators.
\subsection{Solving the correlators}
The genus-0 or higher genus contributions to the correlators of single trace operators can be deduced from the free energy of the matrix model involving higher genus contributions. Now, the precise relation that connects the free energy and the correlators are given by,
\begin{align}
    \begin{split}
        \frac{d}{d t_6}\underbrace{\log Z_{g}}_{\text{free energy}}=-N \bigg\langle \text{Tr}(M^6)\bigg\rangle_{g}=-N\,\bigg[N\,m_6^{(0)}+\frac{1}{N}\,m_6^{(1)}+\cdots\bigg]
    \end{split}
\end{align}
where, $m_6^{(h)}$ is the genus-$h$ contribution to the correlator of the single trace operator.
\noindent
Below, we list some of the correlators of single-trace operators.

\noindent
{\subsubsection{Genus-0 correlators}}
\noindent
The genus-0 contribution from the correlators of the single trace operator can be computed using the moment integral as defined below,\\
\begin{align}
\begin{split}
   \bigg\langle\frac{1}{N} \text{Tr}(M^6)\bigg\rangle_{g=0}&=m_6^{(0)}\\&
= \frac{1}{2\pi}\int_{a^2}^{b^2}du \,u^3(t_6u+c)\sqrt{(b^2-u)(u-a^2)}; \,\,\,\,\,\,\,u=x^2\\&
=\frac{\mathbfcal{D}}{8\pi}
\int_{-1}^{1} dy\,
\left(\frac{\mathbfcal{S}}{2}+\frac{\sqrt\mathbfcal{D}}{2}y\right)^{3}
\left[t_{6}\left(\frac{\mathbfcal{S}}{2}+\frac{\sqrt\mathbfcal{D}}{2}y\right)+c\right]
\sqrt{1-y^{2}}.\label{2.42e}
\end{split}
\end{align}
\noindent
In writing the final line we used a variable change,
\begin{align}
    \begin{split}
        u=\frac{\mathbfcal{S}}{2}+\frac{\sqrt{\mathbfcal{D}}}{2}y.
    \end{split}
\end{align}
Now, the final result can be casted as,
\begin{align}
    m_6^{(0)}=\frac{\mathbfcal{D}}{2048}\bigg(t_6\mathbfcal{D}^2+18t_6\mathbfcal{D}\mathbfcal{S}^2+12t_4\mathbfcal{D}\mathbfcal{S}+16 t_6 \mathbfcal{S}^4+16t_4\mathbfcal{S}^3\bigg)
    \label{3.44r}
\end{align}
while writing the above result we used the following relations,
\begin{align}
    \int_{-1}^{1}\sqrt{1-y^2}\,dy=\frac{\pi}{2},\,\,\,\,\int_{-1}^{1}y^2\sqrt{1-y^2}\,\,\,dy=\frac{\pi}{8},\,\,\,\,\int_{-1}^{1}y^4\sqrt{1-y^2}\,dy=\frac{\pi}{16}
\end{align}

\noindent
There can be several other examples of single-trace operator correlators, which can be worked out by using the same moment integral method, without the information of the free energy at genus-0. For higher-genus contributions one has to compute the higher-genus free energy and hence compute the correlators as we will discuss below. The moment integral method only computes the genus-0 contributions as also described earlier. Now, we proceed to compute the genus-1 correlators in the next part. \\

\noindent
\textbullet \,\,\,Before proceeding to compute the correlators, we cook up for the free energy computation at genus-1.
The free energy for the two-cut solution can be written as \cite{Chekhov:2005rr},
\begin{align}
    F_{g=1}=-\frac{1}{24}\log\bigg[\bigg(\prod_{i=1}^4 {M}(x_i)\bigg)\Delta_{\text{br}}^4\mathbfcal{Q}^{12}\bigg]+\text{constant}\label{2.46y}
\end{align}
where, ${M}$ is the prefactor polynomial in the definition of the spectral curve and $M(x_i)$ (except the square-root factors) means we evaluate the polynomial at the end points of the spectral curve.
It can be explicitly written as,
\begin{align}
    \bigg(\prod_{i=1}^4M(x_i)\bigg)=a^2b^2\underbrace{(c+a^2t_6)^2(c+b^2 t_6)^2}_{\frac{\mathbfcal{P}^2}{\mathbfcal{L}^2}}
\end{align}
$\text{Det}(Q_{j,i})=\mathbfcal{Q}\,,\forall i,j=1\,\,\text{to} \,\,n-1$ is given by,
\begin{align}
    {Q}_{j,i}=\oint_{A_j}d\lambda\,\frac{\lambda^{i-1}}{\tilde{y}(\lambda)}; \,\,\,\,\text{with} \,\,\tilde{y}(\lambda)=\sigma(\lambda)\bigg|_{\text{square root factor}}
\end{align}
where, $A_j$ encircles the $j$-th cut of the matrix model spectral curve and $\tilde y(\lambda)$ determines the square root part of $\sigma(\lambda)$ except the polynomial. For our two-cut matrix model indicating a genus-1 spectral curve, this factor can be computed by the periodic integral,
\begin{align}
    \mathbfcal{Q}:=\oint_{A} \frac{dz}{\sqrt{(a^2-z^2)(b^2-z^2)}}
\end{align}
and with the substitution of \begin{align}
    z^2=a^2+(b^2-a^2)\sin^2(\theta)
\end{align}
we can write the following result,
\begin{align}
    \mathbfcal{Q}\sim\frac{i}{a}\mathbfcal{K}(1-\frac{b^2}{a^2}), \,\,\,\,\, \text{with}\, \,\,\,\mathbfcal{K}(k^2)=\int_{0}^{\pi/2} \frac{d\theta}{\sqrt{1-k^2 \,\text{sin}^2(\theta)}}
\end{align}

\noindent
Alongwith that, $\Delta_{\text{br}}$ is the \textit{branch point vandermonde factor and is given by},
\begin{align}
\begin{split}
    &\Delta_{\text{br}}=(b-a)(b+a)(2b)(2a)(a+b)(b-a)=4ab\mathbfcal{D}\\&
    \end{split}
\end{align}
Hence, after computing all the ingredients we are ready to proceed for finding the genus-1 free energy. Now, plugging in the all the above expressions in \eqref{2.46y}, the free energy can be written as \cite{Bertoldi:2006gk,Chekhov:2005rr},
\begin{align}
\begin{split}
F_{g=1}&=-\frac{1}{24}\sum_{i=1}^{4}\ln\left|M(x_i)\right|
-\frac{1}{2}\ln|\mathbfcal{K}(\nu)|
-\underbrace{\frac{1}{6}\sum_{i<j}
\ln|x_i-x_j|}_{\text{branch point Vandermonde}}+\frac{1}{4}
\left(\ln|x_1-x_3|+\ln|x_2-x_4|
\right).\\&
=\frac{1}{4}\ln\mathbfcal{ L} -\frac{1}{12}\ln \mathbfcal{P}-\frac{1}{8}\ln {(1-\nu)}-\frac{1}{2}\ln( \mathbfcal{K}(\nu))
\end{split}
\end{align}
with the following definitions,
\begin{align}
\begin{split}
    \mathbfcal{L}:=t_4+t_6 \mathbfcal{S},\,\,\,\,\,\,\,
     \mathbfcal{P}:=\mathbfcal{L}^3-4t_6^2,\,\,\,\,
     \nu=1-\frac{b^2}{a^2}
     \end{split}
\end{align}
\noindent
Now, we proceed to compute the genus-1 contribution of the single trace operator correlator of our chosen matrix model. One should remember that it will be a contribution of the correlator at genus-1.\\
\noindent
{\subsubsection{Genus-1 correlators}}
Substituting the explicit expressions, we get a compact form of the correlators with turning on the $t_4$ and $t_6$ couplings for the wrongsign of the quadratic term and is given by,\footnote{A generic expectation due to ’t Hooft, B\'rezin–Itzykson–Parisi–Zuber is that for any potential $V(M)$ of polynomial type, the generic form of the correlators are given by,

\begin{align}
    \frac{1}{Z}\int [dM]\,e^{-N\, TrV(M)}=\sum_{G}\frac{N^{\chi(G)}}{\text{Aut(G)}}\prod_k t_{2k}^{\text{No. of $k$ vertices}}
\end{align}}

\hfsetfillcolor{gray!5}
\hfsetbordercolor{white}
\begin{align}
\begin{split}
  \tikzmarkin[disable rounded corners=true]{espqo}(0.5,-1.2)(-0.4,0.8)  \langle \text{Tr}(M^6)\rangle_{g=1}&= m_{6}^{(1)}=-6\frac{d}{d t_6}F_{g=1}\\&
   =-\frac{3}{2}\frac{\mathbfcal{L}_{6}}{\mathbfcal{L}}
+\frac{3\mathbfcal{L}^{2}\mathbfcal{L}_{6}-8t_{6}}{2\mathbfcal{P}}-
\frac{3}{2(\mathbfcal{S}+\sqrt\mathbfcal{D})}
\left(\mathbfcal{S}_{6}+\frac{\mathbfcal{S} \mathbfcal{L}_{6}}{2\mathbfcal{L}}
\right)\left[\frac{\mathbfcal{E}(\nu)}{\mathbfcal{K}(\nu)}
-\frac{\mathbfcal{S}}{\mathbfcal{S}-\sqrt\mathbfcal{D}}
\right].\tikzmarkend{espqo}
    \end{split}
\end{align}\\
with the following definitions,
\begin{align}
\begin{gathered}
q=\mathbfcal{L}=t_{4}+t_{6}\mathbfcal{S},
\qquad \mathbfcal{P}=\mathbfcal{L}^{3}-4t_{6}^{2},
\qquad \sqrt\mathbfcal{D}=\frac{4}{\sqrt{\mathbfcal{L}}},
\\[2mm]
\mathbfcal{S}_{6}
=-\frac{\mathbfcal{S}^{2}\mathbfcal{L}^{2}+8t_{4}}{2\mathbfcal{P}},\,\,\,\,
\mathbfcal{L}_{6}
=\mathbfcal{S}-\frac{t_{6}\left(\mathbfcal{S}^{2}\mathbfcal{L}^{2}+8t_{4}\right)}{2\mathbfcal{P}},
\\[2mm]
\nu
=-\frac{2\sqrt\mathbfcal{D}}{\mathbfcal{S}-\sqrt\mathbfcal{D}}
=1-\frac{b^{2}}{a^{2}},\,\,\,\mathbfcal{E}(k^2):=\int_0^{\pi/2} \sqrt{1-k^2 \sin^2(\theta)}\,\,d\theta.\label{3.77u}
\end{gathered}
\end{align}
Now, after computing the matrix model correlators for the two-cut model, in the next section we proceed to compute the correlators using the techniques of \textit{CohFT and Givental's action}, to address the equality between matrix model and its equivalent dual worldsheet theory.\\

\section{A concise review of the Landau-Ginzburg theory} 
\label{sec44}
In this section, we briefly describe the main aspects of the topological LG theory \cite{Vafa:1990mu}. The main interesting aspect is that it does not depend on the worldsheet metric and hence topological. To contain the chiral ring of operators one often use to write down the B-twisted sigma model ($N=2$) on the Calabi-Yau target space ($\mathbfcal{T}_{cal}$) with holomorphic superpotential $W$, which has critical points being non-degenerate. The twisted action on the genus-g worldsheet with metric $h_{\mu\nu}$ is given by,
\begin{align}
\begin{split}
    S=&\int_{\Sigma} d^2\sigma \sqrt{h}\bigg(h^{\mu\nu}G_{\phi\bar\phi}\partial_{\mu}\phi\partial_{\nu}\bar\phi+\frac{1}{8}G^{\phi\bar \phi}(\partial_{\bar\phi}\bar W)(\partial_\phi W)-iG_{\phi\bar\phi}\chi D_{\bar\sigma}\rho_{\sigma}+iG_{\phi\bar\phi}\bar\chi D_{\sigma}\rho_{\bar\sigma}\\&\hspace{6 cm}+(\frac{1}{4}\nabla_{\phi}\partial_{\phi}W)\rho_{\bar\sigma}\rho_{\sigma}+(\frac{1}{4}\nabla_{\bar\phi}\partial_{\bar\phi}W)\chi\bar\chi\bigg)
    \end{split}
\end{align}
where, $\phi$ is the bosonic mode and it maps $\phi:\Sigma\to\mathbfcal{T}_{cal}$ and $G_{\phi\bar\phi}$ is the Ricci flat k\"ahler metric dwelling on the target space \cite{Labastida:1994ss}. The covariant derivatives $\{D,\nabla\}$ are pullbacks of the original derivative on the target space. Along with this $\chi,\bar\chi$ are fermionic scalars and and $\rho $ is an fermionic one-form. Now, an interesting fact of such topological LG theory is that, the worldsheet localizes on the constant maps $\phi(\sigma) =\alpha$, with $dW(\alpha)=0$, which are basically the location of the critical points of the superpotential. The bosonic and fermionic path integral determinants cancel by supersymmetry and the total contribution comes out of the zero modes. The resulting zero modes cotributes a non-vanishing top form $\Omega$ on the target, which sometimes are referred as the Calabi-Yau one form. The integral produces factors of Hessian $H_{\alpha}$ at each critical point, where,
\begin{align}
    H_{\alpha}=\frac{\nabla dW}{\Omega^2}\bigg|_{\alpha}
\end{align} 
This helps in computing all genus correlators along with the partition functions. The form of the correlators are given below,
\begin{align}
    \bigg\langle \prod_{i=1}^n F_{i}\bigg\rangle^{\text{LG}}_g=\sum_{\alpha}\Res_{\alpha}\bigg(\prod_{i=1}^n F_i\bigg)H^{g}\,\frac{\Omega^2}{dW}
\end{align}
Now, the chiral ring of indempotent primary operators satisfy the following correlation,
\begin{align}
     \bigg\langle \prod_{i=1}^n \mathcal{O}_{\alpha_i}\bigg\rangle^{\text{LG}}_g=\sum_{\alpha}\delta_{\alpha_1,\alpha_2\cdots ,\alpha_n,\alpha}\,H_{\alpha}^{g-1}
\end{align}
Finding these correlation functions will be useful later while constructing the integrand on the moduli space of Riemann surfaces. From the matrix model side, the large-$N$ saddles describe the target space of the dual string theory. B-twisting the $N=2$ sigma model with Calabi-Yau target and adding a superpotential defines the theory on the worldsheet. Hence, from the matrix model spectral curve $x(z)$ becomes the input for the worldsheet superpotential and this becomes a crucial ingredient. $dy(z)$ becomes the a top form (Calabi-Yau) and helps in normalising the Hessians. In the next section, we describe the construction of the cohomological field theory using Givental's description.
\section{Givental's action and CohFT construction} 
\label{sec55}
Using the recursive boundary structures for the moduli space of genus-$g$, $n$-punctured Riemann surfaces ($\mathcal{M}_{g,n}$), the Givental's action systematically organises the rotation ($\mathbfcal{R}$) and translation ($\mathbfcal{T}$) matrices using the preliminary basic unit CohFT. CohFT's satisfy three axioms of \textit{unity, gluing and symmetry}. For the reader we briefly describe the basics of CohFT in order to proceed to the construction of them for higher genus cases.
\subsection{Axioms of the Cohomological field theory}
\begin{figure}[b!]
    \centering
\includegraphics[width=0.6\linewidth]{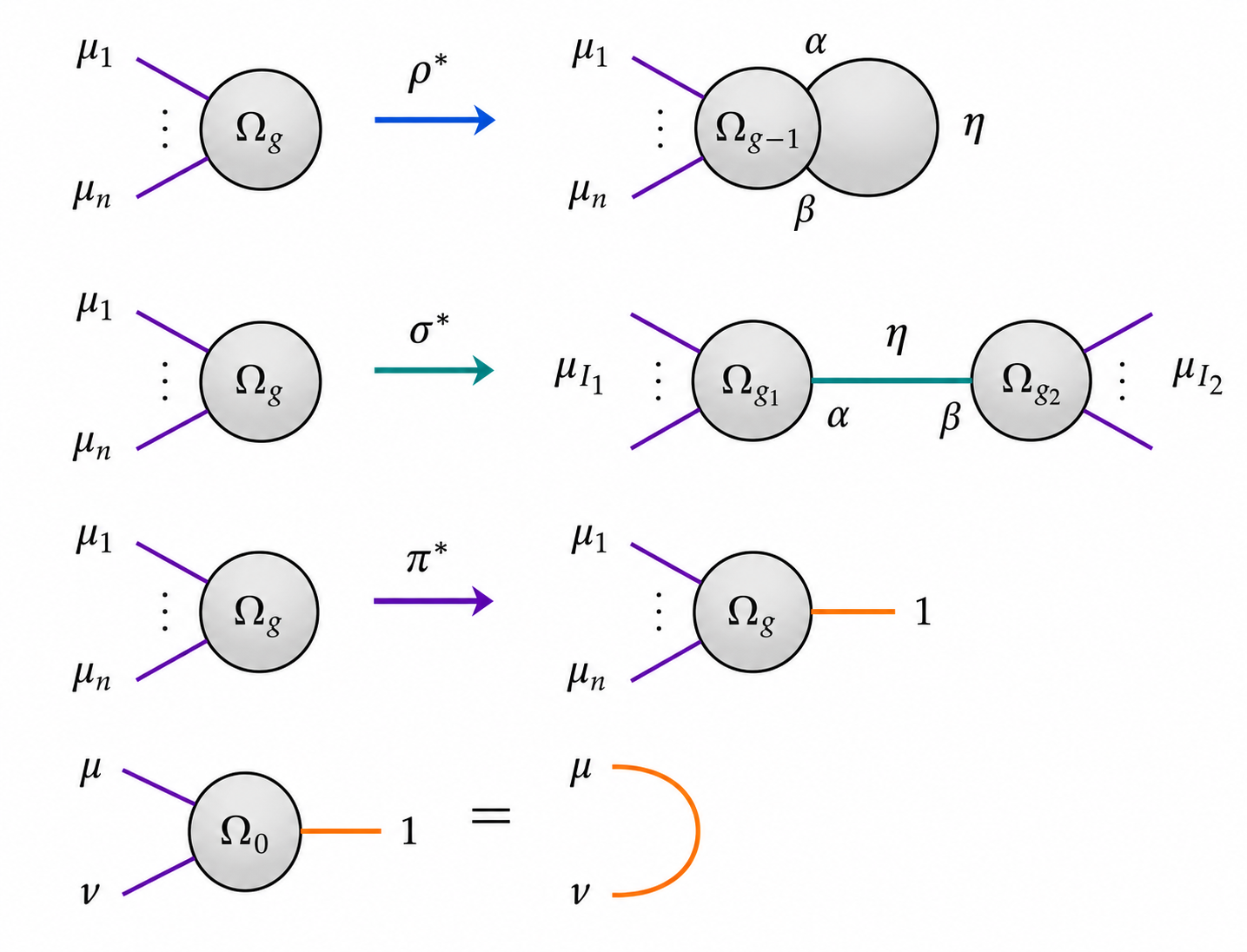}
    \caption{Different set of maps relevant for defining the cohomological field theory, with purple set of lines are \textit{leaves} and \textit{edges} are marked in green.
    These maps are very important for constructing the correlator of matter operators in CohFTs. }
    \label{fig:2}
\end{figure}

A \textit{cohomological field theory} defined on a vector space $(\mathcal{V},\eta)$ consists of a collection of linear maps,
\begin{align}
    \Omega_{g,n}: \mathcal{V}^{\otimes n}\to H^{2\bullet}(\mathcal{M}_{g,n},\mathbb{Q}), \,\,\,\,\,\Omega_{g,n}(e_{\mu_1} \otimes\cdots e_{\mu_n}):=\Omega_{g;\mu_1,\mu_2,\mu_3,\cdots \mu_n,}
\end{align}
The cohomological field theories related to the moduli spaces of Riemann surfaces, satisfy the three basic axioms defined as,\\ 
\par
\noindent
\textbf{Unity:} Due to the forgetful map, (it forgets about the last puncture on the Riemann surface of genus-$g$ and $n+1$ punctures) expressed as,
\begin{align}
   \pi:\,& \overline{\mathcal{M}}_{g,n+1}\to \overline{\mathcal{M}}_{g,n}
\end{align}
we need the following to hold, \begin{align}
    \pi^{*}\Omega_{g;\mu_1,\mu_2,\mu_3,\cdots \mu_n}=\Omega_{g;\mu_1,\mu_2,\mu_3,\cdots \mu_n,1} \,\,\,\,\,\,\,\text{and} \,\,\,\,\,\,\,\Omega_{0,\mu,\nu,1}=\eta_{\mu,\nu}
\end{align}
where,
\begin{align}
    \Omega_{g;\mu_1,\mu_2,\mu_3,\cdots \mu_n}=\Omega_{g,n}(e_{\mu_1}\otimes\cdots \otimes e_{\mu_n})
\end{align}
\noindent
\textbf{Symmetry:} It claims that, under the action of the symmetric permutation $S_n$ it permutes the copies of $V^{\otimes n}$ and the marked points of the moduli space $\overline{\mathcal{M}}_{g,n}$.\\

\noindent
\textbf{Gluing:} For the following gluing maps,
\begin{align}
\begin{split}
    &\rho:\overline{\mathcal{M}}_{g-1,n+2}\to \overline{\mathcal{M}}_{g,n},\\&
    \sigma: \overline{\mathcal{M}}_{g_1,1+|I_1|}\times \overline{\mathcal{M}}_{g_2,1+|I_2|}\to \overline{\mathcal{M}}_{g,n}, 
    \label{4.5t}
    \end{split}
\end{align}
with $g_1+g_2=g,$ $I_1\cup I_2=\{1,\cdots n\}$ requires to satisfy,
\begin{align}
\begin{split}
    &\rho^{*}\Omega_{g;\mu_1\cdots \mu_n}=\eta^{\alpha,\beta}\Omega_{g-1;\mu_1\cdots \mu_n,\alpha,\beta}\\&
    \sigma^{*}\Omega_{g;\mu_1\cdots \mu_n}=\eta^{\alpha,\beta}\Omega_{g_1;\alpha,\mu_{I_1}}\otimes\Omega_{g_2;\beta,\mu_{I_2}}
    \end{split}
\end{align}\\
These operations are demonstrated in Fig.~\eqref{fig:2}. Hence, defining the CohFT in the abstract way we proceed to construct the $\mathbfcal{R}$ and $\mathbfcal{T}$ matrices, which will be extremely relevant in the Givental's construction, in the following section.
\subsection{Validation  of the calculation so far}
The calculation admits a sequence of checks that can be applied independently. First, the planar resolvent must have the analytic behavior at infinity and the correct discontinuities across the two cuts. These conditions are important to fix the endpoint equations and the filling-fraction convention, which we solve explicitly in Appendix~\eqref{appA}. Second, the local series at each branch point must be consistent with the normalization of the square-root coordinate. Third, the extracted rotation data must satisfy its symplectic constraint to the order used in the integral of certain CohFT as depicted in \eqref{7.5u} and \eqref{7.6u}.\\

The worldsheet calculation has an equally clear set of internal tests. Dimension counting determines which powers of the elements of CohFT (rotation, translation and edge matrices) can contribute. The gluing rule determines the signs of the edge term. Known intersection numbers on low-dimensional moduli spaces determine the coefficients multiplying the relevant tautological classes. Once the ingredients are combined, the answer can be simplified into rational and period-dependent pieces (elliptic). Keeping these pieces separate until the last step is often the most effective way to diagnose a mismatch.\\

The matrix-model calculation supplies an independent verification rather than a restatement of the worldsheet result. The resolvent and free-energy derivations use the loop equations and large-$N$ expansion, while the worldsheet derivation uses the CohFT reconstruction and intersection theory. Their overlap is the spectral curve, but the computational routes are
otherwise distinct. A term-by-term equality therefore checks that the spectral data have been transferred correctly through two different formalisms. The elliptic part is especially informative because it is sensitive to the choice of normalized global differential, a datum that cannot be inferred from a single local residue.

\subsection{Constructing $\mathbfcal{R}$ and $\mathbfcal{T}$ matrices of the CohFT}
In this section we demonstrate how to compute the rotation and translation matrices. These are the main ingredients for defining the cohomological field theory in the Givental's description. considering a spectral curve,
\begin{align}
    {S}=(\Sigma,x,y,B)
\end{align}
having $r$ simple ramification points. We can prefer to choose local coordinates \(\zeta_{\mu}\) around a ramification point \(\mu\) such that the spectral curve $x$ coordinate becomes quadratic in the variable $\zeta_{\mu}$ ,
\begin{align}
    x=\zeta_{\mu}^{2}+x(\mu).
\end{align}
where, $\zeta_\mu$ are sometimes called the local Airy coordinates. We can express the auxiliary functions \(\xi^{\mu}\) and the associated meromorphic differentials \({d}\xi^{\mu,k}\), relevant to the spectral curve can be casted as,
\begin{equation}
\xi^{\mu}(z)
=\left.\int^{z}\frac{B(w,\cdot)}{{d}\zeta_{\mu}(w)}
\right|_{w=\mu},
\qquad{d}\xi^{\mu,k}(z)
={d}\left[\left(
-\frac{1}{\zeta_{\mu}}
\frac{{d}}{{d}\zeta_{\mu}}
\right)^{k}
\xi^{\mu}(z)
\right].
\end{equation}
Now, we can set
\begin{align}
 t^{\mu}
=-2\left.
\frac{{d}y(z)}
{{d}\zeta_{\mu}(z)}
\right|_{z=\mu}.
\end{align}
Define the $r$-copies of the \textit{trivial CohFT} on $\mathcal{V}=\bigoplus_{\mu=1}^{r}\mathbb{C}\cdot e_{\mu}.$
Now, after setting $ \eta(e_{\mu},e_{\nu})=\delta_{\mu,\nu}$ we can define $w_{g,n}$'s as,
\begin{align}
w_{g;\mu_{1},\ldots,\mu_{n}}
=\frac{
\delta_{\mu_{1},\ldots,\mu_{n}}}{
\bigl(t^{\mu_{1}}\bigr)^{2g-2+n}
}.
\end{align}
\noindent
One can define the translation $\mathbfcal{T}$ and rotation matrix $\mathbfcal{R}$ by setting,

\begin{align}
\begin{split}
  & \mathbfcal{R}_{\mu}^{\nu}(u)
=-\sqrt{\frac{u }{2\pi}}\int_{\gamma_{\nu}}
\exp\left(-\frac{x-x(\nu)}{2u}
\right)\mathrm{d}\xi^{\mu},\\&
\mathbfcal{T}^{\mu}(u)=\left(u\,t^{\mu}+
\frac{1}{\sqrt{2\pi u}}
\int_{\gamma_{\mu}}
\exp\left(-\frac{x-x(\mu)}{2u}
\right)
\omega_{0,1}\right).
\label{4.9i}
    \end{split}
\end{align}

\noindent
where, $\gamma_\ell$ is the steepest descent path for \(x(z)\) from the ramification point $\ell$, locally which can be taken
along the real axis in the \(\zeta_{\mu}\)-plane.\\

Through the \textit{Givental action}, we can then define a CohFT
\begin{equation}
\Omega_{g,n}
=\mathbfcal{R}\cdot \mathbfcal{T}\cdot w_{g,n}
\colon \mathcal{V}^{\otimes n}
\longrightarrow
H^{2\bullet}
\bigl(\overline{\mathcal{M}}_{g,n},
\mathbb{C}
\bigr).
\end{equation}
Hence, having defined the CohFT, we proceed to construct its components which are the rotation and translation matrices.
\subsubsection{Computing $\mathbfcal{R}$ and $\mathbfcal{T}$ matrices for genus-1 spectral curve}
As described in the above section, the main inputs are the $\mathbfcal{R}$ and $\mathbfcal{T}$ matrices for the CohFT according to the Givental's action. Before proceeding to compute the genus-1 spectral curve, we formally write down the matrices for the genus-0 spectral curve for quartic model as also found in \cite{Giacchetto:2026gpq},
\begin{align}
    \mathbfcal{R}_{\alpha}^{\alpha}= {_2}F_0\left(\frac{1}{2},-\frac{1}{2};;-\frac{u}{4\alpha \gamma}\right)
\end{align}
where, $\alpha$ is the solution defined in \eqref{B.7u}.
Now, similarly the $\mathbfcal{T}$ matrices are also given by,
\begin{align}
    \mathbfcal{T}^{\alpha}(u)=u\bigg(1- {_2}F_0\left(3/2,-1/2;;-\frac{u}{4\alpha\gamma}\right)+\frac{6t_4\alpha\gamma u}{1-6t_4\gamma^2}{_2}F_0\left(5/2,-3/2;;-\frac{u}{4\alpha\gamma}\right)\bigg);  
\end{align}
The matter operators $\mathcal{O}_{\alpha}$ are a part of the vertex operator as shown in \eqref{eq:final-correspondence}. 
Now, the degenerate cyllinder amplitudes are given by the edge term expressed as,
\begin{align}
    E^{\mu,\nu}(u',u''):=\frac{\eta^{\mu,\nu}-\mathbfcal{R}^{\mu}_{\rho}(u')\eta^{\rho,\sigma}\mathbfcal{R}^{\nu}_{\sigma}(u'')}{u'+u''}\label{4.12r}
\end{align}
where, $\eta_{\mu,\nu}:=\langle\mathcal{O}_{\mu}\mathcal{O}_{\nu}\rangle$ and $\eta^{\mu,\nu}$ is its inverse. one can find the expression of it after plugging in the expressions of $\mathbfcal{R}$ easily. 
But, the important point to note here is that we need to rederive the things from the $\mathbfcal{R}$ and $\mathbfcal{T}$ matrices. The reason is, for two cut models the Bergmann kernel has elliptic function inputs. Hence, we proceed to compute these quantities using the definitions of $\mathbfcal{R}$ and $\mathbfcal{T}$ matrices using \eqref{4.9i}. We also need to find the expression of the vertex operators for the two-cut model and worldsheet correspondence. We will discuss and evaluate in the next section.

\section{Dictionary of the matter correlator}\label{sec66}

\subsection{Interpreting the operator dictionary}

The operator dictionary can be read in both directions. Starting from the matrix model, the expansion of a planar resolvent at infinity selects a single-trace observable and its connected correlators. Starting from the worldsheet theory,
the label of an integrated vertex operator selects the same coefficient after the CohFT class has been integrated over the appropriate moduli space of Riemann surfaces as also advocated in the main claim of the paper. The claim is consequently more precise than a correspondence between partition functions. It identifies individual observables, including their dependence on the branch points and on the global periods of the two-cut solution.\\ \par

This interpretation explains why the normalization of a vertex operator must be fixed. A multiplicative change in a matrix-model observable can be absorbed into a convention for the loop insertion operator, while a multiplicative change in a worldsheet insertion can be absorbed into a convention for the state-space basis. The relative normalization becomes meaningful once the dictionary is imposed. The explicit genus-one calculation fixes this relative
normalization. The action of the translation is to introduce extra marked points of the tautological class and integrate over the position over the location of the marked point. Mathematically this can be done by pushforwarding the tautological class insertion (extra marked points) by the help of the forgetful map ($\pi$). This pushforward is reflected by the integration on the worldsheet.\\ \par

Matter operators forms the BRST closed chiral ring (Similar to described in \eqref{ring}). The main theme is to couple matter operators to gravitational descendents, whose effects are localized on the branch points of the spectral curve. In case of different multicut models the localization depends on the critical points of the superpotential, which in turn calibrates the chiral ring. The dictionary is local in the space of matrix-model couplings but global on
the spectral curve. A small variation of a potential coupling changes the resolvent through loop insertion and changes the endpoint positions through the saddle constraints. The latter change propagates into periods and local coordinate coefficients. On the worldsheet side, the same variation changes the numerical coefficients of the CohFT classes while leaving the universal intersection theory unchanged. This clean separation between universal tautological geometry and model-dependent spectral data is one of the main structural benefits of the reconstruction.
\noindent
The matter correlator one point function is given by \cite{Giacchetto:2026gpq},
\begin{align}
\begin{split}
    \bigg\langle\!\!\!\bigg\langle\prod_{i=1}^n \mathcal{O}_{k}\bigg\rangle\!\!\!\bigg\rangle^{LG+grav}_{g=1}&=\sum_{\beta}\mathbfcal{R}_{k}^{\beta}(\psi_1)\bigg(\langle \mathcal{O}_{\beta}\rangle^{LG}_{g=1}+\langle\mathcal{O}_{\beta}\mathcal{O}_{\sigma}\rangle^{LG}_{g=1}\pi_{*}\mathbfcal{T}^{\sigma}(\psi_2)\bigg)\\&\hspace{1.5 cm}+\frac{1}{2}\rho_{*}\sum_{\beta,\mu,\nu}\mathbfcal{R}_{k}^{\beta}(\psi_1)E^{\mu,\nu}(\psi',\psi'')\langle\mathcal{O}_{\beta}\mathcal{O}_{\mu}\mathcal{O}_{\nu}\rangle_{g=0}^{LG}
    \end{split}
\end{align}
where, $\rho$ denotes the gluing map advocated earlier in \eqref{4.5t}. It identifies two branches corresponding to one node. 
The reason of excising the disks around each puncture is the following:
Topological gravity effects can be merged into local marked points and nodes, on the moduli space of Riemann surfaces. As the effects of topological gravity on the moduli space localizes, we can excise small disks around these punctures as also advocated in \cite{Giacchetto:2026gpq}. Hence, gravitational corrections on top of the topological theory comes via disk amplitudes. Now, the effective gluing map and the forgetful map acts in this case as,
\begin{align}
\rho:\overline{\mathcal{M}}_{0,3}\longrightarrow\overline{\mathcal{M}}_{1,1},\,\,\,\,\,\,\,\,\,\,\,\pi:\overline{\mathcal{M}}_{1,2}\longrightarrow\overline{\mathcal{M}}_{1,1}.
\end{align} 
Now to compute the integral of the vertex operators on the compactified moduli space, we proeed to compute the $\mathbfcal{R},\mathbfcal{T}$ matrices, and use them to calculate the expectation value of the the matter primaries in the worldsheet Landau-Ginzburg theory including gravitational effects.
\subsection{Stable graphs and gluing}
The graph expansion in the CohFT can be understood directly from the geometry of stable curves. A stable graph assigns a genus to each vertex, a leg to every external insertion, and an internal edge to every node. \textit{Its total genus is the
sum of the vertex genera plus the first Betti number of the graph.}\\

\begin{figure}[htb!]
    \centering
\includegraphics[width=0.65\linewidth]{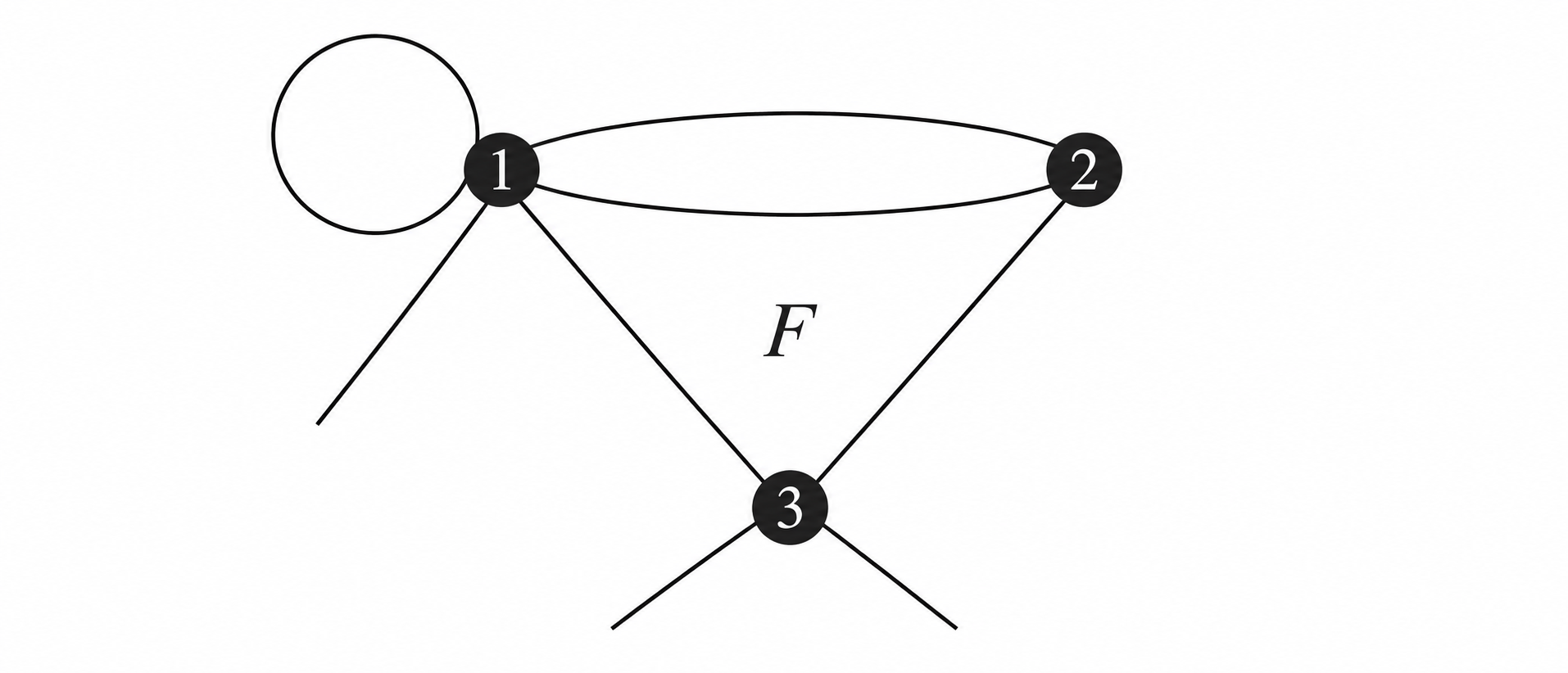}
    \caption{Figure depicting a stable graph with genus-1 surface with non-seperating degeneration is joined with other higher genus surfaces. }
    \label{fig:7}
\end{figure}

The effective genus of the graph is given by,
\begin{align}
    g(F):=\sum_{v\in V(F)}g(v)+\underbrace{h^1(F)}_{\text{Betti number}}
\end{align}
where, $V(F)$ denotes the number of vertices of the stable graph alongwith $h^1(F)$ denotes the first Betti number. The $g(F)$ coincides with the effective genus of the Riemann surface. Now, denoting $E(F)$ as the edge and $n(v)$ as the valency of the vertex for $v\in V(F)$. The betti number is given by $n-v+1$. where, $n$ is the number of internal edges along with $v$ being the independent irreducible vertices of the stable graph as shown in Fig.~\eqref{fig:7}.\\ \par

Dimension counting gives a useful check before integrations are evaluated.
Since $\overline{\mathcal M}_{1,1}$ has complex dimension one, only degree-one
classes can survive in its integral. Terms at excessively high order in a formal local series cannot contribute. Conversely, a term which appears
subleading in a local coordinate may be essential if it has the correct cohomological degree. It is consequently safer to apply dimension selection after identifying the geometric origin of a term, rather than using it to discard terms prematurely.

\section{ Discussion of the CohFT reconstruction}\label{sec77}

The CohFT formulation is valuable because it gives an intrinsic worldsheet answer. Rather than assigning a separate prescription to each correlator, it contains the information of the full collection of correlators into classes on the moduli spaces of stable curves. The gluing axioms ensure that these classes behave consistently when a curve degenerates. This consistency is not an optional condition: it is the geometric counterpart of the factorization expected of the matrix-model topological expansion.\\

The formal series $\mathbfcal R(z)$ obeys a symplectic condition of the form
\begin{equation}
 \mathbfcal R(z)\mathbfcal R^{\mathsf T}(-z)=\mathbf 1.
\end{equation}
This condition has a direct consequence.  It correlates different orders of the expansion and prevents the coefficients from being chosen independently. When computing a finite-order answer, verifying this relation to the same order is an efficient check. The translation series incorporates the effect of unstable data that is not captured by a pure rotation. In the language of stable graphs, it dresses the legs with weights. In local spectral-curve language, it is extracted from the regular part of the relevant expansion. The distinction between a singular local piece and a regular local piece is therefore reflected directly in the
distinction between rotation and translation. This is a particularly useful organizational principle in genus one, where terms that look similar in a local coordinate may contribute to different components of the CohFT data.\\ \par

An internal edge represents a node after normalization. For a separating node, it connects two components with lower number of genus, and for a non-separating node, it joins two marked points on the same component and raises the genus. The latter case is precisely the boundary geometry needed for the first handle contribution. One should note, that the same rotation matrix can be used with the data of the matter operator correlators (two-point) to reconstruct the edge matrices. The moduli-space integral should be understood as a finite computation at each fixed genus and number of insertions. Although the generating series contains
infinitely many descendants, dimension constraints eliminate all but finitely many terms. If $\dim_\mathbb C\overline{\mathcal M}_{g,n}=3g-3+n$, only the component of total cohomological degree matching this dimension contributes. \\

 At genus-one with one marked point, the complex dimension of the moduli is one. Hence, the calculation is especially transparent: only degree-one $\psi$ classes can survive. Nevertheless, the result is not trivial, because several sources can produce degree one, including a first-order leg correction and the boundary contribution associated with a non-separating edge. The matching therefore provides a nice way to check relative coefficients.

\begin{figure}
    \centering
    \includegraphics[width=0.84\linewidth]{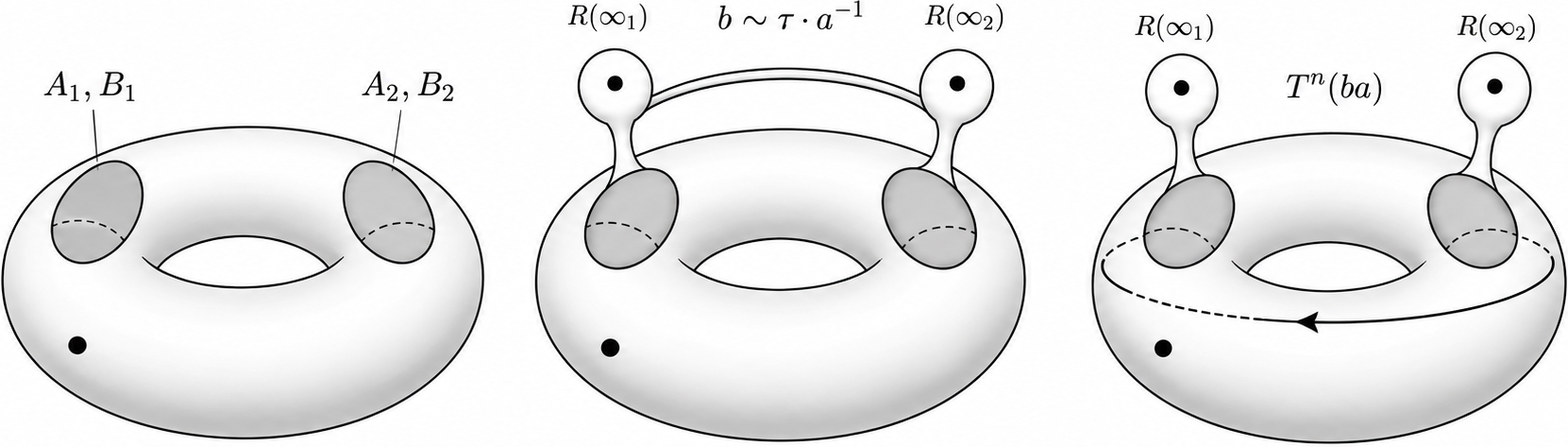}
    \caption{Schematic picture showing the genus-one disk amplitudes. Disk contributions are deep gray shaded. $\mathbfcal{T}$ generates translations and $\mathbfcal{R}$ is the rotation matrix in the CohFT. The gray disks without punctures in the middle (leftmost figure) does have a boundary condition at the circumference and these are technically the $\mathbfcal{T}$-matrices. While the gray disks with punctures in the middle (rightmost figure) does have a boundary condition at the circumference and also puncture in the middle. These are technically the $\mathbfcal{R}$-matrices. }
    \label{fig:8}
\end{figure}

\subsubsection{Computing $\mathbfcal{R}$-matrices for genus-1 curve }
For the Givental's action it is very important to compute the $\mathbfcal{R}$ matrices. We compute it in this section for two-cut model relevant at $g=1$.
The $\mathbfcal{R}$ matrix is given in the $\hbar$ expansion as follows,
\begin{align}
    \mathbfcal{R}=I+\hbar\mathbfcal{R}_1+\hbar^2\mathbfcal{R}_2^2+\mathcal{O}(\hbar^3)
\end{align}
Now, as the moduli space of our integral is $\overline{\mathbfcal{M}}_{1,1}$, only $\mathbfcal{R}_1$ will contribute, because 
\begin{align}
    \text{dim}_C\overline{\mathbfcal{M}}_{1,1}=1.
\end{align}
Now, to compute the $\mathbfcal{R}$-matrix, we use the thimble definition in \eqref{4.9i}, defined as,
\begin{align}
    \mathbfcal{R}_i^{\ i}(-\hbar)
\sim
\exp\!\left(\frac{e_i}{\hbar}\right)
\frac{\hbar}{\sqrt{2\pi\hbar}}
\int_{\Gamma_i}
\exp\!\left(-\frac{x}{\hbar}\right)
d\xi_i.
\end{align}
 In contrast to the genus-0 case, all puncture formula at every order is slightly complicated, and we do not focus on that here. As we will compute the rotation and translation matrices order by order, we take the leverage of replacing $u$ in \eqref{4.9i} by $\hbar$. Now, the CohFT rotation matrix has two components. The diagonal coefficient is derived in \eqref{D.22i} and is given by the following expression,
\begin{align}
(\mathbfcal{R}_1)_{ii}
=A_i
=2\eta_1(\tau)\chi_i^2-\frac{\sigma''(e_i)}{24\sigma'(e_i)}.
\label{7.5u}
\end{align}
where, $\eta_1(\tau)$ is related to the second Eisenstein series.

\noindent
Similarly the non-diagonal coefficient can be written as,
\begin{align}
    (\mathbfcal{R}_1)^{i\neq j }_{ij}=\chi_i\chi_j\bigg[\wp(u_i-u_j|\tau)+2\eta_1(\tau)\bigg]
    \label{7.6u}
\end{align}

\noindent
where, $\wp$ is the Weistress-P function. Now, we proceed to compute the translation matrix which is also a very important component of the calculation.
\subsubsection{Computing $\mathbfcal{T}$ matrix for genus-1 curve}
In this part we compute the translation matrices for genus-1 spectral curve. It is useful to define the action of the translation matrices in terms of the forgetful map. Under the forgetful map, 
\begin{align}
    \pi_r: \overline{\mathcal{M}}_{g,n+r}\to \overline{\mathcal{M}}_{g,n}
\end{align}
it forgets the last $r$ points of the original moduli space. The translation action is defined in terms of these forgetful maps as follows,
\begin{align}
    \hat{\mathbfcal{T}}\Omega_{g,n}=\sum_{r}\frac{1}{r!}(\pi_r)_*\prod_{a=1}^r\mathbfcal{T}(\psi_{n+a})
\end{align}
where, $\psi$ denotes the $\psi$-classes. Hence, the translation operation also pullbacks its action by the forgetful operation. In this way defining the translation matrices can generate the actual CohFT from the identity cohomology.\\

\noindent
Now, we proceed to compute the translation matrix for the genus-1 spectral curve.
The translation components are defined by the thimble integrals as follows,
\begin{align}
\mathbfcal{T}_i(\hbar)
=\hbar-\sqrt{\frac{\hbar H_i}{2\pi}}\int_{\Gamma_i}e^{-(x-e_i)/\hbar}\,dy.
\end{align}

\noindent
Now, expanding the $y$ coordinate locally,
\begin{align}
y(q_i)
=\Omega_i q_i\left(1+A_i q_i^2+B_i q_i^4+O(q_i^6)\right).
\end{align}

\noindent
For the present spectral curve, we get,
\begin{align}
\begin{split}
    A_i
=\frac{p'(e_i)}{2p(e_i)}
+\frac{\sigma''(e_i)}{8\sigma'(e_i)},\,\,\,\,\,\,\,\,\,\,
B_i
=\frac{p''(e_i)}{8p(e_i)}+\frac{\sigma'''(e_i)}{48\sigma'(e_i)}
-\frac{\sigma''(e_i)^2}{128\sigma'(e_i)^2}
+
\frac{p'(e_i)\sigma''(e_i)}
{16p(e_i)\sigma'(e_i)}.
\label{4.20i}
\end{split}
\end{align}

\noindent
Gaussian asymptotics then gives\footnote{For quick substitution in \eqref{4.20i},
\begin{align}
\begin{split}
p(e)=e(t_6e^2+c),
\qquad
p'(e)=3t_6e^2+c,
\qquad
p''(e)=6t_6e,
\end{split}
\end{align}
and
\begin{align}
\sigma'(e)=2e(2e^2-\mathbfcal{S}),
\qquad
\sigma''(e)=12e^2-2\mathbfcal{S},
\qquad
\sigma'''(e)=24e.
\end{align}},
\begin{align}
   \mathbfcal{T}_i(\hbar)
=-3A_i\hbar^2
-15B_i\hbar^3+O(\hbar^4)
\label{7.14u}
\end{align}
Again, in deriving the translation matrix we use the main formula advocated in \eqref{4.9i}.
The $\mathbfcal{T}$-matrices with one index $i$, can be interpreted as equivariant disk partition functions with $i$ representing a chosen boundary condition on the circumference of the disk, where as $\mathbfcal{R}$-matrices  with two indices $i,j$ defines the same equivariant partition function but now with the insertion of the dual matter operator (to $\mathcal{O}_j$) at the center and the same boundary condition $i$ as shown in Fig.~\eqref{fig:8}. Now, as we have the ingredients to compute the edge matrix, we gradually proceed to compute the edge term.
\subsubsection{Computing edge matrix for genus-1 curve}
The edge matrices given in \eqref{4.12r} are combination of elements of the $\mathbfcal{R}$ matrices. Below we compute them for our case explicitly. Clubbing terms of the rotation matrix we can explicitly write down the result for the $E$-edge matrix as follows where,
the diagonal entries are given by,
\begin{align}
E_{ii}^{(0,0)}
=-2E_2(\tau)\chi_i^2
+\frac{\sigma''(e_i)}{24\sigma'(e_i)}.
\end{align}

\noindent
For \(i\neq j\), the off-diagonal entries are
\begin{align}
E_{ij}^{(0,0)}
=
-\chi_i\chi_j
\left[
\wp(u_i-u_j|\tau)+2E_2(\tau)
\right].\end{align}

\noindent
With the following half-period assignment,
\begin{align}
    u_1=\frac{1+\tau}{2},
\qquad u_2=\frac{\tau}{2},\qquad
u_3=\frac12,
\qquad
u_4=0.
\end{align}
Now to simplify our computation we specify the values of the Weistress-P function at these half period points. 
Now defining,
\begin{align}
    F_1:=\wp\!\left(\frac12|\tau\right)+2E_2(\tau),\,
    F_\tau:=\wp\!\left(\frac{\tau}{2}|\tau\right)+2E_2(\tau),\,
F_{1+\tau}:=\wp\!\left(\frac{1+\tau}{2}|\tau\right)+2E_2(\tau)
\end{align}

\noindent
The off-diagonal covariant edge matrix is given by,
\begin{align}
E^{(0,0)}_{\mathrm{off}}
=-\begin{pmatrix}
0 & \chi_1\chi_2F_1 & \chi_1\chi_3F_\tau & \chi_1\chi_4F_{1+\tau}
\\
\chi_1\chi_2F_1 & 0 & \chi_2\chi_3F_{1+\tau} & \chi_2\chi_4F_\tau
\\
\chi_1\chi_3F_\tau & \chi_2\chi_3F_{1+\tau} & 0 & \chi_3\chi_4F_1
\\
\chi_1\chi_4F_{1+\tau} & \chi_2\chi_4F_\tau & \chi_3\chi_4F_1 & 0
\end{pmatrix}.
\end{align}\\

    Now, having all the ingredients of the rotation, translation and edge matrix found above, we proceed to compute the integral of the vertex operators over the moduli space of Riemann surfaces.\\
    
    Now, in the construction of the string side, the target space is assigned with the spectral curve (say $S$) itself, modulo the poles of $x$ and poles and zeros of the differential $dy$. $dy$ is the Calabi-Yau top form and it is holomorphic. One should note a crucial feature i.e the worldsheet localizes on the ramification points (or zeroes of $dx$) which in our case comprises of four points. Extra normalization of the non trivial cycle is fixed by the Bergmann kernel ($B$). For our two-cut spectral curve geometry we have the following identifications \cite{Giacchetto:2026gpq}:
    \begin{align}
      \mathbfcal{T}_{cal}=S/\{pts\},\,\,\,\,\,\, W=x,\,\,\, \,\,\,\Omega=dy,\,\,\,\,\, \,\,\,\,B=\omega_{0,2}
    \end{align}
    with this identification, we first compute the genus-0 correlator for our spectral curve and then generalise it to genus-1 surfaces gradually.
     \subsection{Genus-0 correlator for two-cut spectral curve}
 One can use the constructed $\mathbfcal{R}$ and $\mathbfcal{T}$ matrices above to compute the correlator of vertex operators like $\langle\mathcal{V}_6\mathcal{V}_6\mathcal{V}_6\rangle$ on the sphere. The correlator of the single trace operator in the matrix model side is already found in \eqref{3.44r}. But as the relevant moduli space of Riemann surfaces is given by $\overline{\mathbfcal{M}}_{0,3}$, one has to generalise this with two more insertions as the relevant stable curve from the worldsheet point of view (on sphere) is with three punctures. The relevant diagonal and off-diagonal components of the rotation and translation matrices are already computed in \eqref{7.5u},\eqref{7.6u},\eqref{7.14u} explicitly. We show the matching of the matrix model side computation and worldsheet derivation explicitly below for genus-0 Riemann surfaces. 
We use equation \eqref{A.14r},
\begin{align}
    \mu=\frac{t_4\mathbfcal{S}}{2}+\frac{t_6\mathbfcal{S}^2}{4}+\frac{t_6}{8} \mathbfcal{D}
\end{align}
Now, plugging in the following expression 
\begin{align}
\mathbfcal{D}=\frac{16}{t_4+t_6\mathbfcal{S}},
\end{align}
we can write down the following equation, which implies that for fixed $\mu$ and $t_4 $, the derivative with respect to $t_6$ vanishes,
\begin{align}
    0=\frac{d}{dt_6}\bigg[\frac{t_4}{2}\mathbfcal{S}+\frac{t_6}{4} \mathbfcal{S}^2+2\frac{t_6}{t_4+t_6\mathbfcal{S}}\bigg]
\end{align}
Hence, we get,
\begin{align}
    \mathbfcal{S}_6=\frac{d}{dt_6}\mathbfcal{S}=-\frac{\mathbfcal{S}^2q^2+8t_4}{2\mathbfcal{P}}
\end{align}

\noindent
After putting everything over the common denominator $\mathbfcal P$ (defined in \eqref{3.77u}), one obtains the three point function of single trace operators as follows,
\begin{align}
\frac{d^2 m_6^{(0)}}{d t_6^2}
=
\frac{3}{256\mathbfcal P}
\left[
4\mathbfcal L\mathbfcal S
\left(
12\mathbfcal S^4
+16\mathbfcal S^2\mathbfcal D
+3\mathbfcal D^2
\right)
-t_6\left(2\mathbfcal S^2+\mathbfcal D\right)
\left(4\mathbfcal S^4+16\mathbfcal S^2\mathbfcal D
+\mathbfcal D^2\right)\right].\label{7.32e}
\end{align}

\noindent
\textbf{Checks from worldsheet data.}
 The natural three-point function from the matrix model side  is given by,
\begin{align}
\bigg\langle
\text{Tr} \mathbf{M}^6\,\text{Tr} \mathbf{M}^6\,\text{Tr} \mathbf{M}^6\bigg\rangle_{g=0,c},
\end{align}
where, $c$ denotes the connected three point function of single trace operators. From the worldsheet side of the computation, the worldsheet moduli space is $\overline{\mathcal M}_{0,3}$. Now, the complex dimension,
\begin{align}
    \dim_{\mathbb C}\overline{\mathcal M}_{0,3}=0.
\end{align}
Hence, there are no descendant, $\mathbfcal{R}$, $\mathbfcal{T}$, or edge contributions. In particular,
$v^i_{6,1}$ does not enter and only $v^i_{6,0}$ contributes. The four critical values are mentioned earlier as the edges of the symmetric double cut matrix model,
\begin{align}
    x_i\in\{-a,-b,a,b\}.
\end{align}
 The matter operators obey the following relation,
\begin{align}
    \mathcal{O}_i \mathcal{O}_j=\delta_{ij}\mathcal{O}_i,
\end{align}
with the following three point functions,
\begin{align}
    \left\langle \mathcal{O}_i\mathcal{O}_j\mathcal{O}_k\right\rangle^{\rm LG}_{0}
=\delta_{ij}\delta_{ik}\,H_i^{-1},
\qquad H_i=
\frac{8}{p(x_i)^2\sigma'(x_i)}.\label{7.30u}
\end{align}
where, $H_i $ is the hessian.
Thus, the primary coefficient of the $\Tr M^6$ vertex is (The derivation is given later  in \eqref{4.40i}),
\begin{align}
    v^i_{6,0}
=\frac{3\left(\mathbfcal S^2+4x_i^4\right)}
{8p(x_i)}.\label{7.31w}
\end{align}

\noindent
Therefore, the vertex operator of the worldsheet side is given by the following combination without the contribution of higher $\psi$-classes,
\begin{align}
    \mathcal V_6\big|_{\overline{\mathcal M}_{0,3}}
=\sum_{i=1}^4 v^i_{6,0}\mathcal{O}_i.
\end{align}
and finally the result can be shown to be equal to,
\begin{align}
\begin{split}
C^{(0)}_{666}
=\frac{27}{64\mathcal P}\Bigg\{&4\mathbfcal L\mathbfcal S
\left[12\mathbfcal S^4+16\mathbfcal S^2\mathbfcal D+3\mathbfcal D^2\right]
\\
&-t_6\left(2\mathbfcal S^2+\mathbfcal D\right)\left[4\mathbfcal S^4
+16\mathbfcal S^2\mathbfcal D
+\mathbfcal D^2\right]\Bigg\}.
\end{split}
\end{align}
The derivation is given in Appendix~\eqref{appD1}. Now one can check that, the result agrees with what obtained from purely matrix model side in \eqref{7.32e} apart from the over all normalization of 36. This factor of 36 is missing in matrix model side. This is easy to fix because one should remember that every derivative should be multiplied with a factor of 6, to make it a correlator of $\text{Tr}(M^6)$ and not $\frac{1}{6}\text{Tr}(M^6)$. Hence it becomes an exact match. Now, we proceed to verify the computation for the genus-1 surface, where elliptic uniformization becomes important.

\subsection{Explicit crosschecks on Matrix model and worldsheet data for genus-1 surface}
We will show one explicit example of matching the matrix model single trace correlator with the insertion of the operator $\text{Tr}(M^6)$, with the string theory vertex operator correlator $\mathcal{V}_6$. The definition of the vertex primaries in terms of the matter operators are given by,
\begin{align}
    \mathcal{V}_k=\sum_i\sum_{m=0}^{k-1} v^i_{k,m}\mathcal{O}_i\psi^{m}\label{4.34r}
\end{align}
where, the expansion coefficients are given by the residue formula,
\begin{align}
    v^i_{k,m}=\textbf{Res}_{x=\infty}\frac{k!}{(k-m)!}x^{k-m}\theta_i
\end{align}
Now, for $g=1$ only $m=0,1$ can contribute in \eqref{4.34r}. Hence,
\begin{align}
    \mathcal{V}_6=\sum_{i=1}^4 v^i_{6,0}\mathcal{O}_i+v^i_{6,1}\mathcal{O}_i\psi_1\label{4.36t}
\end{align}
Now, we compute the residues to figure out the expansion coefficients in \eqref{4.36t} using residue computation. They are the most crucial element for figuring out the basis. We will calculate the residues using the local Airy coordinates.    
\subsubsection{Direct residue computation of $v^i_{6,0}$ and $v^i_{6,1}$}

We use the physical-sheet puncture above $x=\infty$ and set $t=\frac{1}{x}$. Since,
\begin{equation}
\eta=\sqrt{\sigma(x)}\sim x^2\sqrt{1-\mathbfcal S t^2+\Pi t^4},
\end{equation}
we obtain the differential to be given by,
\begin{align}
    \begin{split}
       du&=
-\frac{dt}{\omega_A}
\left[
1+\frac{\mathbfcal S}{2}t^2
+\left(
\frac{3\mathbfcal S^2}{8}-\frac{\Pi}{2}
\right)t^4
+O(t^6)
\right],\\& 
u-u_\infty
=-\frac{1}{\omega_A}\left[
t+\frac{\mathbfcal S}{6}t^3
+\left(\frac{3\mathbfcal S^2}{40}-\frac{\Pi}{10}
\right)t^5
+O(t^7)
\right].   
    \end{split}
\end{align}
Now, with this in hand we compute the residues to determine the expansion coefficients in the vertex operator.
Define,
\begin{equation}
F_i(z):=\wp(z-u_i)+2\eta_1.
\end{equation}
\noindent
A straightforward Laurent expansion around $u=u_\infty$ gives
\begin{equation}
v^i_{6,0}
=-\frac{1}{30\omega_A^7 p_i\sigma'_i}
\left[
F_i^{(5)}
+20\mathbfcal S\omega_A^2F_i^{(3)}
+8(8\mathbfcal S^2-9\Pi)\omega_A^4F_i^{(1)}
\right]_{u=u_\infty}
\label{4.40i}
\end{equation}
where, $F_i^{(n)}$ symbolises the $`n$'-th order derivative with respect to $u$ alongwith $p_i=p(x_i)
=x_i\left(t_6x_i^2+t_4+\frac{t_6\mathbfcal{S}}{2}
\right),$ and $\sigma'_i=\sigma'(x_i)$, with\footnote{While differentiating we use, \begin{align}
    \bigl(\wp'(z)\bigr)^2 = 4\bigl(\wp(z)\bigr)^3 - g_2\,\wp(z) - g_3.
\end{align}}
\begin{equation}
v^i_{6,1}
=\frac{1}{\omega_A^6p_i\sigma'_i}
\left[
F_i^{(4)}
+10\mathbfcal S\omega_A^2F_i^{(2)}
+(9\mathbfcal S^2-12\Pi)\omega_A^4F_i
\right]_{u=u_\infty}
\label{4.41i}
\end{equation}

\noindent
These two equations are the direct residue result, no genus-one free-energy formula has been used. Now, as clear from \eqref{4.40i}, \eqref{4.41i} we simplify the derivatives of the Weierstrass-P function below, using the definitions of the elliptic curve in terms of  Weierstrass-P.\\ \par
\noindent

\subsection{Global elliptic data and limiting checks}

The elliptic functions in the calculation are best viewed as coordinate descriptions of the same torus. A Weierstrass function emphasizes the period lattice and is efficient for quasi-period identities. A term algebraic in a quadratic coordinate can involve complete elliptic integrals in a Weierstrass coordinate, and conversely. The complete integrals $\mathcal K$ and $\mathcal E$ are geometric period data, not accidental special functions. Their ratio is dimensionless and often remains after common scales cancel. In a symmetric two-cut solution, it measures the response of the normalized differential to the separation of the cuts. As endpoints approach, the torus approaches a degeneration, and the period ratio tracks that change continuously. This is the description of the non-rational or elliptic part of the result.\\ \par

The choice of cycles must be held fixed throughout the comparison. Choosing the $A$-cycle fixes the normalization of the holomorphic differential, while the dual $B$-cycle determines the modular parameter. An integral change of cycle basis changes the presentation of periods but not the underlying curve. Several limits offer independent checks. When neighboring branch points coalesce, the period integrals have the expected degeneration asymptotics and the free energy should display matching singular behavior. In the symmetric
limit, odd moments vanish whenever the contour and filling fractions respect $x\mapsto-x$. Finally, when the two-cut geometry degenerates to a one-cut curve, the elliptic expressions reduce to the rational one-cut data after an appropriate choice of degeneration variable. These checks test endpoint normalization, symmetry, and period dependence independently of the main
correlator comparison.

\subsubsection{Evaluation of the Weierstrass data at the puncture}
In this section, we compute the residue calculus and use the half period values to match the worldsheet data to the matrix model computations.
Though it seems we choose a different uniformization, they are connected by Landen transform and they give isomorphic unifromization.
We give the details in appendix~\eqref{appe}.
Now as the $x(u)$ of the spectral curve data, is a degree-two elliptic function, it is a M\"obius transform
of $w(u):=\wp(u|\tau)$:
\begin{equation}
x(u)
=
b-\frac{C}{w(u)+\delta_0},\label{7.42i}
\end{equation}
where, comparison with
\begin{equation}
\left(\frac{dx}{du}\right)^2
=\omega_A^2\sigma(x)\label{7.43o}
\end{equation}
fixes the constants as follows,
\begin{equation}
C=-\frac{1}{2}\omega_A^2b\Delta,
\qquad
\delta_0
=
\frac{\omega_A^2}{12}(a^2-5b^2).
\end{equation}

\noindent
Now, the Weierstrass-P function is evaluated at the half periods to extract the expansion coefficients of the chiral ring. The half-period values are given as follows,
\begin{align}
\begin{split}
e^{\wp}_1
:&=\wp\!\left(\frac{1}{2}\right)
=-\frac{\omega_A^2}{12}
(a^2+6ab+b^2),\,\,\,\,
e^{\wp}_2:=\wp\!\left(\frac{\tau}{2}\right)
=\frac{\omega_A^2}{6}(a^2+b^2),\\&
\,\,\,\,\,\,\,\,\,e^{\wp}_3
:=\wp\!\left(\frac{1+\tau}{2}\right)
=-\frac{\omega_A^2}{12}
(a^2-6ab+b^2).\label{7.45i}
\end{split}
\end{align}

\noindent
Now, using the equation of motion for a standard elliptic curve, the corresponding invariants are given by,\footnote{To fix the invariants we use \begin{align}
\begin{split}
    &\wp(u)'^2=4\wp(u)^3-g_2\wp(u)-g_3
    =4(\wp-e_1^{\wp})(\wp-e_2^{\wp})(\wp-e_3^{\wp})
    \end{split}
\end{align}. Now, from here one can extract, \begin{align}
\begin{split}
    &g_2=-4(e_1^{\wp}e_2^{\wp}+e_2^{\wp}e_3^{\wp}+e_1^{\wp}e_3^{\wp}),\,\,\,\,\,
    g_3=4(e_1^{\wp}e_2^{\wp}e_3^{\wp})
    \end{split}
\end{align}}
\begin{equation}
g_2
=\frac{\omega_A^4}{12}
(\mathbfcal S^2+12\Pi),
\qquad
g_3=
\frac{\omega_A^6}{216}
\mathbfcal S(\mathbfcal S^2-36\Pi).
\end{equation}

\noindent
At the physical infinity one has $w(u_\infty)=-\delta_0$. Using the
half-period shift formula for $\wp$, one obtains the particularly useful
identities
\begin{align}
\begin{split}
&\wp(u_\infty-u_i)
=\frac{\omega_A^2}{12}
(6x_i^2-\mathbfcal S),\,\,\,\,\,\,\,\,\,\,\,\,
\wp'(u_\infty-u_i)
=-\frac{\omega_A^3}{4}\sigma'(x_i).
\end{split}
\end{align}

\noindent
Now use the derivative property of the Weierstrass function,
\begin{align}
\begin{split}
&\wp''=
6\wp^2-\frac{g_2}{2},
\,\,\wp'''
=12\wp\wp',\\&
\wp^{(4)}
=120\wp^3-18g_2\wp-12g_3,\,\,
\wp^{(5)}
=(360\wp^2-18g_2)\wp'.
\end{split}
\end{align}

\noindent
Substitution into the residue formulas collapses $v^i_{6,0}$ to
\begin{equation}
v^i_{6,0}
=\frac{3(\mathbfcal{S}^2+4x_i^4)}{8p(x_i)}.
\end{equation}

\noindent
For $v^i_{6,1}$ define the dimensionful quasi-period combination, $\epsilon
:=\frac{2\eta_1(\tau)}{\omega_A^2}.$ Then the coefficient becomes,
\begin{equation}
v^i_{6,1}
=\frac{1}{
4p(x_i)\sigma'(x_i)
}
\bigg(30(3\mathbfcal{S}^2-4\Pi)x_i^2
-\mathbfcal{S}(3\mathbfcal{S}^2+92\Pi)+
\underbrace{12(3\mathbfcal{S}^2-4\Pi)\epsilon}_{\text{Elliptic sector contributions}}\bigg)
\end{equation}\\
Now, we focus on reproducing the elliptic sector factor we obtained from the matrix model while computing the correlator of single-trace operator $\text{Tr}(M^6)$ in the sextic model.
\subsubsection{Replacing the quasi-period by $\mathcal{E}/\mathcal{K}$}

Introducing,
\begin{equation}
\nu
=1-\frac{b^2}{a^2}
=-\frac{2\Delta}{\mathbfcal{S}-\Delta}.
\end{equation}

\noindent
The Legendre modulus naturally associated with the chosen Weierstrass half-period ordering is,
$\lambda
=-\frac{(b-a)^2}{4ab}.$ The standard quasi-period relation and the quadratic transformation
\begin{align}
\mathcal{K}(\lambda)
=
\sqrt{\frac{b}{a}}\,
\mathcal{K}(\nu),\,\,\,\,\,\,\
\mathcal{E}(\lambda)
=\frac{\mathcal{E}(\nu)+(b/a)\mathcal{K}(\nu)}{
2\sqrt{b/a}}
\end{align}
gives,
\begin{equation}
\epsilon
=\frac{\mathbfcal{S}}{12}
-\frac{a^2}{2}
\frac{\mathcal{E}(\nu)}{\mathcal{K}(\nu)}.
\end{equation}

\noindent
Hence, the descendant coefficient can be written directly in matrix-model
variables as
\begin{equation}
v^i_{6,1}
=
\frac{3}{2p(x_i)\sigma'(x_i)}
\left[
5(3\mathbfcal{S}^2-4\Pi)x_i^2
-16\mathbfcal{S}\Pi
-a^2(3\mathbfcal{S}^2-4\Pi)
\frac{\mathcal{E}(\nu)}{\mathcal{K}(\nu)}
\right].
\end{equation}

\noindent
For the ordering $(-a,-b,a,b)$ the coefficients are therefore given by,
\begin{align}
\begin{split}
&v^1_{6,0}=-V_a,
\qquad
v^2_{6,0}=-V_b,
\qquad
v^3_{6,0}=V_a,
\qquad
v^4_{6,0}=V_b,\\&
v^1_{6,1}=U_a,
\qquad
v^2_{6,1}=U_b,
\qquad
v^3_{6,1}=U_a,
\qquad
v^4_{6,1}=U_b,
\end{split}
\end{align}
with
\begin{equation}
V_a
=
\frac{3(\mathbfcal{S}^2+4a^4)}
{8a(t_6a^2+c)},
\qquad
V_b
=
\frac{3(\mathbfcal{S}^2+4b^4)}
{8b(t_6b^2+c)},
\end{equation}
\begin{align}
\begin{split}
&U_a
=-\frac{3}
{4a^2\Delta(t_6a^2+c)}
\left[
5(3\mathbfcal{S}^2-4\Pi)a^2
-16\mathbfcal{S}\Pi
-a^2(3\mathbfcal{S}^2-4\Pi)
\frac{\mathcal{E}}{\mathcal{K}}
\right],\\&
U_b
=\frac{3}
{4b^2\Delta(t_6b^2+c)}
\left[
5(3\mathbfcal{S}^2-4\Pi)b^2
-16\mathbfcal{S}\Pi
-a^2(3\mathbfcal{S}^2-4\Pi)
\frac{\mathcal{E}}{\mathcal{K}}
\right].
\end{split}
\end{align}
Next, we proceed to compute the main ingredients to define the CohFT from the unital one, using Givental's prescription. These translation, rotation and edge matrices appears naturally while studying recursive boundary structures for a Riemann surface with fixed genus and number marked points. Boundary recursive structures defines the \textit{strata} ($\mathbfcal{B}_{g,n}$), which in turn fixes the stable graphs. The strata comes out of degenerations of the initial moduli space. Hence, the complex dimension of it is always smaller than the original moduli space. 

\subsection{$\mathbfcal{R}$-matrix, translation and edge at the required order}
In this section we compute and explicitly demonstrate the rotation and translation matrices upto specific order, relevant for our computation of vertex operator correlator on the worldsheet to match with the matrix model single trace correlator.
The thimble convention is most cleanly written as,
\begin{equation}
\widehat{\mathbfcal{R}}(-\hbar)
=I-\hbar\widehat{\mathbfcal{R}}_1+O(\hbar^2).
\end{equation}

\noindent
The normalized coefficient is given by,
\begin{align}
\begin{split}
&(\widehat{\mathbfcal{R}}_1)_{ii}
=2\eta_1\chi_i^2
-\frac{\sigma''(x_i)}{24\sigma'(x_i)},\,\,\,\,\,\,
(\widehat{\mathbfcal R}_1)_{ij}
=\chi_i\chi_j
\left[\wp(u_i-u_j)+2\eta_1\right],
\qquad
i\neq j.
\end{split}
\end{align}
where, the diagonal part and the non diagonal part of the rotation ($\mathbfcal R$-matrix) matrix are given in the above equation. 
\noindent
For the covariant action on insertion labels used below,
\begin{equation}
(\mathbfcal R_1)^j{}_i
=\sqrt{\frac{H_j}{H_i}}\,
(\widehat{\mathbfcal R}_1)_{ji}.
\end{equation}

\noindent
It is convenient to divide every half-period value by $\omega_A^2$. Define,
\begin{equation}
f_{ij}:=\frac{\wp(u_i-u_j)+2\eta_1}{
\omega_A^2}.
\end{equation}

\noindent
Then the non diagonal and diagonal entries of the rotation matrix are given by,
\begin{align}
\begin{split}
&(\mathbfcal{R}_1)^j{}_i=\frac{2p(x_i)}{p(x_j)\sigma'(x_j)}
f_{ij},\,\,\,\,\,\,(\mathbfcal{R}_1)^i{}_i=\frac{2\epsilon}{\sigma'(x_i)}-\frac{\sigma''(x_i)}{24\sigma'(x_i)}.
\end{split}
\end{align}

\noindent
Now, the three distinct half-period values are given by,
\begin{align}
\begin{split}
&
f_{12}=f_{34}
=\epsilon-\frac{\mathbfcal{S}+6ab}{12},\\&
f_{13}=f_{24}
=\epsilon+\frac{\mathbfcal{S}}{6},\,\,
f_{14}=f_{23}
=\epsilon-\frac{\mathbfcal{S}-6ab}{12}.
\label{3.69u}
\end{split}
\end{align}

\noindent
For the translation, writing locally,
\begin{equation}
y(q_i)
=\Omega_i q_i\left(1+A_iq_i^2+O(q_i^4)
\right).\end{equation}

\noindent
The thimble Gaussian integral gives
\begin{equation}
\mathbfcal{T}_i(\hbar)
=\mathbfcal{T}^i_2\hbar^2+O(\hbar^3),
\qquad \mathbfcal{T}^i_2=-3A_i
,
\end{equation}
 with the following definition,
\begin{equation}
A_i=\frac{p'(x_i)}{2p(x_i)}+\frac{\sigma''(x_i)}{8\sigma'(x_i)}.
\end{equation}

Because, the leg is dressed by $\mathbfcal{R}(-\psi)$, the degree-one leg term is
$-\mathbfcal{R}_1\psi$. With the same convention the constant edge is $+\mathbfcal{R}_1$.
This sign is important for the final $11/24$ diagonal coefficient shown in the next section.

\section{The $\overline{\mathcal{M}}_{1,1}$ integral}
\label{sec88}
In this section we compute the integral over the compactified moduli space of genus-1 Riemann surface with a puncture.
The standard integral of the intersections are given by following,
\begin{equation}
\int_{\overline{\mathcal{M}}_{1,1}}\psi_1
=\frac{1}{24},
\qquad\int_{\overline{\mathcal{M}}_{1,1}}\kappa_1
=
\frac{1}{24},
\qquad
\int_{\overline{\mathcal{M}}_{1,1}}
[\partial\overline{\mathcal{M}}_{1,1}]
=
\frac{1}{2}.
\end{equation}

\noindent
Now, we proceed to compute the integral of the vertex operator which consists of coefficients having both the rational and elliptic contributions.
Therefore, there are four contributions:
\begin{itemize}
    \item the descendant $v^i_{6,1}\psi_1$
    \item the $R(-\psi)$ dressing of the external leg
    \item one forgotten $T_2$ marking, giving $T_2\kappa_1$
    \item the nonseparating edge for genus-1 on the boundary divisor
\end{itemize}
\noindent
Using the TFT correlators above, the integrated amplitude is,
\begin{equation}
\begin{aligned}
\mathcal{I}_6
&:=
\int_{\overline{\mathcal{M}}_{1,1}}
\left\langle\!\left\langle
\mathcal{V}_6
\right\rangle\!\right\rangle^{\mathrm{LG+grav}}_{g=1}
\\[2mm]
&=
\frac{1}{24}\sum_i v^i_{6,1}
+\frac{1}{24}\sum_i v^i_{6,0}\mathbfcal{T}^i_2
-\frac{1}{24}
\sum_{i\neq j}
v^i_{6,0}(\mathbfcal{R}_1)^j{}_i
+
\frac{11}{24}
\sum_i
v^i_{6,0}(\mathbfcal{R}_1)^i{}_i .
\end{aligned}
\end{equation}

\noindent
The coefficient $11/24$ does simply come out of,
\begin{equation}
-\frac{1}{24}
+\frac{1}{2}
=
\frac{11}{24}.
\end{equation}
the first term is the diagonal part of the external $R(-\psi)$ leg and the
second is the constant non-separating edge integrated over the boundary
divisor.

\subsection{Matching the exactly to the matrix-model answer}

Substitute the explicit expressions for $v^i_{6,0}$, $v^i_{6,1}$,
$T^i_2$ and $R_1$ into the boxed moduli-space integral. The calculation
now consists only of a finite sum over $x_i\in\{-a,-b,a,b\}.$
Using, 
\begin{align}
    \begin{split}
&\sigma'(x)
=2x(2x^2-\mathbfcal{S}),
\qquad
\sigma''(x)
=12x^2-2\mathbfcal{S},\\&
p(x)
=x\left(t_6x^2+t_4+\frac{t_6\mathbfcal{S}}{2}
\right),
\end{split}
\end{align}
and the three half-period values $f_{12},f_{13},f_{14}$ given above in \eqref{3.69u}. All dependence on $ab$ cancels from the final scalar amplitude. The endpoint variation combinations appearing in the matrix-model result are,
\begin{align}
    \begin{split}
     &\mathbfcal{S}_6
=-\frac{
\mathbfcal{S}^2\mathcal{L}^2+8t_4
}{2\mathbfcal P},\,\,\,\,\,\,\mathcal{L}_6=\mathbfcal{S}+t_6\mathbfcal{S}_6
=\mathbfcal{S}-\frac{
t_6(\mathbfcal{S}^2\mathcal{L}^2+8t_4)}{
2\mathbfcal P}.   
    \end{split}
\end{align}

\noindent
After imposing only the planar endpoint relation,
\begin{equation}
\Delta^2\mathbfcal{L}
=
16,
\end{equation}
the finite CohFT sum simplifies exactly to,
\begin{equation}
\begin{aligned}
\mathcal{I}_6
=
-\frac{3\mathbfcal{L}_6}{2\mathbfcal{L}}
+
\frac{
3\mathbfcal{L}^2\mathbfcal{L}_6-8t_6
}{2\mathbfcal P}-
\frac{3}{2(\mathbfcal{S}+\Delta)}
\left(\mathbfcal{S}_6
+\frac{\mathbfcal S\mathbfcal{L}_6}{2\mathbfcal{L}}\right)
\left[\frac{\mathbfcal{E}(\nu)}{\mathbfcal{K}(\nu)}
-\frac{\mathbfcal{S}}{\mathbfcal{S}-\Delta}\right].
\end{aligned}
\end{equation}

\noindent
Since,
\begin{equation}
\nonumber
\Delta
=\sqrt{\mathbfcal D}
=\frac{4}{\sqrt{\mathbfcal{L}}},
\end{equation}
this is precisely the matrix model correlator,
\begin{equation}
\begin{aligned}
\left\langle
\operatorname{Tr}M^6
\right\rangle_{g=1}
={}&
-\frac{3\mathbfcal{L}_6}{2\mathbfcal{L}}
+\frac{3\mathbfcal{L}^2\mathbfcal{L}_6-8t_6}{
2\mathbfcal P}
\\[1mm]
&-\frac{3}{2(\mathbfcal{S}+\sqrt{\mathbfcal D})}\left(\mathbfcal{S}_6
+\frac{\mathbfcal S\mathbfcal{L}_6}{2\mathbfcal{L}}\right)\left[
\frac{\mathbfcal{E}(\nu)}{\mathbfcal{K}(\nu)}
-\frac{\mathbfcal{S}}{\mathbfcal{S}-\sqrt{\mathbfcal D}}
\right].
\end{aligned}
\end{equation}

Thus,
\begin{equation}
\int_{\overline{\mathcal{M}}_{1,1}}
\left\langle\!\left\langle
\mathcal{V}_6
\right\rangle\!\right\rangle^{\mathrm{LG+grav}}_{1}
=\left\langle
\operatorname{Tr}M^6
\right\rangle_{g=1}=m^{(1)}_6.
\end{equation}

\noindent
Hence, it shows that one can match the correlators of the matrix model single trace operators to the integral of vertex operators on the worldsheet of different genus. We also emphasize that taking the $a\to 0$ for the symmetric sextic well does not automatically reproduce the genus-0 spectral curve result as it seems for the merging of cuts. It produces some divergences in the elliptic sector for the correlator of single-trace operators like $\text{Tr}(M^6)$. One should note that it only matches under the double scaling limit adapted in a proper way.

\section{Additional checks and limiting regimes}
\label{sec99}

Several limiting regimes provide useful checks on the construction.

\begin{itemize}

\item The first is the degeneration in which two neighboring branch points approach each other. For a genus-0 Riemann surface with four punctures, it leads to degenerations into different channel field theory diagrams. In case of the torus the elliptic modulus then approaches a boundary value, and the curve develops a pinched cycle. In this limit, the period integrals have well-known logarithmic or regular asymptotics depending on the cycle under consideration. A correctly normalized matrix-model result and a correctly normalized CohFT result must display the same asymptotic behavior. Though there are subtlties in reducing the double cut result to a single cut as mentioned earlier, there should be a precise double scaling limit. Though, our calculations are ambitious without the calibration of double-scaling, it is important to verify the results in the double-scaling limit.

\item The second check is the symmetric limit. Because the potential and the cut
configuration are invariant under $x\mapsto-x$, odd moments vanish when the integration contour and filling fractions preserve this symmetry. This gives a simple way for intermediate formulas: any  expression for an odd observable should either vanish or be accompanied by an explicitly symmetry-breaking choice of data. Such checks are valuable because elliptic identities can otherwise disguise a sign error as a reparametrization.

\item The third check is the reduction to the one-cut situation. When a cut shrinks or when the global period degenerates appropriately, the genus-one geometry reduces to a rational curve. The local branch-point analysis must then agree
with the familiar one-cut formulas. This does not mean that every elliptic expression has a finite term-by-term limit in an arbitrary parameterization, one must first express it in variables suited to the degeneration. After this
is done, the disappearance of the elliptic sector is an important consistency test of the extension.
\end{itemize}

\section{Conclusion and future outlooks}
Mainly motivated by the aspects of Topological QFT and Gravity, in this paper, we discuss the generalization of the genus-0 computation of the integral of the worldsheet vertex operators for the genus-1 case. Genus-1 spectral data is much richer and contains elliptic sectors on top of the rational contributions. Below, we summarize the main findings of the paper,   
\begin{itemize}
\item We compute the rotation ($\mathbfcal{R}$) and translation ($\mathbfcal{T}$) matrices associated with genus-one, two-cut matrix models. In contrast to the one-cut case, these matrices have both the genuinely elliptic contributions arising from the higher-genus spectral curve and the rational sector. We explicitly evaluate these contributions and demonstrate that the resulting expressions reproduce the correlators of single-trace operators computed directly in the matrix model without invoking the double scaled limit.

\item We show that the corresponding chiral ring is enlarged to a four-dimensional space of operators. Accordingly, the rotation and translation matrices are rank four. This enhancement is intrinsically tied to the two-cut structure of the matrix model and, equivalently, to the richer geometry of the associated genus-one spectral curve. The additional chiral-ring sectors capture information that is absent in the genus-zero, one-cut construction.

\item We use the spectral-curve data as the fundamental input for constructing the corresponding Landau--Ginzburg (LG) theory, in particular for determining its superpotential and the appropriately normalized disk differentials. These objects provide the basic ingredients entering the worldsheet description. We further develop in detail the dictionary relating matter correlators in the LG theory before coupling to gravity to the corresponding observables of the two-cut matrix model. Upon coupling the matter theory to topological gravity, the gravitational descendants and the action of the $\mathbfcal{R}$- and $\mathbfcal{T}$-matrices systematically dress these matter correlators, thereby providing a worldsheet realization of the matrix-model observables.

\item Finally, for the genus-one sector, we establish an explicit correspondence between correlators of single-trace matrix-model operators and correlation functions of integrated vertex operators on the worldsheet. In particular, we identify separately the rational and elliptic components of the vertex operators defined in terms of the CohFT and show that their combined contribution precisely reproduces the genus-one matrix-model correlators. This provides a nontrivial extension of the matrix-model/worldsheet correspondence from the genus-zero, one-cut setting to spectral curves with nontrivial elliptic geometry and more than a single-cut structure.

\end{itemize}
\textbf{Future outlooks.} Below we mention several future outlooks worth investigating,
\begin{itemize}

\item \textbf{Multi-matrix models and their worldsheet duals.}
A natural and potentially very interesting extension of the present construction is to multi-matrix models. While the worldsheet description of the single-matrix model can be formulated in terms of an appropriate Landau--Ginzburg (LG) theory whose superpotential is determined by the spectral-curve data, the corresponding structure for multi-matrix models is expected to be considerably richer. In particular, one should identify the appropriate generalization of the worldsheet LG theory, potentially involving several interacting LG fields, a multi-component superpotential, and a higher-rank chiral ring. It would be important to understand how the spectral data of multi-matrix models are encoded in such a worldsheet description and whether the associated topological recursion admits a direct interpretation in terms of gravitational descendants and intersection theory. Establishing this dictionary would provide an important step toward extending the matrix-model/worldsheet correspondence beyond the single-matrix universality class.

\item \textbf{Localization on moduli space and the Belyi-map description.}
Another important direction is to make the proposed localization on the moduli space of Riemann surfaces more explicit in the presence of gravitational effects. A concrete way of framing this question would be to construct the relevant Belyi maps for two-cut matrix models and study how the corresponding branched coverings encode the localization of the worldsheet and figure the strebel graphs. Such an analysis could clarify how the elliptic data associated with the genus-one spectral curve are translated into the geometry of the worldsheet moduli space. In particular, it would be useful to verify explicitly whether the matrix-model amplitudes localize onto distinguished subspaces of $\overline{\mathcal{M}}_{g,n}$ and to understand how this localization is modified by the higher-rank CohFT data appearing in the two-cut problem.
\item \textbf{Instanton contributions.}As there is a duality of the LG theory on the worldsheet and the two matrix model, one can study the instanton analysis from the matrix model side and worldsheet side in full fledge along the lines of \cite{Aniceto:2018bis,Baldino:2022aqm} and also the study the resurgent analysis. Usually the transition from one well of the potential to the other well gives the instanton action. It is quite interesting to study non-perturbative contributions from the matrix model side and match it with brane (FZZT,ZZ) type contributions from the worldsheet side, as we have control over the worldsheet theory of the corresponding matrix model. 

\item \textbf{Higher-genus worldsheets and a larger number of punctures.}
It would also be very interesting to test the proposed duality in more general and geometrically critical settings. In particular, one should extend the analysis to worldsheets of higher-genus and to amplitudes involving a larger number of punctures at a fixed genus. On the matrix-model side, this requires computing higher-genus multi-point correlators and understanding their decomposition into the rational and genuinely higher-genus genera of the spectral curve. On the worldsheet side, these quantities should be reproduced by integrals over the corresponding moduli spaces $\overline{\mathcal{M}}_{g,n}$ with insertions of the relevant matter operators, gravitational descendants, and CohFT classes.

\item \textbf{Beyond the genus-one two-cut geometry.}
More generally, it would be worthwhile to study the spectral curves with a larger number of cuts and correspondingly higher genus. In such cases, the period matrix, normalized differentials, and higher-dimensional space of cycles become essential ingredients of the topological recursion. From the worldsheet perspective, this should translate into a further enlargement of the matter sector and of the associated chiral ring. Understanding how the rank of the $\mathbfcal{R}$- and $\mathbfcal{T}$-matrices grows with the number of cuts may therefore provide a wayout for understanding the CohFT relevant to that case.
\end{itemize}
\section*{Acknowledgements} We thank Saptaswa Ghosh and Arpan Bhattacharyya for useful discussions. S.P (PMRF ID: 1703278) is supported by the Prime Minister’s Research Fellowship of the Government of India. We acknowledge the use of ChatGPT 5.6 Sol-Ultra for language editing and presentation of the manuscript. The calculations are done solely by the author and few verifications for CohFT constructions are done using ChatGPT 5.6.

\appendix
\section{Details on determining spectral edges}
\label{appA}
In this section we introduce some of the variables which will be useful to express various quantities.
Before solving the equation to find the spectral edges, we introduce the symmetric combinations as follows,
\begin{equation}
\label{eq:SP-def}
\mathbfcal{S}=a^2+b^2,
\qquad
\Pi=a^2b^2,
\end{equation}
and we also define,
\begin{equation}
\label{eq:D-def}
\mathbfcal{D}={\mathbfcal{S}^2-4\Pi}=(b^2-a^2)^2.
\end{equation}
The square-root factor for a two-cut model can then be written as,
\begin{equation}
\sigma_2(z)
:=\sqrt{(z^2-a^2)(z^2-b^2)}
=z^2\sqrt{1-\frac{\mathbfcal{S}}{z^2}+\frac{\Pi}{z^4}}.
\end{equation}
Now, after using the binomial expansion
\begin{equation}
\sqrt{1+u}
=1+\frac{u}{2}-\frac{u^2}{8}+\frac{u^3}{16}+O(u^4),\,\,\,\text{with}\,\,\,
u=-\frac{\mathbfcal{S}}{z^2}+\frac{\Pi}{z^4},
\end{equation}
we obtain,
\begin{align}
\begin{split}
\sigma_2(z)
&=z^2\left[
1-\frac{\mathbfcal{S}}{2z^2}
+\left(\frac{\Pi}{2}-\frac{\mathbfcal{S}^2}{8}\right)\frac{1}{z^4}
+\left(\frac{\mathbfcal{S}\Pi}{4}-\frac{\mathbfcal{S}^3}{16}\right)\frac{1}{z^6}
+O(z^{-8})
\right] \\
&=z^2-\frac{\mathbfcal{S}}{2}-\frac{\mathbfcal{D}}{8z^2}-\frac{\mathbfcal{S}\mathbfcal{D}}{16z^4}
+O(z^{-6}).
\label{eq:R-expansion}
\end{split}
\end{align}
In the second line, we used
\begin{equation}
\frac{\Pi}{2}-\frac{\mathbfcal{S}^2}{8}
=-\frac{\mathbfcal{S}^2-4\Pi}{8}
=-\frac{\mathbfcal{D}}{8},
\,\,\,\text{and}\,\,\,
\frac{\mathbfcal{S}\Pi}{4}-\frac{\mathbfcal{S}^3}{16}
=-\frac{\mathbfcal{S}(\mathbfcal{S}^2-4\Pi)}{16}
=-\frac{\mathbfcal{S}\mathbfcal{D}}{16}.
\end{equation}
\subsection{Large-$z$ expansion of the spectral curve}
To find the large-z expansion of the spectral curve, we multiply \eqref{eq:R-expansion} by $t_6z^3+c z$, and we get the following expression,
\begin{align}
\begin{split}
y(z)
&=(t_6z^3+c z)
\left(
 z^2-\frac{\mathbfcal{S}}{2}-\frac{\mathbfcal{D}}{8z^2}-\frac{\mathbfcal{S}\mathbfcal{D}}{16z^4}
 +O(z^{-6})
\right) \\
&=t_6z^5
+\left(c-\frac{t_6\mathbfcal{S}}{2}\right)z^3
-\left(\frac{c\mathbfcal{S}}{2}+\frac{t_6\mathbfcal{D}}{8}\right)z
-\frac{\mathbfcal{D}}{16}\left(t_6\mathbfcal{S}+2c\right)\frac{1}{z}
+O(z^{-3}).
\label{eq:y-expansion}
\end{split}
\end{align}
Since,
\begin{equation}
\overline{W}_{0,1}(z)=\frac{1}{2}\bigl(V'(z)-y(z)\bigr),
\end{equation}
By the substitution of the derivative of the potential as,
\begin{equation}
V'(z)=t_6z^5+t_4z^3-\mu z
\end{equation}
into \eqref{eq:y-expansion} gives
\begin{align}
\overline{W}_{0,1}(z)
=\frac{1}{2}\Bigg[&
\left(t_4-c+\frac{t_6\mathbfcal{S}}{2}\right)z^3
+\left(-\mu+\frac{c\mathbfcal{S}}{2}+\frac{t_6\mathbfcal{D}}{8}\right)z+\frac{\mathbfcal{D}}{16}\left(t_6\mathbfcal{S}+2c\right)\frac{1}{z}+O(z^{-3})\Bigg].
\label{A.10i}
\end{align}
Using the asymptotic condition,
\begin{equation}
\overline{W}_{0,1}(z)=\frac{1}{z}+\mathcal{O}(z^{-3})
\end{equation}
requires the coefficients of $z^3$ and $z$ to vanish and the coefficient of $z^{-1}$ to be one. Now, we compute the constraints, on \eqref{A.10i} as follows,
\subsection*{\textit{First constraint: cancellation of the $z^3$ term}}
From the coefficient of $z^3$ in \eqref{A.10i},
\begin{equation}
t_4-c+\frac{t_6\mathbfcal{S}}{2}=0.\implies 
c=t_4+\frac{t_6}{2}\mathbfcal{S}.
\label{eq:c-solution}
\end{equation}

\subsection*{\textit{Second constraint: cancellation of the $z$ term}}

The coefficient of $z$ in \eqref{A.10i} gives,
\begin{equation}
-\mu+\frac{c\,\mathbfcal{S}}{2}+\frac{t_6\mathbfcal{D}}{8}=0
\implies
\mu=\frac{c\,\mathbfcal{S}}{2}+\frac{t_6\mathbfcal{D}}{8}.
\label{eq:mu-before-c}
\end{equation}
Substituting \eqref{eq:c-solution} Using $\mathbfcal{D}=\mathbfcal{S}^2-4\Pi$, this may equivalently be written as,
\begin{equation}
\mu=\frac{t_4\,\mathbfcal{S}}{2}+\frac{t_6}{8}\left(3\,\mathbfcal{S}^2-4\,\Pi\right).\label{A.14r}
\end{equation}

\subsection*{\textit{Third constraint: normalization of the resolvent}}

The coefficient of $z^{-1}$ in \eqref{A.10i} must equal one. Therefore,
\begin{equation}
\frac{1}{2}\cdot\frac{\mathbfcal{D}}{16}\left(t_6\mathbfcal{S}+2c\right)=1.
\implies 
\mathbfcal{D}\left(t_6\mathbfcal{S}+2c\right)=32.
\label{eq:norm-before-c}
\end{equation}
Using \eqref{eq:c-solution},
\begin{equation}
t_6\mathbfcal{S}+2c
=t_6\mathbfcal{S}+2t_4+t_6\mathbfcal{S}
=2\left(t_4+t_6\mathbfcal{S}\right).
\end{equation}
Thus the normalization condition reduces to,
\begin{equation}
\mathbfcal{D}\left(t_4+t_6\mathbfcal{S}\right)=16.
\label{eq:norm-SD}
\end{equation}
Since, $\mathbfcal{D}=(b^2-a^2)^2$ and $\mathbfcal{S}=a^2+b^2$, this is
\begin{equation}
(b^2-a^2)^2
\bigg[t_4+t_6(a^2+b^2)\bigg]=16.
\end{equation}
Equations \eqref{A.14r} and \eqref{eq:norm-SD} are the two independent endpoint constraints. Now, taking the leverage to combine the constraints in a single equation we write down the following cubic equation in $q$ and thereafter specify the endpoints.

\subsection*{{Reduction to a single cubic equation}} Now, we try to define a single equation which has the information of both the equation.
We define,
\begin{equation}
q=t_4+t_6\mathbfcal{S}.
\label{eq:q-def}
\implies
\mathbfcal{S}=\frac{q-t_4}{t_6},
\end{equation}
and the normalization equation \eqref{eq:norm-SD} becomes,
\begin{equation}
\mathbfcal{D}=\frac{16}{q}.
\label{eq:D-q}
\end{equation}
Since, $\mathbfcal{D}=(b^2-a^2)^2$, one has
\begin{equation}
b^2-a^2=\frac{4}{\sqrt{q}},
\qquad q>0.
\label{eq:gap-q}
\end{equation}
Now substituting,
\begin{equation}
\mathbfcal{S}=\frac{q-t_4}{t_6},
\qquad
\mathbfcal{D}=\frac{16}{q}
\end{equation}
into \eqref{eq:mu-before-c}, we obtain
\begin{align}
\begin{split}
\mu
&=\frac{t_4(q-t_4)}{2t_6}
+\frac{(q-t_4)^2}{4t_6}
+\frac{2t_6}{q}.
\end{split}
\end{align}
The first two terms can be combined and can be written as,
\begin{align}
\begin{split}
\frac{t_4(q-t_4)}{2t_6}
+\frac{(q-t_4)^2}{4t_6}
&=\frac{2t_4(q-t_4)+(q-t_4)^2}{4t_6} \\
&=\frac{q^2-t_4^2}{4t_6}.
\end{split}
\end{align}
Therefore,
\begin{equation}
\mu=\frac{q^2-t_4^2}{4t_6}+\frac{2t_6}{q}.
\label{eq:mu-q}
\end{equation}
Hence, $q$ obeys the cubic equation as follows,
\begin{equation}
q^3-\left(t_4^2+4\mu t_6\right)q+8t_6^2=0.
\label{eq:q-cubic}
\end{equation}
\subsection*{{The two-cut endpoints}}
Once a physical positive root $q$ of \eqref{eq:q-cubic} is selected, the sum and difference of the squared endpoints are,
\begin{equation}
a^2+b^2=\mathbfcal{S}=\frac{q-t_4}{t_6},\,\,\,\text{and}\,\,\,
b^2-a^2=\frac{4}{\sqrt{q}}.
\end{equation}
Adding and subtracting these two equations yields,
\begin{equation}
a^2=\frac{q-t_4}{2t_6}-\frac{2}{\sqrt{q}},
\,\,\,\,\,b^2=\frac{q-t_4}{2t_6}+\frac{2}{\sqrt{q}}.
\label{eq:b-square}
\end{equation}

\section{General operator correspondence at genus-0}

The matrix moments are obtained from the large-$x$ expansion of the
resolvent,
\begin{equation}
    \Tr\!\left(\frac{d x}{x-M}\right)
    =\sum_{k\geq 0}
    \Tr M^k\,\frac{d x}{x^{k+1}}.
\end{equation}
After identifying the genus-0 spectral-curve data with the Landau-Ginzburg(LG) data,
\begin{equation}
    W=x,\qquad
    \Omega=d y,
    \qquad
    B=\omega_{0,2},
\end{equation}
the single-trace operator insertion ($\Tr M^k$) is mapped to the worldsheet insertion of the following vertex operator,
\begin{equation}
    \text{Tr} M^k
    \longleftrightarrow
    V_k
    =\sum_{\alpha\in\mathrm{Crit}(W)}
    \sum_{m=0}^{k-1}
    \mathcal{O}_\alpha\,\underbrace{\psi^m}_{\text{Gravitational Descendents}}\,
    \text{Res}_{W=\infty}
    \frac{k!}{(k-m)!}\,
    W^{k-m}\theta_\alpha ,
    \label{eq:general-dictionary}
\end{equation}
where,
\begin{equation}
    \theta_\alpha(\phi)
    =\left.
    \frac{B(\phi,\phi')}{\Omega(\phi')}
    \right|_{\phi'=\alpha}.
\end{equation}
Here $\mathcal{O}_\alpha$ denotes the idempotent matter primary associated with
the critical point \(\alpha\) and \(\psi^m\) is its $m$-th gravitational descendant.\\

\noindent
\textbf{{One-cut spectral curve}}

\noindent
For a one-cut matrix model, as introduced earlier the Joukowsky uniformization for genus-0 spectral curve can be written as follows,
\begin{equation}
    W(\phi)=x(\phi)
    = \gamma\left(\phi+\frac{1}{\phi}\right)+\delta.
    \label{eq:Joukowsky}
\end{equation}
Now, the critical points are determined by,
\begin{equation}
    \frac{d W}{d\phi}
    =
    \gamma\left(1-\frac{1}{\phi^2}\right)=0,
\end{equation}
and hence
\begin{equation}
    \alpha=\pm1\label{B.7u}
\end{equation}
are the critical points. The corresponding idempotent matter primaries may be written as,
\begin{equation}
    \mathcal{O}_\alpha=\frac{1+\alpha\phi}{2},
    \qquad
    \alpha=\pm1.\label{ring}
\end{equation}
The two branch points of the cut are given by following,
\begin{equation}
    x_\pm=\delta\pm2\gamma.
\end{equation}

\noindent
For the genus-zero spectral curve the Bergman kernel is given by,
\begin{equation}
    B(\phi,\phi')
    =\frac{d\phi\,d\phi'}{(\phi-\phi')^2}.
\end{equation}
Writing
\begin{equation}
    \Omega(\phi)=\omega(\phi)\,d\phi,
    \qquad
    \Omega_\alpha
    \equiv
    \left.\frac{\Omega}{d\phi}\right|_{\phi=\alpha}
    =
    \omega(\alpha),
\end{equation}
we obtain the following definition,
\begin{align}
    \theta_\alpha(\phi)
    &=\left.
    \frac{
        \dfrac{d\phi\,d\phi'}{(\phi-\phi')^2}
    }{
        \omega(\phi')\,d\phi'
    }
    \right|_{\phi'=\alpha}
    = \frac{d\phi}{\Omega_\alpha(\phi-\alpha)^2}.
    \label{eq:theta-alpha}
\end{align}
Now, we proceed to compute the residue generating function we defined in \eqref{eq:general-dictionary} in the next subsection.
\subsection{Generating function for the residues}

Define
\begin{equation}
    R_{\alpha,n}
    \equiv
    \Res_{W=\infty}W^n\theta_\alpha,
    \qquad n\geq1.
\end{equation}
Introduce the exponential generating function
\begin{equation}
    \mathscr{R}_\alpha(u)
    \equiv
    \sum_{n=1}^{\infty}
    \frac{u^n}{n!}R_{\alpha,n}.
\end{equation}
Using the definition of \(R_{\alpha,n}\),
\begin{align}
    \mathscr{R}_\alpha(u)
    &=\Res_{W=\infty}
    \left(e^{uW}-1\right)\theta_\alpha=\Res_{W=\infty}
    e^{uW}\theta_\alpha,
\end{align}
because the constant term does not contribute to the residue. Substituting \eqref{eq:Joukowsky} and \eqref{eq:theta-alpha}, one finds
\begin{equation}
    \mathscr{R}_\alpha(u)
    =\frac{e^{\delta u}}{\Omega_\alpha}
    \Res_{\phi=0}
    e^{\gamma u(\phi+\phi^{-1})}
    \frac{d\phi}{(\phi-\alpha)^2}.
    \label{eq:residue-generating-function}
\end{equation}
The matrix-model point at infinity is represented by the puncture \(\phi=0\). Since $\alpha^2=1$,
\begin{equation}
    \frac{1}{(\phi-\alpha)^2}
    =
    \frac{1}{(1-\alpha\phi)^2}
    =
    \sum_{r=0}^{\infty}(r+1)\alpha^r\phi^r.
    \label{eq:geometric-expansion}
\end{equation}
The generating function of the modified Bessel functions is given by,
\begin{equation}
    e^{x(\phi+\phi^{-1})}
    =
    \sum_{\ell=-\infty}^{\infty}
    I_\ell(2x)\phi^\ell,
    \qquad
    x=\gamma u.
    \label{eq:bessel-generating-function}
\end{equation}
Therefore, the complete residue can be compactly written as,
\begin{align}
    &\Res_{\phi=0}
    e^{x(\phi+\phi^{-1})}
    \frac{d\phi}{(\phi-\alpha)^2}
    \nonumber\\
    &\qquad=
    [\phi^{-1}]
    \left[
    \sum_{\ell\in\mathbb{Z}}I_\ell(2x)\phi^\ell
    \sum_{r\geq0}(r+1)\alpha^r\phi^r
    \right]
    \nonumber\\
    &\qquad=
    \sum_{r=0}^{\infty}
    (r+1)\alpha^rI_{-r-1}(2x)
    = \sum_{n=1}^{\infty}
    n\alpha^{n-1}I_n(2x) \,\,\,\,\,\text{as\,} 
    I_{-n}(z)=I_n(z).\label{B.19u}
\end{align}
Now, apply the recurrence relation
\begin{equation}
    nI_n(2x)
    = x\left[I_{n-1}(2x)-I_{n+1}(2x)\right].
\end{equation}
Thus, the equality in \eqref{B.19u} cn be written as,
\begin{align}
    \sum_{n=1}^{\infty}
    n\alpha^{n-1}I_n(2x)
    &=
    x\sum_{n=1}^{\infty}
    \alpha^{n-1}
    \left[I_{n-1}(2x)-I_{n+1}(2x)\right]
    \nonumber\\
    &=
    x\left[I_0(2x)+\alpha I_1(2x)\right],
\end{align}
where all terms with \(I_n\), \(n\geq2\), cancel beautifully because
\(\alpha^2=1\). Substituting \(x=\gamma u\) into
\eqref{eq:residue-generating-function}, we obtain,
\begin{equation}
    \mathscr{R}_\alpha(u)
    =\frac{\gamma u\,e^{\delta u}}{\Omega_\alpha}
    \left[
    I_0(2\gamma u)+\alpha I_1(2\gamma u)
    \right]
    \label{eq:R-generating-final}
\end{equation}
Equating, coefficients of $u^n$ gives,
\begin{equation}
    R_{\alpha,n}
    =\frac{n!\gamma}{\Omega_\alpha}
    \left[
    e^{\delta u}
    \left(
    I_0(2\gamma u)+\alpha I_1(2\gamma u)
    \right)
    \right]_{n-1} ,
    \label{eq:R-coefficient}
\end{equation}
where $[F(u)]_p$ denotes the coefficient of $u^p$ in the power-series expansion of \(F(u)\).\\

\noindent
\textbf{{Substitution into the operator dictionary}}

\noindent
 In \eqref{eq:general-dictionary}, Setting
\begin{equation}
    n=k-m.
\end{equation}
Now, using \eqref{eq:R-coefficient}, we find the following
\begin{align}
    V_k
    &=  \sum_{\alpha=\pm1}
    \sum_{m=0}^{k-1}
    \mathcal{O}_\alpha\psi^m
    \frac{k!}{(k-m)!}
    R_{\alpha,k-m}
    \nonumber\\[1mm]
    &=k!\gamma\sum_{\alpha=\pm1}
    \frac{\mathcal{O}_\alpha}{\Omega_\alpha}
    \sum_{m=0}^{k-1}
    \psi^m
    \left[
    e^{\delta u}
    \left(
    I_0(2\gamma u)+\alpha I_1(2\gamma u)
    \right)
    \right]_{k-m-1}.
    \label{eq:Vk-before-resummation}
\end{align}

\noindent
For an arbitrary power series \(F(u)\),
\begin{equation}
    \sum_{m=0}^{k-1}
    \psi^m[F(u)]_{k-m-1}
    =
    \left[
    \frac{uF(u)}{1-u\psi}
    \right]_k,
    \label{eq:descendant-resummation}
\end{equation}
because we know the following relation holds true,
\begin{equation}
    \frac{1}{1-u\psi}
    =\sum_{m=0}^{\infty}u^m\psi^m.
\end{equation}
Taking
\begin{equation}
    F_\alpha(u)
    = e^{\delta u}
    \left[
    I_0(2\gamma u)+\alpha I_1(2\gamma u)
    \right],
\end{equation}
equations \eqref{eq:Vk-before-resummation} and
\eqref{eq:descendant-resummation} gives the following expression for the vertex operator,
\begin{equation}
    \text{Tr} M^k
    \longleftrightarrow
    V_k:=
    \left[
    \frac{k!\gamma u e^{\delta u}}{1-u\psi}
    \sum_{\alpha=\pm1}
    \mathcal{O}_\alpha
    \frac{
        I_0(2\gamma u)+\alpha I_1(2\gamma u)
    }{
        \Omega_\alpha
    }
    \right]_k.
    \label{eq:final-correspondence}
\end{equation}
\section{Contribution of the sepertating and non-seperating edges}
\tikzset{
  stable vertex/.style={circle,draw,inner sep=0pt,minimum size=5mm,line width=.65pt},
  graph edge/.style={line width=.65pt},
  leg label/.style={font=\small}
}
In this appendix section, we briefly describe how to compute the generic contribution of rotation, translation and edge matrices and how to determine themselves. The contribution in a graph of boundary recursive structure is given by,
\begin{equation}
\Cont_{\Gamma;\mu_1,\ldots,\mu_n}
=\left(
  \prod_{v\in V(\Gamma)}
  \Omega_{g(v);(\nu_h)_{h\rightsquigarrow v}}
\right)
\left(
  \prod_{e=(h,h')\in E(\Gamma)}
  E^{\nu_h,\nu_{h'}}(\psi_h,\psi_{h'})
\right)
\left(
  \prod_{i=1}^{n} \mathbfcal{R}^{\nu_i}_{\mu_i}(\psi_i)
\right).
\end{equation}
Now, though the expressions
$E^{\nu_h,\nu_{h'}}(\psi_h,\psi_{h'})$ and
$\mathbfcal{R}^{\nu_i}_{\mu_i}(\psi_i)$ have, a priori, infinitely many terms, they terminate due to the cohomological degree of the moduli space of Riemann surfaces. 
Now, defining $\mathbfcal{R}\Omega_{g;\mu_1,\ldots,\mu_n}$ to be the sum of contributions of all stable graphs, after pushforward to the moduli space weighted by automorphism factors:
\begin{equation}
\mathbfcal{R}\Omega_{g;\mu_1,\ldots,\mu_n}
=\sum_{\substack{\Gamma\ \text{of type }(g,n)}}
\frac{1}{|\Aut(\Gamma)|}\,
\xi_{\Gamma *}\Cont_{\Gamma;\mu_1,\ldots,\mu_n}.
\end{equation}

Now, we discuss it in some examples relevant for genus-0,1.
\begin{itemize}
\item $\mathbfcal{R}\Omega_{0,3}$. There is only one stable graph of type $(0,3)$, and for
dimensional specifications, the contribution of $R(\psi_i)$ at the leaves is simply the
identity. Thus, as the cohomological field theory is defined by the action of matrices on the identity CohFT, here this becomes quite trivial. Thus, we find
\begin{equation}
\mathbfcal{R}\Omega_{0,3}=\Omega_{0,3}.
\end{equation}

\item $\mathbfcal{R}\Omega_{0,4}$. In this case the Riemann surface structure is basically the sphere with four punctures. As it has a moduli, the decorations of its leaves and edges are subtle. The stable graphs of type $(0,4)$ are the following:
\begin{equation}
\Gamma_0=
\begin{tikzpicture}[baseline=-.65ex,x=6mm,y=6mm]
  \node[stable vertex] (v) at (0,0) {$0$};
  \draw[graph edge] (v)--(-1,1) node[leg label,above left=-1pt] {$1$};
  \draw[graph edge] (v)--(1,1) node[leg label,above right=-1pt] {$4$};
  \draw[graph edge] (v)--(-1,-1) node[leg label,below left=-1pt] {$2$};
  \draw[graph edge] (v)--(1,-1) node[leg label,below right=-1pt] {$3$};
\end{tikzpicture}
\qquad
\Gamma_{ij\mid k\ell}=
\begin{tikzpicture}[baseline=-.65ex,x=7mm,y=6mm]
  \node[stable vertex] (a) at (0,0) {$0$};
  \node[stable vertex] (b) at (1.8,0) {$0$};
  \draw[graph edge] (a)--(b);
  \draw[graph edge] (a)--(-.7,.9) node[leg label,above left=-1pt] {$i$};
  \draw[graph edge] (a)--(-.7,-.9) node[leg label,below left=-1pt] {$j$};
  \draw[graph edge] (b)--(2.5,.9) node[leg label,above right=-1pt] {$k$};
  \draw[graph edge] (b)--(2.5,-.9) node[leg label,below right=-1pt] {$\ell$};
\end{tikzpicture}.
\end{equation}
\noindent
Here $ij\mid k\ell\in\{12\mid34,13\mid24,14\mid23\}$. The contribution of
the first stable graph $\Gamma_0$ is given by the folllowing expression,
\begin{align}
\Cont_{\Gamma_0;\mu_1,\mu_2,\mu_3,\mu_4}
={}&\Omega_{0;\mu_1,\mu_2,\mu_3,\mu_4}
+\Omega_{0;\alpha,\mu_2,\mu_3,\mu_4}(\mathbfcal{R}_1)^\alpha_{\mu_1}\psi_1
+\Omega_{0;\alpha,\mu_1,\mu_3,\mu_4}(\mathbfcal{R}_1)^\alpha_{\mu_2}\psi_2
\notag\\
&+\Omega_{0;\alpha,\mu_1,\mu_2,\mu_4}(\mathbfcal{R}_1)^\alpha_{\mu_3}\psi_3
+\Omega_{0;\alpha,\mu_1,\mu_2,\mu_3}(\mathbfcal{R}_1)^\alpha_{\mu_4}\psi_4.
\end{align}
In the second stable graph, the sphere with four punctures breaks into two spheres with punctures on which joining at a nodal singularity. The contribution of the stable graph $\Gamma_{ij\mid k\ell}$ is given by,
\begin{equation}
\Cont_{\Gamma_{ij\mid k\ell};\mu_1,\mu_2,\mu_3,\mu_4}
=\Omega_{0;\mu_i,\mu_j,\alpha}\,
E_{0,0}^{\alpha,\beta}\,
\Omega_{0;\beta,\mu_k,\mu_\ell}.
\end{equation}
It can be shown that
$\xi_{\Gamma_{ij\mid k\ell},*}1=[\Gamma_{ij\mid k\ell}]=\kappa_1$, hence the action of the $\mathbfcal{R}$-matrices are given by,
\begin{align}
\mathbfcal{R}\Omega_{0;\mu_1,\mu_2,\mu_3,\mu_4}
={}&\Omega_{0;\mu_1,\mu_2,\mu_3,\mu_4}
+\Omega_{0;\alpha,\mu_2,\mu_3,\mu_4}(\mathbfcal{R}_1)^\alpha_{\mu_1}\psi_1
+\Omega_{0;\alpha,\mu_1,\mu_3,\mu_4}(\mathbfcal{R}_1)^\alpha_{\mu_2}\psi_2
\notag\\
&+\Omega_{0;\alpha,\mu_1,\mu_2,\mu_4}(\mathbfcal{R}_1)^\alpha_{\mu_3}\psi_3
+\Omega_{0;\alpha,\mu_1,\mu_2,\mu_3}(\mathbfcal{R}_1)^\alpha_{\mu_4}\psi_4
\notag\\
&+\left(
\sum_{ij\mid k\ell}
\Omega_{0;\mu_i,\mu_j,\alpha}E_{0,0}^{\alpha,\beta}
\Omega_{0;\beta,\mu_k,\mu_\ell}
\right)\kappa_1.
\end{align}

\item $\mathbfcal{R}\Omega_{1,1}$. For genus-1 surface with a single puncture on it, there are two stable graphs of type $(1,1)$ corresponding to the generic and non-seperating representation respectively as follows:
\begin{equation}
\Gamma=
\begin{tikzpicture}[baseline=-.65ex,x=7mm,y=7mm]
  \node[stable vertex] (v) at (1,0) {$1$};
  \draw[graph edge] (v)--(0,0) node[leg label,left] {$1$};
\end{tikzpicture}
\qquad\qquad
\Gamma'=
\begin{tikzpicture}[baseline=-.65ex,x=7mm,y=7mm]
  \node[stable vertex] (v) at (1,0) {$0$};
  \draw[graph edge] (v)--(0,0) node[leg label,left] {$1$};
  \draw[graph edge] (v) .. controls (1.35,.85) and (2.25,.85) .. (2.25,0)
                    .. controls (2.25,-.85) and (1.35,-.85) .. (v);
\end{tikzpicture}.
\end{equation}
The contribution of the generic diagram (without degenerations), $\Gamma$ is
\begin{equation}
\Cont_{\Gamma;\mu}=\Omega_{1;\mu}
+\Omega_{1;\nu}(\mathbfcal{R}_1)^\nu_{\mu}\psi_1.
\end{equation}
\noindent
For the non-seperating degenerations, $\Gamma'$, we find
\begin{equation}
\Cont_{\Gamma';\mu}=\Omega_{0;\mu,\alpha,\beta}E_{0,0}^{\alpha,\beta}.
\end{equation}
It can be shown that
$\tfrac12\xi_{\Gamma',*}1=[\Gamma']=12\psi_1$, so that
\begin{equation}
\mathbfcal{R}\Omega_{1;\mu}=\Omega_{1;\mu}
+\left(\Omega_{1;\nu}(\mathbfcal{R}_1)^\nu_{\mu}
+12\Omega_{0;\mu,\alpha,\beta}E_{0,0}^{\alpha,\beta}\right)\psi_1.
\end{equation}
\end{itemize}
\section{Derivation of the diagonal coefficient $(\mathbfcal{R}_1)_{ii}$}
\label{appCC}
Let us define the function $F(z)$ as follows,
\begin{align}
    F(z):=\wp(z|\tau)+2\eta_1(\tau).
\end{align}
Then the genus-one Bergman kernel is
\begin{align}
    B(p,q)
=F\!\left(u(p)-u(q)\right)\,du(p)\,du(q).
\end{align}
This kernel is general and complicated than the genus-0 case. In genus-1 scenario it contains the elliptic sector, on top of their rational counterpart.

\noindent
The corresponding normalized disk differentials associated with the $i$-th ramification point is given explictly as,
\begin{align}
    d\xi_i(p)
=-\chi_i\,F\!\left(u_i-u(p)\right)\,du(p).
\label{4.18u}\end{align}

\noindent
Now, near each of the branch point $e_i$ of our spectral curve, choose the local Airy coordinate $q_i$ given by,
\begin{align}
    x=e_i+\frac{q_i^2}{2}.\label{4.19u}
\end{align}
This coordinate is chosen such that the $\mathbfcal{R},\mathbfcal{T}$ matrices feature locally gaussian measures in the integral \eqref{4.9i}.
\noindent
Now, one can define the generic form of the spectral curve in these airy coordinate. Hence near $x=e_i$, $\sigma(x)$ is given by,
\begin{align}
    \sigma(x)
=\sigma'(e_i)(x-e_i)
+\frac12\sigma''(e_i)(x-e_i)^2+\cdots .
\end{align}
Now, using the equation \eqref{4.19u} we obtain the following relation,
\begin{align}
    \sigma(x)=\frac{\sigma'(e_i)}{2}q_i^2
\left[
1+
\frac{\sigma''(e_i)}{4\sigma'(e_i)}q_i^2+\cdots
\right].
\end{align}
\noindent
Therefore, the parameters are given by, 
\begin{align}
   \eta=\sqrt{\sigma(x)}
=\kappa_i q_i
\left[
1+s_iq_i^2+\cdots
\right], \,\,\,\,\kappa_i=\sqrt{\frac{\sigma'(e_i)}{2}},
s_i=\frac{\sigma''(e_i)}{8\sigma'(e_i)}
\end{align}
Now, after introducing the Abel coordinate defined by,
\begin{align}
du=\frac{dx}{\omega_A\eta}.
\end{align}
and after plugging in the values of the parameters determined above we get,
\begin{align}
du
=\frac{q_i\,dq_i}
{\omega_A\kappa_i q_i(1+s_iq_i^2+\cdots)}
=\chi_i(1-s_iq_i^2+\cdots)dq_i,\,\,\,\,\chi_i:=\frac{1}{\omega_A\kappa_i},\,\,\,dx=q_i\,dq_i.
\label{D.9i}
\end{align}
Hence, after integrating \eqref{D.9i} one obtains,
\begin{align}
    u(p)-u_i
=\chi_i
\left[
q_i-\frac{s_i}{3}q_i^3+\cdots
\right].
\end{align}
\noindent
Now, using the local expansion of the Weierstrass-P function given by, $\wp(z)=\frac1{z^2}+O(z^2)$,
we can cast the elliptic Bergmann kernel in the following form, 
\begin{align}
    F(u_i-u(p))
=\wp(u_i-u(p))+2\eta_1.
\end{align}
One can see directly, that the Bergmann kernel has elliptic sectors on top of the rational counterparts available also in genus-0 data. Hence,\footnote{Since
$u_i-u(p)
=-\chi_iq_i\left[
1-\frac{s_i}{3}q_i^2+\cdots\right],$
we have the following,
\begin{align}
\frac{1}{(u_i-u(p))^2}
=\frac{1}{\chi_i^2q_i^2}
\left[
1+\frac{2s_i}{3}q_i^2+\cdots
\right].
\end{align}}
\begin{align}
F(u_i-u(p))
=
\frac{1}{\chi_i^2q_i^2}
+
\frac{2s_i}{3\chi_i^2}
+2\eta_1
+O(q_i^2).
\end{align}

\noindent
Substituting this into \eqref{4.18u} we get the following expression of the disk diffrential,
\begin{align}
    d\xi_i(p)
=-\chi_iF(u_i-u(p))\,du(p),
\end{align}
and using the expression obtained in \eqref{D.9i},
\begin{align}
    du(p)=\chi_i(1-s_iq_i^2+\cdots)dq_i,
\end{align}

we find that,
\begin{align}
d\xi_i(p)
=-\chi_i^2
\left[
\frac{1}{\chi_i^2q_i^2}
+\frac{2s_i}{3\chi_i^2}
+2\eta_1+\cdots\right](1-s_iq_i^2+\cdots)dq_i.
\end{align}

\noindent
Keeping the singular and constant terms give,
\begin{align}
\begin{split}
    d\xi_i(p)
&=-\left[
\frac{1}{q_i^2}+\left(\frac{2s_i}{3}-s_i
\right)
+2\eta_1\chi_i^2+\cdots\right]dq_i.\\&
=-\frac{dq_i}{q_i^2}
-\left[2\eta_1(\tau)\chi_i^2
-\frac{\sigma''(e_i)}{24\sigma'(e_i)}
\right]dq_i+\cdots. \,\,\,\,\,\,\, \text{with} \,\,\,s_i=\frac{\sigma''(e_i)}{8\sigma'(e_i)},
\end{split}
\end{align}
Now, defining 
\begin{align}
    \mathcal{A}_i:=
2\eta_1(\tau)\chi_i^2
-\frac{\sigma''(e_i)}{24\sigma'(e_i)},
\end{align}
Then, finally the differential can be casted into a compact form as follows,
\begin{align}
    d\xi_i(p)
=-\frac{dq_i}{q_i^2}
-\mathcal{A}_i\,dq_i+\cdots 
\end{align}
\noindent
So, the diagonal components of the rotation matrix is given by following,
\begin{align}
    \widehat {\mathbfcal R}_i^{\ i}(-\hbar)
\sim \frac{\hbar}{\sqrt{2\pi\hbar}}
\int_{\Gamma_i}
e^{-q_i^2/(2\hbar)}
\left(
-\frac{dq_i}{q_i^2}
-\mathcal{A}_i\,dq_i+\cdots
\right).
\end{align}
\noindent
The finite part of the singular Gaussian integral gives,\footnote{The regular part gives
\begin{align}
\frac{\hbar}{\sqrt{2\pi\hbar}}
\int_{\Gamma_i}
e^{-q_i^2/(2\hbar)}
(-\mathcal{A}_i\,dq_i)
=-\mathcal{A}_i\hbar.
\end{align}}
\begin{align}  
\frac{\hbar}{\sqrt{2\pi\hbar}}
\operatorname{FP}
\int_{\Gamma_i}
e^{-q_i^2/(2\hbar)}
\left(
-\frac{dq_i}{q_i^2}
\right)
=1.
\end{align}
then the diagonal coefficient of the rotation matrix is,
\begin{align}
(\widehat{\mathbfcal{R}}_1)_{ii}
=\mathcal{A}_i
=2\eta_1(\tau)\chi_i^2
-\frac{\sigma''(e_i)}{24\sigma'(e_i)}.\label{D.22i}
\end{align}
Thus, the term
$2\eta_1(\tau)\chi_i^2$ comes from the regular part of the elliptic Bergman kernel, while
$-\frac{\sigma''(e_i)}{24\sigma'(e_i)}$ comes from the nonlinear relation between the Abel coordinate $u$ and the local Airy coordinate $q_i$.
\subsection{Worldsheet correlators for genus-0 surface}
\label{appD1}
In this part of the appendix we compute the three point correlator of the vertex operator on the genus-0 Riemann surface. The correlator of the vertex operators on the moduli space of Riemann surfaces including the boundary is given by,

\begin{align}
\begin{split}
C^{(0)}_{666}
&:=
\left\langle
\mathcal V_6\mathcal V_6\mathcal V_6
\right\rangle_0\\&=
\sum_{i,j,k}
v^i_{6,0}v^j_{6,0}v^k_{6,0}
\left\langle \mathcal{O}_i\mathcal{O}_j\mathcal{O}_k\right\rangle_0=
\sum_{i=1}^4
\left(v^i_{6,0}\right)^3H_i^{-1}.
\end{split}
\end{align}
In writing down the last line we substituted \eqref{7.30u}.
Now, substituting $v^i_{6,0}$ and $H_i^{-1}$ already obtained in \eqref{7.30u},\eqref{7.31w} we get,
\begin{align}
\begin{split}
C^{(0)}_{666}
&=
\sum_i
\frac{27(\mathbfcal S^2+4x_i^4)^3}
{512\,p(x_i)^3}
\frac{p(x_i)^2\sigma'(x_i)}8
\\
&=
\frac{27}{4096}
\sum_{x_i=-a,-b,a,b}
\frac{\sigma'(x_i)}{p(x_i)}
\left(\mathbfcal S^2+4x_i^4\right)^3.
\end{split}
\end{align}

\noindent
Now as,
\begin{align}
    \sigma'(x)=2x(2x^2-\mathbfcal S),
\end{align}
\noindent
so evaluating it on four symmetric branch points of the spectral curve gives the following expression,
\begin{align}
\begin{split}
    &\sigma'(a)=-2a\sqrt{\mathbfcal D},
\qquad
\sigma'(b)=2b\sqrt{\mathbfcal D},\\&
\sigma'(-a)=2a\sqrt{\mathbfcal D},
\qquad
\sigma'(-b)=-2b\sqrt{\mathbfcal D}.
\end{split}
\end{align}
\noindent
Now as already computed earlier, for the sextic double well potential, the polynomial part of the density of state is given by,
\begin{align}
    p(x)
=x\left(t_6x^2+t_4+\frac{t_6}{2}\mathbfcal S
\right).
\end{align}

\noindent Now, Using the following relation,
\begin{align}
a^2=\frac{\mathbfcal S-\sqrt{\mathbfcal D}}2,
\qquad
b^2=\frac{\mathbfcal S+\sqrt{\mathbfcal D}}2,
\end{align}
we obtain
\begin{align}
    \frac{p(a)}{a}
=\mathbfcal L-\frac{t_6}{2}\sqrt{\mathbfcal D},\,\,\,\,\,\,\,\,\,
    \frac{p(b)}{b}
=\mathbfcal L+\frac{t_6}{2}\sqrt{\mathbfcal D}.
\end{align}

\noindent
Because both satisfies,
\begin{align}
    \frac{\sigma'(-x)}{p(-x)}
=\frac{\sigma'(x)}{p(x)}
\end{align}
\noindent
and $x^4$ are even, the $\pm a,\pm b$ contributions pair. Thus, the three point function of the closed string vertex operators on the genus-0 Riemann surface is given by the following expression,

\begin{align}
\begin{split}
C^{(0)}_{666}
=\frac{27\sqrt{\mathbfcal D}}{1024}
\Bigg[
&\frac{(\mathbfcal S^2+4b^4)^3}
{\mathbfcal L+\frac{t_6}{2}\sqrt{\mathbfcal D}}-\frac{(\mathbfcal S^2+4a^4)^3}
{\mathbfcal L-\frac{t_6}{2}\sqrt{\mathbfcal D}}
\Bigg].\label{7.46y}
\end{split}
\end{align}

\noindent
This is already the explicit three-point correlator.
Now eliminate $a,b$. Since,
\begin{align}
    4b^4=
\left(\mathbfcal S+\sqrt{\mathbfcal D}\right)^2,
\qquad
4a^4=\left(\mathbfcal S-\sqrt{\mathbfcal D}\right)^2,
\end{align}

\noindent
After combining the two fractions in\footnote{we have
\begin{align}
    \mathbfcal S^2+4b^4
=2\mathbfcal S^2
+
2\mathbfcal S\sqrt{\mathbfcal D}
+\mathbfcal D,\,\,\,\,\,\,\,\,\,\,\,\,
\mathbfcal S^2+4a^4
=2\mathbfcal S^2
-2\mathbfcal S\sqrt{\mathbfcal D}
+\mathbfcal D.
\end{align}} \eqref{7.46y},
\begin{align}
\begin{split}
C^{(0)}_{666}
=\frac{27}{64\mathbfcal P}\Bigg\{&4\mathbfcal L\mathbfcal S
\left[12\mathbfcal S^4+16\mathbfcal S^2\mathbfcal D+3\mathbfcal D^2\right]
\\
&-
t_6\left(2\mathbfcal S^2+\mathbfcal D\right)\left[4\mathbfcal S^4
+16\mathbfcal S^2\mathbfcal D
+\mathbfcal D^2\right]\Bigg\}.
\end{split}
\end{align}
\section{Uniformization for two-cut spectral curve}
\label{appe}
The M\"obius transform of $\wp$ and the Jacobi $\dn$ parametrization describe the same elliptic curve. Their direct conversion introduces a Landen-transformed parameter, keeping the two parameters distinct fixes the period normalization, which we describe below.
Let $0<a<b$, and use the curve and normalized coordinate,
\begin{equation}
 \eta^2=\sigma(x)=(x^2-a^2)(x^2-b^2),\qquad
 d u=\frac{d x}{\omega_A \sqrt{\sigma(x)}},\qquad u(b)=0.
 \label{eq:curve}
\end{equation}
Thus $(d x/d u)^2=\omega_A^2\sigma(x)$, as in the equation \eqref{7.43o}. Define
\begin{equation}
 m=1-\frac{a^2}{b^2},\qquad \Delta=b^2-a^2=b^2m,
 \qquad v=i b\omega_Au.
 \label{eq:m}
\end{equation}
Here $m$ is the elliptic \emph{parameter}, the modulus is $\sqrt m$. Substitute,
\begin{equation}
 x=b\sqrt{1-m\sin^2\phi}.
 \label{eq:angle}
\end{equation}
Hence, the two quadratic factors become,
\begin{equation}
 x^2-a^2=b^2m\cos^2\phi,\qquad
 x^2-b^2=-b^2m\sin^2\phi.
 \label{eq:factors}
\end{equation}
On $0<\phi<\pi/2$, choose the sheet $\eta=Y=-i b^2m\sin\phi\cos\phi$. Then
\begin{equation}
 d x=-\frac{bm\sin\phi\cos\phi}{\sqrt{1-m\sin^2\phi}}d\phi,
 \qquad
 \frac{d x}{Y}=-\frac{i}{b}
 \frac{d\phi}{\sqrt{1-m\sin^2\phi}}.
 \label{eq:differential}
\end{equation}
Equations \eqref{eq:curve} and \eqref{eq:m} therefore implies the following,
\begin{equation}
 v=F(\phi\mid m):=\int_0^\phi
 \frac{d\theta}{\sqrt{1-m\sin^2\theta}},
 \qquad \phi=\am(v\mid m).
 \label{eq:amplitude}
\end{equation}
Since $\sn(v\mid m)=\sin\phi$ and $\dn^2(v\mid m)=1-m\sn^2(v\mid m)$, the inversion gives
{Jacobi uniformization} as,
\begin{equation}
 x(u)=b\,\dn(i b\omega_Au\mid m),\qquad
 \phi(u)=\am(i b\omega_Au\mid m).
 \label{eq:dn}
\end{equation}
The initial value is $x(0)=b$. The square root in \eqref{eq:angle} is continued with the sheet, the $\dn$ expression defines the meromorphic continuation. Also $\partial_v\am(v\mid m)=\dn(v\mid m)$.
\subsection{Converting the Weierstrass representation}
In this section, we describe the relation between Jacobi uniformization and the Weierstrass data. Keeping precisely the half-period labels of the equations \eqref{7.42i}--\eqref{7.45i}, write $w(u)=\wp(u\mid\tau)$ with period lattice $\mathbb Z+\tau\mathbb Z$. The map is given by,
\begin{equation}
 x(u)=b-\frac{C}{w(u)+\delta_0},\qquad
 C=-\frac12\omega_A^2b\Delta,\qquad
 \delta_0=\frac{\omega_A^2}{12}(a^2-5b^2).
 \label{eq:mobius}
\end{equation}
Its half-period values are
\begin{align}
 \eA&=\wp\!\left(\frac12\middle|\tau\right)
 =-\frac{\omega_A^2}{12}(a^2+6ab+b^2),\nonumber\\
 \eB&=\wp\!\left(\frac\tau2\middle|\tau\right)
 =\frac{\omega_A^2}{6}(a^2+b^2),\nonumber\\
 \eAB&=\wp\!\left(\frac{1+\tau}{2}\middle|\tau\right)
 =-\frac{\omega_A^2}{12}(a^2-6ab+b^2).
 \label{eq:roots}
\end{align}

\noindent
\textbf{Root labels.} In the standard Jacobi--Weierstrass convention, the roots at $1/2$, $(1+\tau)/2$ and $\tau/2$ are called $e_1,e_2,e_3$, respectively. Thus the standard $(e_1,e_2,e_3)$ corresponds here to $(\eA,\eAB,\eB)$. Define the scale and the new parameter,
\begin{equation}
 \alpha=\frac{i\omega_A(a+b)}2,\qquad
 \kappa=\frac{b-a}{b+a},\qquad \lambda=\kappa^2.
 \label{eq:lambda}
\end{equation}
Direct subtraction of \eqref{eq:roots} gives
\begin{equation}
 \alpha^2=\eA-\eB=-\frac{\omega_A^2}{4}(a+b)^2,
 \qquad
 \lambda=\frac{\eAB-\eB}{\eA-\eB}
 =\left(\frac{b-a}{b+a}\right)^2.
 \label{eq:rootratios}
\end{equation}
The Jacobi--Weierstrass identity, expressed with these labels, is
\begin{equation}
 \wp(u\mid\tau)=\eB+
 \frac{\alpha^2}{\sn^2(\alpha u\mid\lambda)}.
 \label{eq:wp-sn}
\end{equation}
One can derive it directly from the cubic differential equation. Set
\begin{equation}
 s(u)=\frac{\alpha^2}{w(u)-\eB},\qquad
 w'^2=4(w-\eA)(w-\eAB)(w-\eB).
 \label{eq:cubic}
\end{equation}
Since $w=\eB+\alpha^2/s$, substitution gives
\begin{equation}
 s'^2=4\alpha^2s(1-s)(1-\lambda s).
 \label{eq:s-equation}
\end{equation}
With $t=\sqrt s$, this becomes $t'^2=\alpha^2(1-t^2)(1-\lambda t^2)$. The Laurent behavior $w(u)\sim u^{-2}$ fixes $t(u)\sim\alpha u$ at the origin. Hence $t(u)=\sn(\alpha u\mid\lambda)$ and \eqref{eq:wp-sn} follows.
\subsection{The M\"obius map and the Landen transformation}
Now, insert \eqref{eq:wp-sn} into \eqref{eq:mobius}. Because
\begin{equation}
 \eB+\delta_0=-\frac{\omega_A^2\Delta}{4},\qquad
 w+\delta_0=-\frac{\omega_A^2}{4}
 \left(\Delta+\frac{(a+b)^2}{s}\right),
 \label{eq:denominator}
\end{equation}
then the map simplifies as,
\begin{align}
 x(u)&=b-\frac{2b\Delta s}{(a+b)^2+\Delta s}\nonumber\\
 &=b\,\frac{1-\kappa s}{1+\kappa s},
 \qquad s=\sn^2(\alpha u\mid\kappa^2),
 \label{eq:rational}
\end{align}
where $\Delta/(a+b)^2=(b-a)/(b+a)=\kappa$. Thus this becomes
\begin{align}
    x(u)=b\,
 \frac{1-\kappa\sn^2(\alpha u\mid\kappa^2)}
 {1+\kappa\sn^2(\alpha u\mid\kappa^2)}.
 \label{eq:jacobi-rational}
\end{align}
Equivalently, for the second amplitude $\Phi=\am(\alpha u\mid\kappa^2)$,
\begin{equation}
 x(u)=b\,\frac{1-\kappa\sin^2\Phi}{1+\kappa\sin^2\Phi}.
 \label{eq:second-amplitude}
\end{equation}
To relate this expression to \eqref{eq:dn}, note that we have,
\begin{equation}
 m=\frac{4\kappa}{(1+\kappa)^2},\qquad
 \kappa=\frac{1-\sqrt{1-m}}{1+\sqrt{1-m}},\qquad
 \alpha u=\frac{v}{1+\kappa}.
 \label{eq:landen-parameters}
\end{equation}
These are exactly the descending Landen transformation. Its $\dn$ identity is,
\begin{equation}
 \dn(v\mid m)=
 \frac{1-\kappa\sn^2\!\left(v/(1+\kappa)\mid\kappa^2\right)}
 {1+\kappa\sn^2\!\left(v/(1+\kappa)\mid\kappa^2\right)}.
 \label{eq:landen-identity}
\end{equation}
In terms of the amplitudes $\phi=\am(v\mid m)$ and $\Phi=\am(v/(1+\kappa)\mid\kappa^2)$, the corresponding $\sn$ identity is given by,
\begin{equation}
 \sin\phi=\frac{(1+\kappa)\sin\Phi}{1+\kappa\sin^2\Phi}.
 \label{eq:amplitude-connection}
\end{equation}
Two checks fix the scale and sign. First, differentiating $x=b\dn(v\mid m)$ gives
\begin{equation}
 \frac{d x}{d u}=-i\omega_A b^2m\sn(v\mid m)\cn(v\mid m),
 \quad
 \sigma(x)=-b^4m^2\sn^2(v\mid m)\cn^2(v\mid m),
 \label{eq:check-ode}
\end{equation}
so $(d x/d u)^2=\omega_A^2\sigma(x)$. Second, near $u=0$,
\begin{equation}
 x(u)=b+\frac{\omega_A^2b\Delta}{2}u^2+O(u^4)
 =b-Cu^2+O(u^4),
 \label{eq:check-local}
\end{equation}
which agrees with the M\"obius map because $\wp(u\mid\tau)=u^{-2}+O(u^2)$.
\subsubsection{Normalized periods and branch points}
Use parameter notation for the complete integral,
\begin{equation}
 K(m):=F\!\left(\frac\pi2\middle|m\right)
 =\int_0^{\pi/2}\frac{d\theta}{\sqrt{1-m\sin^2\theta}}.
 \label{eq:complete}
\end{equation}
Choose the $A$-cycle to encircle one cut with the sheet used earlier. Its period is $\omega_A=\oint_Ad x/Y=-2i K(m)/b$. Consequently,
\begin{equation}
 v=2K(m)u,\qquad
 \alpha=\frac{(a+b)K(m)}{b}=2K(\lambda),\qquad
 K(\lambda)=\frac{K(m)}{1+\kappa}.
 \label{eq:normalization}
\end{equation}
The last equality is also the complete-integral Landen relation. Reversing the cycle changes the sign of $\omega_A$ and of $v$, $x$ is unchanged because $\dn$ is even.

The full periods of $\dn(v\mid m)$ are $2K(m)$ and $4i K(1-m)$. Therefore the period lattice in the normalized $u$ coordinate is,
\begin{equation}
 \begin{gathered}
 u\sim u+1\sim u+\tau,\qquad
 \tau=\frac{2i K(1-m)}{K(m)}
      =\frac{i K(1-\lambda)}{K(\lambda)},\\[4pt]
 x(u)=b\dn(2K(m)u\mid m)
 =b\frac{1-\kappa\sn^2(2K(\lambda)u\mid\lambda)}
 {1+\kappa\sn^2(2K(\lambda)u\mid\lambda)}.
 \end{gathered}
 \label{eq:normalized-result}
\end{equation}
The factor of two in the first expression for $\tau$ comes from the imaginary period of $\dn$. In particular, $\dn(v+2i K(1-m)\mid m)=-\dn(v\mid m)$,this shift sends $x$ to $-x$ rather than returning to the same point.

\bibliography{REFM}

@article{Maldacena:1998im,
    author = "Maldacena, Juan Martin",
    title = "{Wilson loops in large N field theories}",
    eprint = "hep-th/9803002",
    archivePrefix = "arXiv",
    reportNumber = "HUTP-98-A014",
    doi = "10.1103/PhysRevLett.80.4859",
    journal = "Phys. Rev. Lett.",
    volume = "80",
    pages = "4859--4862",
    year = "1998"
}

@article{Beisert:2010jr,
    author = "Beisert, Niklas and others",
    title = "{Review of AdS/CFT Integrability: An Overview}",
    eprint = "1012.3982",
    archivePrefix = "arXiv",
    primaryClass = "hep-th",
    reportNumber = "AEI-2010-175, CERN-PH-TH-2010-306, HU-EP-10-87, HU-MATH-2010-22, KCL-MTH-10-10, UMTG-270, UUITP-41-10",
    doi = "10.1007/s11005-011-0529-2",
    journal = "Lett. Math. Phys.",
    volume = "99",
    pages = "3--32",
    year = "2012"
}

@article{Atiyah:1984px,
    author = "Atiyah, M. F. and Bott, R.",
    title = "{The moment map and equivariant cohomology}",
    doi = "10.1016/0040-9383(84)90021-1",
    journal = "Topology",
    volume = "23",
    pages = "1--28",
    year = "1984"
}

@article{Eniceicu:2022xvk,
    author = "Eniceicu, Dan Stefan and Mahajan, Raghu and Maity, Pronobesh and Murdia, Chitraang and Sen, Ashoke",
    title = "{The ZZ annulus one-point function in non-critical string theory: A string field theory analysis}",
    eprint = "2210.11473",
    archivePrefix = "arXiv",
    primaryClass = "hep-th",
    doi = "10.1007/JHEP12(2022)151",
    journal = "JHEP",
    volume = "12",
    pages = "151",
    year = "2022"
}

@article{Kutasov:2004fg,
    author = "Kutasov, David and Okuyama, Kazumi and Park, Jong-won and Seiberg, Nathan and Shih, David",
    title = "{Annulus amplitudes and ZZ branes in minimal string theory}",
    eprint = "hep-th/0406030",
    archivePrefix = "arXiv",
    reportNumber = "EFI-04-19, PUPT-2121",
    doi = "10.1088/1126-6708/2004/08/026",
    journal = "JHEP",
    volume = "08",
    pages = "026",
    year = "2004"
}

@article{Collier:2026pxi,
    author = "Collier, Scott and Eberhardt, Lorenz and Rodriguez, Victor A.",
    title = "{$c=1$ strings as a matrix integral}",
    eprint = "2604.06301",
    archivePrefix = "arXiv",
    primaryClass = "hep-th",
    month = "4",
    year = "2026"
}

@article{Knapp:2020oba,
    author = "Knapp, Johanna and Romo, Mauricio and Scheidegger, Emanuel",
    title = "{D-Brane Central Charge and Landau{\textendash}Ginzburg Orbifolds}",
    eprint = "2003.00182",
    archivePrefix = "arXiv",
    primaryClass = "hep-th",
    doi = "10.1007/s00220-021-04042-w",
    journal = "Commun. Math. Phys.",
    volume = "384",
    number = "1",
    pages = "609--697",
    year = "2021"
}

@article{Eynard:2007kz,
    author = "Eynard, Bertrand and Orantin, Nicolas",
    title = "{Invariants of algebraic curves and topological expansion}",
    eprint = "math-ph/0702045",
    archivePrefix = "arXiv",
    reportNumber = "SPHT-07-021",
    doi = "10.4310/CNTP.2007.v1.n2.a4",
    journal = "Commun. Num. Theor. Phys.",
    volume = "1",
    pages = "347--452",
    year = "2007"
}

@article{Chekhov:2006cb,
    author = "Chekhov, L.",
    title = "{Matrix models with hard walls: Geometry and solutions}",
    eprint = "hep-th/0602013",
    archivePrefix = "arXiv",
    reportNumber = "ITEP-TH-27-05",
    doi = "10.1088/0305-4470/39/28/S06",
    journal = "J. Phys. A",
    volume = "39",
    pages = "8857--8894",
    year = "2006"
}

@article{Guffin:2008pi,
    author = "Guffin, Josh and Sharpe, Eric",
    title = "{A-twisted heterotic Landau-Ginzburg models}",
    eprint = "0801.3955",
    archivePrefix = "arXiv",
    primaryClass = "hep-th",
    reportNumber = "VPI-IPNAS-08-03, ILL-TH-08-2",
    doi = "10.1016/j.geomphys.2009.07.013",
    journal = "J. Geom. Phys.",
    volume = "59",
    pages = "1581--1596",
    year = "2009"
}

@article{Closset:2015rna,
    author = "Closset, Cyril and Cremonesi, Stefano and Park, Daniel S.",
    title = "{The equivariant A-twist and gauged linear sigma models on the two-sphere}",
    eprint = "1504.06308",
    archivePrefix = "arXiv",
    primaryClass = "hep-th",
    doi = "10.1007/JHEP06(2015)076",
    journal = "JHEP",
    volume = "06",
    pages = "076",
    year = "2015"
}

@article{Garavuso:2022nac,
    author = "Garavuso, Richard S.",
    title = "{R-Symmetries and Curvature Constraints in A-Twisted Heterotic Landau{\textendash}Ginzburg Models}",
    eprint = "2205.15533",
    archivePrefix = "arXiv",
    primaryClass = "hep-th",
    doi = "10.3390/particles6030047",
    journal = "Particles",
    volume = "6",
    number = "3",
    pages = "746--761",
    year = "2023"
}

@article{Baldino:2022aqm,
    author = "Baldino, Salvatore and Schiappa, Ricardo and Schwick, Maximilian and Vega, Roberto",
    title = "{Resurgent Stokes data for Painlev{\'e} equations and two-dimensional quantum (super) gravity}",
    eprint = "2203.13726",
    archivePrefix = "arXiv",
    primaryClass = "hep-th",
    doi = "10.4310/CNTP.2023.v17.n2.a5",
    journal = "Commun. Num. Theor. Phys.",
    volume = "17",
    number = "2",
    pages = "385--552",
    year = "2023"
}

@article{Aniceto:2018bis,
    author = "Aniceto, In{\^e}s and Basar, Gokce and Schiappa, Ricardo",
    title = "{A Primer on Resurgent Transseries and Their Asymptotics}",
    eprint = "1802.10441",
    archivePrefix = "arXiv",
    primaryClass = "hep-th",
    reportNumber = "NSF-ITP-17-153",
    doi = "10.1016/j.physrep.2019.02.003",
    journal = "Phys. Rept.",
    volume = "809",
    pages = "1--135",
    year = "2019"
}

@article{Chekhov:2005rr,
    author = "Chekhov, L. and Eynard, B.",
    title = "{Hermitean matrix model free energy: Feynman graph technique for all genera}",
    eprint = "hep-th/0504116",
    archivePrefix = "arXiv",
    reportNumber = "SPHT-T05-046",
    doi = "10.1088/1126-6708/2006/03/014",
    journal = "JHEP",
    volume = "03",
    pages = "014",
    year = "2006"
}

@article{Bertoldi:2006gk,
    author = "Bertoldi, Gaetano",
    title = "{Double scaling limits and twisted non-critical superstrings}",
    eprint = "hep-th/0603075",
    archivePrefix = "arXiv",
    reportNumber = "SWAT-06-458",
    doi = "10.1088/1126-6708/2006/07/006",
    journal = "JHEP",
    volume = "07",
    pages = "006",
    year = "2006"
}

@article{Borot:2024zwn,
    author = {Borot, Ga{\"e}tan and Giacchetto, Alessandro and Umer, Giacomo},
    title = "{Symmetries of F-Cohomological Field Theories and F-Topological Recursion}",
    eprint = "2406.06304",
    archivePrefix = "arXiv",
    primaryClass = "math-ph",
    doi = "10.1007/s00220-025-05347-w",
    journal = "Commun. Math. Phys.",
    volume = "406",
    number = "10",
    pages = "248",
    year = "2025"
}

@article{Bertoldi:2006eh,
    author = "Bertoldi, Gaetano and Hollowood, Timothy J. and Miramontes, J. Luis",
    title = "{Double scaling limits in gauge theories and matrix models}",
    eprint = "hep-th/0603122",
    archivePrefix = "arXiv",
    reportNumber = "SWAT-06-459",
    doi = "10.1088/1126-6708/2006/06/045",
    journal = "JHEP",
    volume = "06",
    pages = "045",
    year = "2006"
}

@article{Labastida:1994ss,
    author = "Labastida, J. M. F. and Marino, M.",
    title = "{Type B topological matter, Kodaira-Spencer theory, and mirror symmetry}",
    eprint = "hep-th/9405151",
    archivePrefix = "arXiv",
    reportNumber = "US-FT-7-94",
    doi = "10.1016/0370-2693(94)90158-9",
    journal = "Phys. Lett. B",
    volume = "333",
    pages = "386--395",
    year = "1994"
}

@article{Dubrovin:1994hc,
    author = "Dubrovin, Boris",
    editor = "Francaviglia, M. and Greco, S.",
    title = "{Geometry of 2-D topological field theories}",
    eprint = "hep-th/9407018",
    archivePrefix = "arXiv",
    reportNumber = "SISSA-89-94-FM",
    doi = "10.1007/BFb0094793",
    journal = "Lect. Notes Math.",
    volume = "1620",
    pages = "120--348",
    year = "1996"
}

@article{McGreevy:2003kb,
    author = "McGreevy, John and Verlinde, Herman L.",
    title = "{Strings from tachyons: The c=1 matrix reloaded}",
    eprint = "hep-th/0304224",
    archivePrefix = "arXiv",
    reportNumber = "PUPT-2083",
    doi = "10.1088/1126-6708/2003/12/054",
    journal = "JHEP",
    volume = "12",
    pages = "054",
    year = "2003"
}

@article{Johnson:2021owr,
    author = "Johnson, Clifford V. and Rosso, Felipe and Svesko, Andrew",
    title = "{Jackiw-Teitelboim supergravity as a double-cut matrix model}",
    eprint = "2102.02227",
    archivePrefix = "arXiv",
    primaryClass = "hep-th",
    doi = "10.1103/PhysRevD.104.086019",
    journal = "Phys. Rev. D",
    volume = "104",
    number = "8",
    pages = "086019",
    year = "2021"
}

@article{Gaberdiel:2020ycd,
    author = "Gaberdiel, Matthias R. and Gopakumar, Rajesh and Knighton, Bob and Maity, Pronobesh",
    title = "{From symmetric product CFTs to AdS$_{3}$}",
    eprint = "2011.10038",
    archivePrefix = "arXiv",
    primaryClass = "hep-th",
    doi = "10.1007/JHEP05(2021)073",
    journal = "JHEP",
    volume = "05",
    pages = "073",
    year = "2021"
}

@article{Maity:2021wsi,
    author = "Maity, Pronobesh",
    title = "{Scattering and Strebel graphs}",
    eprint = "2108.09458",
    archivePrefix = "arXiv",
    primaryClass = "hep-th",
    doi = "10.21468/SciPostPhys.13.1.010",
    journal = "SciPost Phys.",
    volume = "13",
    number = "1",
    pages = "010",
    year = "2022"
}

@article{Gaiotto:2005gd,
    author = "Gaiotto, Davide",
    title = "{Long strings condensation and FZZT branes}",
    eprint = "hep-th/0503215",
    archivePrefix = "arXiv",
    reportNumber = "HUTP-05-A0014",
    month = "3",
    year = "2005"
}

@article{Ishibashi:2005zf,
    author = "Ishibashi, Nobuyuki and Kuroki, Tsunehide and Yamaguchi, Atsushi",
    title = "{Universality of nonperturbative effects in c{\ensuremath{<}}1 noncritical string theory}",
    eprint = "hep-th/0507263",
    archivePrefix = "arXiv",
    reportNumber = "UTHEP-508, RIKEN-TH-51, KEK-TH-1031",
    doi = "10.1088/1126-6708/2005/09/043",
    journal = "JHEP",
    volume = "09",
    pages = "043",
    year = "2005"
}

@article{Eniceicu:2022nay,
    author = "Eniceicu, Dan Stefan and Mahajan, Raghu and Murdia, Chitraang and Sen, Ashoke",
    title = "{Normalization of ZZ instanton amplitudes in minimal string theory}",
    eprint = "2202.03448",
    archivePrefix = "arXiv",
    primaryClass = "hep-th",
    doi = "10.1007/JHEP07(2022)139",
    journal = "JHEP",
    volume = "07",
    pages = "139",
    year = "2022"
}

@article{Okuyama:2021eju,
    author = "Okuyama, Kazumi and Sakai, Kazuhiro",
    title = "{FZZT branes in JT gravity and topological gravity}",
    eprint = "2108.03876",
    archivePrefix = "arXiv",
    primaryClass = "hep-th",
    doi = "10.1007/JHEP09(2021)191",
    journal = "JHEP",
    volume = "09",
    pages = "191",
    year = "2021"
}

@article{Marino:2008vx,
    author = "Marino, Marcos and Schiappa, Ricardo and Weiss, Marlene",
    title = "{Multi-Instantons and Multi-Cuts}",
    eprint = "0809.2619",
    archivePrefix = "arXiv",
    primaryClass = "hep-th",
    reportNumber = "CERN-PH-TH-2008-186",
    doi = "10.1063/1.3097755",
    journal = "J. Math. Phys.",
    volume = "50",
    pages = "052301",
    year = "2009"
}

@article{Chakrabhavi:2024szk,
    author = "Chakrabhavi, Vivek and Eniceicu, Dan Stefan and Mahajan, Raghu and Murdia, Chitraang",
    title = "{Normalization of ZZ instanton amplitudes in type 0B minimal superstring theory}",
    eprint = "2406.16867",
    archivePrefix = "arXiv",
    primaryClass = "hep-th",
    doi = "10.1007/JHEP09(2024)114",
    journal = "JHEP",
    volume = "09",
    pages = "114",
    year = "2024"
}

@article{Guffin:2008kt,
    author = "Guffin, Josh and Sharpe, Eric",
    title = "{A-twisted Landau-Ginzburg models}",
    eprint = "0801.3836",
    archivePrefix = "arXiv",
    primaryClass = "hep-th",
    reportNumber = "VPI-IPNAS-08-01, ILL-TH-08-1",
    doi = "10.1016/j.geomphys.2009.07.014",
    journal = "J. Geom. Phys.",
    volume = "59",
    pages = "1547--1580",
    year = "2009"
}

@article{Dijkgraaf:1991qh,
    author = "Dijkgraaf, Robbert",
    editor = "Froehlich, J. and 't Hooft, Gerard and Jaffe, A. and Mack, G. and Mitter, P. K. and Stora, R.",
    title = "{Intersection theory, integrable hierarchies and topological field theory}",
    eprint = "hep-th/9201003",
    archivePrefix = "arXiv",
    reportNumber = "IASSNS-HEP-91-91",
    journal = "NATO Sci. Ser. B",
    volume = "295",
    pages = "95--158",
    year = "1992"
}

@article{Lazaroiu:2003zi,
    author = "Lazaroiu, C. I.",
    title = "{On the boundary coupling of topological Landau-Ginzburg models}",
    eprint = "hep-th/0312286",
    archivePrefix = "arXiv",
    reportNumber = "MAD-TH-03-6",
    doi = "10.1088/1126-6708/2005/05/037",
    journal = "JHEP",
    volume = "05",
    pages = "037",
    year = "2005"
}

@article{Kapustin:2003ga,
    author = "Kapustin, Anton and Li, Yi",
    title = "{Topological correlators in Landau-Ginzburg models with boundaries}",
    eprint = "hep-th/0305136",
    archivePrefix = "arXiv",
    doi = "10.4310/ATMP.2003.v7.n4.a5",
    journal = "Adv. Theor. Math. Phys.",
    volume = "7",
    number = "4",
    pages = "727--749",
    year = "2003"
}

@article{Morozov:1994hh,
    author = "Morozov, A.",
    title = "{Integrability and matrix models}",
    eprint = "hep-th/9303139",
    archivePrefix = "arXiv",
    doi = "10.1070/PU1994v037n01ABEH000001",
    journal = "Phys. Usp.",
    volume = "37",
    pages = "1--55",
    year = "1994"
}

@article{Gholson:2018zrl,
    author = "Gholson, Sean M. and Melnikov, Ilarion V.",
    title = "{Small Landau-Ginzburg theories}",
    eprint = "1811.06105",
    archivePrefix = "arXiv",
    primaryClass = "hep-th",
    doi = "10.1007/JHEP04(2019)132",
    journal = "JHEP",
    volume = "04",
    pages = "132",
    year = "2019"
}

@article{Gu:2020nub,
    author = "Gu, Wei",
    title = "{Correlation functions in massive Landau-Ginzburg orbifolds and tests of dualities}",
    eprint = "2001.10562",
    archivePrefix = "arXiv",
    primaryClass = "hep-th",
    doi = "10.1007/JHEP12(2020)180",
    journal = "JHEP",
    volume = "12",
    pages = "180",
    year = "2020"
}

@article{Vafa:1990mu,
    author = "Vafa, Cumrun",
    title = "{Topological Landau-Ginzburg models}",
    reportNumber = "HUTP-90-A064",
    doi = "10.1142/S0217732391000324",
    journal = "Mod. Phys. Lett. A",
    volume = "6",
    pages = "337--346",
    year = "1991"
}

@article{Aharony:1999ti,
    author = "Aharony, Ofer and Gubser, Steven S. and Maldacena, Juan Martin and Ooguri, Hirosi and Oz, Yaron",
    title = "{Large N field theories, string theory and gravity}",
    eprint = "hep-th/9905111",
    archivePrefix = "arXiv",
    reportNumber = "CERN-TH-99-122, HUTP-99-A027, LBNL-43113, RU-99-18, UCB-PTH-99-16, LBL-43113",
    doi = "10.1016/S0370-1573(99)00083-6",
    journal = "Phys. Rept.",
    volume = "323",
    pages = "183--386",
    year = "2000"
}

@article{Szabo:1996md,
    author = "Szabo, Richard J.",
    title = "{Equivariant localization of path integrals}",
    eprint = "hep-th/9608068",
    archivePrefix = "arXiv",
    reportNumber = "OUTP-96-51-P",
    month = "8",
    year = "1996"
}

@article{Cordes:1994fc,
    author = "Cordes, Stefan and Moore, Gregory W. and Ramgoolam, Sanjaye",
    title = "{Lectures on 2-d Yang-Mills theory, equivariant cohomology and topological field theories}",
    eprint = "hep-th/9411210",
    archivePrefix = "arXiv",
    reportNumber = "YCTP-P11-94",
    doi = "10.1016/0920-5632(95)00434-B",
    journal = "Nucl. Phys. B Proc. Suppl.",
    volume = "41",
    pages = "184--244",
    year = "1995"
}

@article{Vonk:2005yv,
    author = "Vonk, Marcel",
    title = "{A Mini-course on topological strings}",
    eprint = "hep-th/0504147",
    archivePrefix = "arXiv",
    reportNumber = "UUITP-06-05",
    month = "4",
    year = "2005"
}

@article{Jiang:2022awu,
    author = "Jiang, Shuhan",
    title = "{Mathematical structures of cohomological field theories}",
    eprint = "2202.12425",
    archivePrefix = "arXiv",
    primaryClass = "math-ph",
    doi = "10.1016/j.geomphys.2022.104744",
    journal = "J. Geom. Phys.",
    volume = "185",
    pages = "104744",
    year = "2023"
}

@article{Jiang:2024nnk,
    author = {Jiang, Shuhan and Jost, J{\"u}rgen},
    title = "{Cohomological field theories and generalized Seiberg-Witten equations}",
    eprint = "2407.04019",
    archivePrefix = "arXiv",
    primaryClass = "math-ph",
    month = "7",
    year = "2024"
}

@inproceedings{Moore:2025tmt,
    author = "Moore, Gregory W. and Saxena, Vivek and Freed, with an appendix by Daniel S.",
    title = "{TASI Lectures On Topological Field Theories And Differential Cohomology}",
    booktitle = "{Theoretical Advanced Study Institute in Elementary Particle Physics 2023}: {Aspects of Symmetry}",
    eprint = "2510.07408",
    archivePrefix = "arXiv",
    primaryClass = "hep-th",
    reportNumber = "YITP-SB-2025-04",
    month = "10",
    year = "2025"
}

@article{Witten:1990bs,
    author = "Witten, Edward",
    title = "{Introduction to cohomological field theories}",
    reportNumber = "IASSNS-HEP-90-57",
    doi = "10.1142/S0217751X91001350",
    journal = "Int. J. Mod. Phys. A",
    volume = "6",
    pages = "2775--2792",
    year = "1991"
}

@ARTICLE{2017arXiv171202528P,
       author = {{Pandharipande}, Rahul},
        title = "{Cohomological field theory calculations}",
      journal = {arXiv e-prints},
         year = 2017,
        month = dec,
          eid = {arXiv:1712.02528},
        pages = {arXiv:1712.02528},
          doi = {10.48550/arXiv.1712.02528},
archivePrefix = {arXiv},
       eprint = {1712.02528},
 primaryClass = {math.AG},
       adsurl = {https://ui.adsabs.harvard.edu/abs/2017arXiv171202528P}
}

@article{Eynard:2015aea,
    author = "Eynard, Bertrand and Kimura, Taro and Ribault, Sylvain",
    title = "{Random matrices}",
    eprint = "1510.04430",
    archivePrefix = "arXiv",
    primaryClass = "math-ph",
    month = "10",
    year = "2015"
}

@article{Witten:1998qj,
    author = "Witten, Edward",
    title = "{Anti-de Sitter space and holography}",
    eprint = "hep-th/9802150",
    archivePrefix = "arXiv",
    reportNumber = "IASSNS-HEP-98-15",
    doi = "10.4310/ATMP.1998.v2.n2.a2",
    journal = "Adv. Theor. Math. Phys.",
    volume = "2",
    pages = "253--291",
    year = "1998"
}

@article{Maldacena:1997re,
    author = "Maldacena, Juan Martin",
    title = "{The Large N limit of superconformal field theories and supergravity}",
    eprint = "hep-th/9711200",
    archivePrefix = "arXiv",
    reportNumber = "HUTP-97-A097, HUTP-98-A097",
    doi = "10.4310/ATMP.1998.v2.n2.a1",
    journal = "Adv. Theor. Math. Phys.",
    volume = "2",
    pages = "231--252",
    year = "1998"
}

@article{Collier:2023cyw,
    author = "Collier, Scott and Eberhardt, Lorenz and Muehlmann, Beatrix and Rodriguez, Victor A.",
    title = "{The Virasoro minimal string}",
    eprint = "2309.10846",
    archivePrefix = "arXiv",
    primaryClass = "hep-th",
    doi = "10.21468/SciPostPhys.16.2.057",
    journal = "SciPost Phys.",
    volume = "16",
    number = "2",
    pages = "057",
    year = "2024"
}

@article{Eberhardt:2026hfh,
    author = "Eberhardt, Lorenz",
    title = "{The Super Virasoro Minimal String from 3d Supergravity}",
    eprint = "2604.26038",
    archivePrefix = "arXiv",
    primaryClass = "hep-th",
    month = "4",
    year = "2026"
}

@article{Akemann:1995im,
    author = "Akemann, G.",
    title = "{Loop equations for multicut matrix models}",
    eprint = "hep-th/9503185",
    archivePrefix = "arXiv",
    reportNumber = "ITP-UH-11-95, DESY-95-066",
    month = "3",
    year = "1995"
}

@article{tHooft:1973alw,
    author = "'t Hooft, Gerard",
    editor = "Taylor, J. C.",
    title = "{A Planar Diagram Theory for Strong Interactions}",
    reportNumber = "CERN-TH-1786",
    doi = "10.1016/0550-3213(74)90154-0",
    journal = "Nucl. Phys. B",
    volume = "72",
    pages = "461",
    year = "1974"
}

@article{Gopakumar:1998ki,
    author = "Gopakumar, Rajesh and Vafa, Cumrun",
    editor = "Vafa, Cumrun and Yau, S. -T.",
    title = "{On the gauge theory / geometry correspondence}",
    eprint = "hep-th/9811131",
    archivePrefix = "arXiv",
    reportNumber = "HUTP-98-A078",
    doi = "10.4310/ATMP.1999.v3.n5.a5",
    journal = "Adv. Theor. Math. Phys.",
    volume = "3",
    pages = "1415--1443",
    year = "1999"
}

@article{Gopakumar:2011ev,
    author = "Gopakumar, Rajesh",
    title = "{What is the Simplest Gauge-String Duality?}",
    eprint = "1104.2386",
    archivePrefix = "arXiv",
    primaryClass = "hep-th",
    month = "4",
    year = "2011"
}

@article{Gopakumar:2012ny,
    author = "Gopakumar, Rajesh and Pius, Roji",
    title = "{Correlators in the Simplest Gauge-String Duality}",
    eprint = "1212.1236",
    archivePrefix = "arXiv",
    primaryClass = "hep-th",
    doi = "10.1007/JHEP03(2013)175",
    journal = "JHEP",
    volume = "03",
    pages = "175",
    year = "2013"
}

@article{Gopakumar:2022djw,
    author = "Gopakumar, Rajesh and Mazenc, Edward A.",
    title = "{Deriving the Simplest Gauge-String Duality -- I: Open-Closed-Open Triality}",
    eprint = "2212.05999",
    archivePrefix = "arXiv",
    primaryClass = "hep-th",
    month = "12",
    year = "2022"
}

@article{Gopakumar:2024jfq,
    author = "Gopakumar, Rajesh and Kaushik, Rishabh and Komatsu, Shota and Mazenc, Edward A. and Sarkar, Debmalya",
    title = "{Strings from Feynman diagrams}",
    eprint = "2412.13397",
    archivePrefix = "arXiv",
    primaryClass = "hep-th",
    doi = "10.1007/JHEP05(2026)054",
    journal = "JHEP",
    volume = "05",
    pages = "054",
    year = "2026"
}

@article{Mazenc:2026yrm,
    author = "Mazenc, Edward A. and Sarkar, Debmalya",
    title = "{Open-Closed-Open Triality Beyond Matrix Models}",
    eprint = "2605.02885",
    archivePrefix = "arXiv",
    primaryClass = "hep-th",
    month = "5",
    year = "2026"
}

@article{Giacchetto:2026gpq,
    author = "Giacchetto, Alessandro and Gopakumar, Rajesh and Mazenc, Edward A.",
    title = "{Worldsheet Duals to One-Matrix Models}",
    eprint = "2604.03126",
    archivePrefix = "arXiv",
    primaryClass = "hep-th",
    month = "4",
    year = "2026"
}

@article{Rangamani:2025wfa,
    author = "Rangamani, Mukund and Zheng, Jianming",
    title = "{Towards the super Virasoro minimal string}",
    eprint = "2505.08892",
    archivePrefix = "arXiv",
    primaryClass = "hep-th",
    doi = "10.1007/JHEP09(2025)061",
    journal = "JHEP",
    volume = "09",
    pages = "061",
    year = "2025"
}

@article{Johnson:2026ugu,
    author = "Johnson, V., Clifford",
    title = "{Supersymmetric Virasoro Minimal Strings: Unified Perturbative and Non-Perturbative Treatment}",
    eprint = "2608.20529",
    archivePrefix = "arXiv",
    primaryClass = "hep-th",
    month = "8",
    year = "2026"
}
\bibliographystyle{utphysmodb}
\end{document}